\documentclass[12pt]{article}
\usepackage{latexsym}
\usepackage{amsmath}
\usepackage{amssymb}
\usepackage{epsfig,graphics}
\usepackage{booktabs}
\usepackage{multirow}

\newcommand{\be}{\begin{equation}}
\newcommand{\ee}{\end{equation}}
\newcommand{\otoprule}{\midrule[\heavyrulewidth]}

\begin{document}

\begin{titlepage}
\begin{flushright} DESY 11-045\end{flushright}

\vspace*{0.45in}
 
\begin{center}
{\large\bf Closed flux tubes and their string description\\
in D=2+1 SU(N) gauge theories\\}
\vspace*{0.60in}
{Andreas Athenodorou$^{a}$, Barak Bringoltz$^{b}$  and Michael Teper$^{c}$\\
\vspace*{.2in}
$^{a}$DESY, Platanenallee 6, 15738 Zeuthen, Germany \\
\vspace*{.1in}
$^{b}$IIAR, The Israeli Institute for Advanced Research, Rehovot, Israel\\
\vspace*{.1in}
$^{c}$Rudolf Peierls Centre for Theoretical Physics, University of Oxford,\\
1 Keble Road, Oxford OX1 3NP, UK
}
\end{center}

\vspace*{0.3in}

\begin{center}
{\bf Abstract}
\end{center}

We carry out lattice calculations of the spectrum of confining 
flux tubes that wind around a spatial torus of variable length 
$l$, in 2+1 dimensions. We compare the energies of the lowest $\sim 30$ 
states to the free string Nambu-Goto model and to recent results on 
the universal properties of effective string actions. Our most useful
calculations are in SU(6) at a small lattice spacing, which we check 
is very close to the $N\to \infty$ continuum limit. We find that
the energies, $E_n(l)$, are remarkably close to the predictions of the 
free string Nambu-Goto model, even well below the critical length at which 
the expansion of the Nambu-Goto energy in powers of $1/l^2$ diverges 
and the series needs 
to be resummed. Our analysis of the ground state supports the universality
of the $O(1/l)$ and the $O(1/l^3)$ corrections to $\sigma l$, and we find 
that the deviations from Nambu-Goto at small $l$ prefer a leading correction 
that is $O(1/l^7)$, consistent with theoretical expectations. We find
that the low-lying states that contain a single phonon excitation are also 
consistent with the leading $O(1/l^7)$ correction dominating down to the 
smallest values of $l$. By contrast our analysis of the other light 
excited states clearly shows that for these states the 
corrections at smaller $l$ resum to a much smaller effective power.
Finally, and in contrast to our recent
calculations in $D=3+1$, we find no evidence for the presence 
of any non-stringy states that could indicate the excitation 
of massive flux tube modes.

\vspace*{0.45in}

\leftline{{\it E-mail:} andreas.athenodorou@desy.de,  barak.bringoltz@gmail.com, m.teper1@physics.ox.ac.uk}

\end{titlepage}

\setcounter{page}{1}
\newpage
\pagestyle{plain}

\tableofcontents

\section{Introduction}
\label{section_intro}

In this paper we calculate the energy spectrum of closed flux tubes 
in $D=2+1$ SU($N$) lattice gauge theories.
These flux tubes are stabilised by being wound around a spatial
torus and we calculate the energies of the lightest few eigenstates as 
a function of the flux tube length $l$, for various quantum numbers.
This work greatly extends and supersedes that published in our earlier 
brief letter
\cite{AABBMT_d3k1},
and is part of a larger project which has included the calculation
of the spectrum in $D=3+1$
\cite{AABBMT_d4k1},
as well as the spectrum and string tensions
\cite{AABBMT_d3k}
of flux tubes with flux in some higher representations, e.g. $k$-strings.
The most significant new calculations in this paper are at larger $N$, 
i.e. SU(6), and at a small lattice spacing, $a \simeq 0.086/\surd\sigma$, 
where $\sigma$ is the string tension. The main purpose of these   
calculations is to learn about the effective string theory that 
describes closed flux tubes at $N=\infty$, and possibly at smaller
$N$ as well. The details of the analysis in our earlier $D=2+1$  calculation
\cite{AABBMT_d3k1}
have been rendered out of date by a great deal of recent 
analytic progress 
\cite{OA}
towards determining the universal terms in the derivative expansion 
of this string action, which makes new predictions for the low-lying 
spectrum of long flux tubes, $l\surd\sigma \gg 1$. Our lattice 
calculations are largely complementary in that they concentrate on flux 
tubes that range from the very short to the moderately long, i.e. 
$l\surd\sigma \sim 1$ to $\sim 6$. So together with the analytic work 
they may tell us something about the effective string action over 
the whole range of $l$. 

In the next Section we begin with some general remarks about closed 
flux tubes in $D=2+1$, describe their quantum numbers, and how they 
differ from those in $D=3+1$. We describe in some detail
the spectrum of the free string theory (Nambu-Goto in
flat space time) since the most striking result of our earlier lattice
calculations is how close the actual spectrum is to this Nambu-Goto spectrum,
even for very small values of $l$ where the flux tube is hardly longer than 
it is wide and naively should `look' nothing like an ideal thin string. 
We then give a brief summary of the current status of the analytic
study of the effective string action, and point to some very 
recent lattice and analytic calculations relevant to our work. In 
Section~\ref{section_background} we describe some details of 
our lattice calculation of the spectrum, with the focus on the operators 
we use and how well we control our systematic errors. We briefly discuss 
the large-$N$ limit and show how our calculations of the string tension 
provide rather precise evidence for the conventional large-$N$ counting. 
In Section~\ref{section_results} we present and analyse our numerical 
results for the spectrum. We start with the absolute ground state
and then move on to the excited states. We perform detailed fits to
see how far we can confirm the established universality results, and what 
we can learn about the corrections to the universal terms at smaller $l$. 
We summarise and conclude in Section~\ref{section_conclusion}. Finally
we list in an Appendix the energy eigenvalues from our new SU(6) 
calculation, so as to allow the interested reader to extend the present 
analysis as further theoretical progress is made. 

We have kept the discussion in this paper relatively brief, since
most of the relevant issues are discussed at greater length in our 
recent paper on the flux tube spectrum in $D=3+1$
\cite{AABBMT_d4k1}
and in a recent set of lectures by one of the authors
\cite{MT_Krakow},
to which we refer the interested reader.

\section{Flux tubes and strings}
\label{section_tubesandstrings}

We begin with some general comments about closed flux tubes in 2+1 
dimensions. We then describe in detail the spectrum in the free string 
theory, as given by the Nambu-Goto action in flat space-time, 
since this turns out to describe the numerically determined
flux tube spectrum remarkably well. We then briefly summarise 
recent analytic progress on the form of the effective
string action describing very long flux tubes. We also point to
some lattice and analytic work that has appeared since our
earlier papers, and discuss how the new results in this paper modify the 
conclusions of our earlier work. For a more complete but slightly less 
up-to-date discussion of many of these topics we refer the reader to
\cite{AABBMT_d4k1,MT_Krakow}.

\subsection{Closed flux tubes in D=2+1}
\label{subsection_tubesd3}

We assume that we are in the confining phase of the gauge theory. In this 
phase a closed flux tube carrying fundamental flux cannot break, but it 
can contract. To stabilise such a flux tube at a given (minimal) length
$l$, we make the $x$ direction periodic with period $l$ and we close the 
flux tube around this spatial torus. In our lattice calculations the other 
Euclidean directions will also be periodic tori, but these will be chosen 
large enough that they are effectively infinite.
Such a winding flux tube will have a spectrum of states, which is a 
function of its length $l$, and it is this that we wish to 
calculate numerically. In this paper we will be able to calculate 
the energies of $O(30)$ of the lighter states of the spectrum.

One naively expects the flux tube to have some `intrinsic' width
which is $\sim 1/\surd\sigma$. For a very long flux tube,
$l\gg 1/\surd\sigma$, the flux tube should appear string-like
and the low-lying excitations should be the massless modes along the
string that describe its transverse fluctuations. 
These are quantised, by the periodicity of the flux tube,
to have momenta and energies $k\pi/l$, with $k$ an integer. 
(This is just the Goldstone mode arising from the spontaneous breaking 
of the translation invariance transverse to the flux tube, with discrete
rather than continuous momenta.) Thus the energies of the lightest 
excited states, $E_i(l)$, will converge to the absolute ground state 
energy, $E_0(l)$, at large $l$:
\begin{equation}
E_i(l)  \stackrel{l\to\infty}{=} E_0(l) + O\left(\frac{\pi}{l}\right).
\label{eqn_masslessmode}
\end{equation}
If, on the other hand, we excite a massive mode, e.g. one associated
with the intrinsic width of the flux tube, then we would expect a 
finite gap above the ground state:
\begin{equation}
E_j(l)  = E_0(l) + O(\surd\sigma).
\label{eqn_massivemode}
\end{equation}
To easily locate a massive mode excitation it needs to be amongst the lightest 
few states and so we need to be looking at smaller values of $l$ where
$\pi/l \sim O(\surd\sigma)$, and the gaps between the lightest
states are not small.

As we reduce $l$, we eventually
encounter a phase transition to a phase where we no longer have a 
confining flux tube. This occurs at a critical length $l=1/T_c$
where $T_c$ is the deconfining temperature of the gauge theory. 
If we were to view $x$ as our Euclidean 
time coordinate then this would be nothing but the
usual finite temperature deconfining transition. Of course, we
view $x$ as a spatial coordinate, but a change of name 
cannot influence the presence of the phase transition, although it 
does affect how we interpret it. We will loosely refer to it
as a finite-volume deconfining transition, although it is in 
fact only deconfining in the $(x,t)$ plane: Wilson loops in
the $(y,t)$ plane continue to display an area law (just like the
`spatial' Wilson loops in the usual deconfined phase). 
Thus we can discuss the spectrum of our closed flux tubes only
for $l > l_c = 1/T_c$. We recall
\cite{Tcd3}
that $T_c \sim \surd\sigma$ for $D=2+1$ SU($N$) gauge theories,
so this lower bound on $l$ is $l_c\surd\sigma \sim 1$.

The eigenstates of such a closed flux tube can be labelled by a 
number of quantum numbers. Some of these we will not explore.
For example, we could consider flux in representations 
other than the fundamental, e.g. $k$-strings
\cite{BLMT_d3k,AABBMT_d3k},
but we will not do so here. Again, our flux tube could wind 
around the $x$-torus any number $w$ of times: but we shall restrict
ourselves to $w=1$. It could simultaneously wind around more than
one spatial torus,  but we do not analyse this case. Our flux tube 
could have an arbitrary transverse momentum $p_\perp$, but we expect 
that this would merely lead to $E^2(p_\perp) =  E^2(0) + p^2_\perp$, 
so we will confine ourselves to states with $p_\perp = 0$. 
For $N>2$ we have charge-conjugation, $C$, which reverses the 
direction of the flux. Since a flux tube cannot reverse
the direction of the flux as it evolves in time, states with $C=\pm$
will be degenerate and this quantum number is not interesting for
our purposes. 

The quantum numbers we do explore are as follows.\\
$\bullet$ The longitudinal momentum $p$ along the flux tube, i.e. in
the $x$-direction.
By periodicity this is quantised, $p=2\pi q/l$ where $q$ is an integer.
We expect that the absolute ground state, with energy $E_0(l)$, is
invariant under longitudinal translations, and so must have 
longitudinal momentum $p=0$. To have $p\neq 0$ a flux tube must have 
a deformation so that it is not invariant under longitudinal translations. 
That is to say, it must be excited in some non-trivial way. Thus we do
not simply have $E_0^2(p) = E_0^2(0) + p^2$, and the calculated value of 
$E(p)$ carries non-trivial dynamical information.\\
$\bullet$ The 2 dimensional parity operation $P$: 
$(x,y) \stackrel{P}{\to} (x,-y)$.
We expect that the absolute ground state, with energy $E_0(l)$, 
is invariant under reflection in $y$, and so will have $P=+$ (with the
$P=-$ linear combination being null). The lightest non-null 
$P=-$ state must involve a flux-tube with a non-trivial deformation,
and so $P$ is also an interesting quantum number.\\
$\bullet$ We can consider rotations in the $(x,y)$ plane. Since we
are on a spatial 2-torus we are at most interested in rotations that
are an integer multiple of $\pi/2$. Moreover, since the orthogonal
$y$-torus is effectively infinite, we are only interested in the 
rotation by $\pi$, i.e. $R_\pi$. Amongst other things this will reverse 
the direction of the flux, but this we can undo using charge conjugation.
If we also apply $P$ then all this corresponds to a reflection in $x$,
i.e. $(x,y) {\to} (-x,y)$, followed by $C$. We shall call this
our reflection parity, $P_r$. It clearly  reverses the longitudinal 
momentum, and so is only useful for states with $p=0$. 

The main difference between closed flux tubes in $D=2+1$ and $D=3+1$ 
is that the latter also carry angular momentum. Another difference is
that in $D=2+1$ the deconfining transition is second order for SU(2) and 
SU(3), weakly first order for SU(4), and only becomes robustly first 
order for $N\geq 5$
\cite{Tcd3}, 
whereas in $D=3+1$ it is already first order for SU(3)
\cite{Tcd4}. 
Since the behaviour of flux tubes of length $l$ will be governed
by the critical exponents of the second order transition as $l$ approaches 
$l_c=1/T_c$, and these are given by the universality class of a spin model
in one lower dimension, we need to consider at least $N\geq 4$ or possibly 
$N\geq 5$ if we wish to investigate the large-$N$ stringy behaviour
of flux tubes down to values of $l$ that are close to $l_c$.   
A further, but minor, difference is that the critical 
deconfining length scale is larger, $l_c\surd\sigma \sim 1.5$,
in  $D=3+1$
\cite{Tcd4}
than it is in $D=2+1$
\cite{Tcd3},
where $l_c\surd\sigma \sim 1$. So in  $D=2+1$ we can access 
significantly shorter flux tubes than in $D=3+1$. We also recall
that in $D=2+1$ the coupling, $g^2$, has dimensions of mass. So 
the perturbative expansion parameter on the length scale $l$  will
be $l g^2$. Thus the theory is strongly coupled in the infrared and 
becomes rapidly free in the ultraviolet. 
This also has the consequence that the static potential
is already confining, logarithmically, in perturbation theory.
Indeed it also has a linear perturbative piece at $O(g^4)$,
but the value 
\cite{d3_PT}
of this perturbative `string tension' is not very close to the observed
lattice value
\cite{BBMT_d3k1}.
This is no surprise given that yet higher orders in $g^2$ lead to 
yet higher powers in $l$, which is unphysical
\cite{seiler},
and so this perturbative expression clearly cannot be used once 
$lg^2N > 1$.

At low $N$ the spectrum will be complicated by mixing and decay.
For example, a flux tube can emit and absorb a virtual glueball. 
In terms of the string world sheet swept out by the evolution of a 
flux tube, this means that we have to include surfaces of higher 
genus, with handles on all length scales. An effective string
action for such world sheets is much more challenging 
\cite{PS}
than one for world sheets of minimal topology, with fluctuations only on 
long wavelengths. The latter occurs for long flux tubes at large $N$,
where the glueball emission vertex vanishes, flux tube states do not 
mix and there are no interactions between flux tubes. Thus we will
attempt to calculate the closed flux tube spectrum at large $N$.
In particular, the new calculations described in this paper are for
SU(6) which for our purposes is `close to' $N=\infty$.

\subsection{Nambu-Goto spectrum}
\label{subsection_NG}

The simplest string theory is Nambu-Goto (in flat space-time) which is 
just a theory of free strings. While not consistent in 2+1 or 3+1
dimensions, its anomalies do not appear to affect the spectrum
of long strings (see e.g. 
\cite{Olesen}).
Moreover it is simple enough that the energy spectrum has been long known
\cite{Arvis}.
It is of particular interest to us because, as we have seen in our
earlier work 
\cite{AABBMT_d3k1,AABBMT_d3k},
the flux tube spectrum is described remarkably well by its predictions,
even when the flux tube length $l$ is not much greater than the
minimum, deconfining length $l_c$.
Here we briefly summarise the aspects of the Nambu-Goto spectrum that will
be useful for us in this paper.  

The only degrees of freedom are the massless transverse fluctuations.
Let $h(x,t)$ label the transverse displacement of the string at 
position $x$ and at time $t$ (i.e. we work in `static gauge'). 
We write a Fourier decomposition of these transverse fluctuations 
and then quantise, thus promoting the Fourier coefficients to
creation and annihilation operators. These represent `phonons' running
along the string in the +ve or -ve $x$-direction. We denote by $a_{\pm k}$ 
the creation operator for a phonon of momentum $p=\pm 2\pi k/l$ 
with $k$ a positive integer. (Recall $h(x)$ has periodicity $l$.) The 
energy of the phonon is $\omega=|p|=2\pi k/l$, since the mode is massless. 
The absolute ground state $|0\rangle$ has no phonons, but its energy acquires 
a correction from the zero mode contributions of all these oscillators. 

The spectrum is then as follows. Call the positive momenta left-moving (L) 
and the negative ones right-moving (R). Let $n_{L(R)}(k)$ be the number 
of left(right) moving phonons of momentum $|p|=2\pi k/l$. If we define the 
total energy of the left(right) moving phonons to be $2\pi N_{L(R)}/l$,
then:
\begin{equation}
N_L = \sum_k  n_L(k)k, \qquad
N_R = \sum_k  n_R(k)k.
\label{eqn_NLR}
\end{equation}
If we define $p=2\pi q/l$ to be the total longitudinal momentum of the 
string then, since it is the phonons that provide the momentum, we have
\begin{equation}
N_L - N_R = q.
\label{eqn_NLRmom}
\end{equation}
We can now write down the expression for the energy levels of the 
Nambu-Goto string in $D=2+1$ as
\begin{equation}
E^2_{N_L,N_R}(q,l)
=
(\sigma l)^2 
+
8\pi\sigma \left(\frac{N_L+N_R}{2}-\frac{1}{24}\right)
+
\left(\frac{2\pi q}{l}\right)^2
\label{eqn_EnNG}
\end{equation}
where the $1/24$ term arises from the oscillator zero-point energies. These 
energy levels have, in general, a degeneracy which depends on the number 
of ways the  particular values of $N_L$ and $N_R$ can be formed from 
the $n_L$ and $n_R$ in eqn(\ref{eqn_NLR}).

Under our parity $(x,y) \to (x,-y)$, so $h(x) \to -h(x)$ and $a_k \to -a_k$. 
Thus the parity of a state is simply given by the total number of phonons:
\begin{equation}
P = (-1)^{number\ of\ phonons}.
\label{eqn_Pd3}
\end{equation}
Under $P_r$, the symmetry that combines a reflection in $x$ with charge 
conjugation, the individual phonon momenta are reversed, as is the overall 
momentum. Thus this quantum number is only useful in the $p=0$ sector
and here the lightest non-null pair of states with $P_r=\pm$ is
$\{a_{2}a_{-1}a_{-1} \pm a_{1}a_{1}a_{-2}\}|0\rangle$ and is quite
heavy. In practice this means that this quantum number is of minor utility 
in our calculations and we shall ignore it in the labelling
of our states (but will return to it later).

In Table~\ref{table_NGstates} we list a number of the lightest states of the
Nambu-Goto model, labelling them by their momentum, $p=2\pi q/l$, and 
parity, $P$. Note that in the $q=0$ sector the very lightest states have 
reflection parity $P_r=+$, with the corresponding $P_r=-$ linear combinations 
being null. In the case of the heavier states, with $N_L=N_R \geq 2$, 
some can be paired into non-null linear combinations with  $P_r=\pm$, and
then it is these states that one should compare to the numerically determined 
spectrum. (This only has relevance when analysing corrections to 
Nambu-Goto that split the degeneracy of such an energy level.)

\begin{table}[htp]
\begin{center}
\centering{\scalebox{1.02}{\tiny
\begin{tabular}{c||c|c||c} \otoprule \otoprule 
\ \ \ \ \ \ \ \ \ \ \ \ \ \ \ \ $N_L,N_R$ \ \ \ \ \ \ \ \ \ \ \ \ \ \ \ \ & \ \ \ \ $q$ \ \ \ \ & \ \ \ \ $P$ \ \ \ \  & \ \ \ \ \ \ \ \ \ \ String State \ \ \ \ \ \  \ \ \ \ \\ \otoprule \otoprule 
 $N_L=N_R=0$ & $0$ &  $+$ & $| 0 \rangle$ \\ \midrule 
 $N_L=1,N_R=0$ & $1$ &  $-$ & $a_{1}| 0 \rangle$ \\ \midrule 
$N_L=N_R=1$ & $0$ &  $+$ & $a_{1} a_{-1} | 0 \rangle$ \\ \midrule 
\multirow{2}*{$N_L=2,N_R=0$} & \multirow{2}*{$2$} &  $+$ & $a_{1} {a}_{1} | 0 \rangle$ \\
&  &  $-$ & $a_{2} | 0 \rangle$ \\ \midrule 
\multirow{2}*{$N_L=2,N_R=1$} & \multirow{2}*{$1$} &  $+$ & $a_{2} a_{-1}| 0 \rangle$ \\
&  &  $-$ & $a_{1} a_{1} a_{-1} | 0 \rangle$ \\\midrule
\multirow{3}*{$N_L=3,N_R=0$} & \multirow{3}*{$3$} &  $+$ & $a_{2} a_{1}| 0 \rangle$ \\
&  &  $-$ & $a_{3} | 0 \rangle$ \\
&  &  $-$ & $a_{1} a_{1} a_{1} | 0 \rangle$ \\\midrule
\multirow{4}*{$N_L=N_R=2$} & \multirow{4}*{$0$} &  $+$ & $a_{2} a_{-2} | 0 \rangle$ \\
&  &  $+$ & $a_{1} a_{1} a_{-1} a_{-1} | 0 \rangle$ \\
&  &  $-$ & $a_{2}a_{-1} a_{-1} | 0 \rangle$ \\
&  &  $-$ & $a_{1}a_{1} a_{-2}  | 0 \rangle$ \\\midrule 
\multirow{3}*{$N_L=3,N_R=1$} & \multirow{3}*{$2$} &  $+$ & $a_{3} a_{-1} | 0 \rangle$ \\
&  &  $+$ & $a_{1} a_{1} a_{1} a_{-1} | 0 \rangle$ \\
&  &  $-$ & $a_{2}a_{1} a_{-1} | 0 \rangle$ \\\midrule
\multirow{5}*{$N_L=4,N_R=0$} & \multirow{5}*{$4$} &  $+$ & $a_{3} a_{1} | 0 \rangle$ \\
&  &  $+$ & $a_{2} a_{2} | 0 \rangle$ \\
&  &  $+$ & $a_{1}a_{1} a_{1} a_{1}| 0 \rangle$ \\
&  &  $-$ & $a_{4} | 0 \rangle$ \\
&  &  $-$ & $a_{2}a_{1} a_{1}  | 0 \rangle$ \\\midrule  
\multirow{6}*{$N_L=3,N_R=2$} & \multirow{6}*{$1$} &  $+$ & $a_{3} a_{-2} | 0 \rangle$ \\
&  &  $+$ & $a_{2} a_{1} a_{-1} a_{-1}| 0 \rangle$ \\
&  &  $+$ & $a_{1} a_{1} a_{1} a_{-2} | 0 \rangle$ \\
&  &  $-$ & $a_{3}a_{-1} a_{-1} | 0 \rangle$ \\
&  &  $-$ & $a_{2} a_{1} a_{-2}| 0 \rangle$ \\
&  &  $-$ & $a_{1} a_{1} a_{1} a_{-1} a_{-1}| 0 \rangle$ \\\midrule
\multirow{5}*{$N_L=4,N_R=1$} & \multirow{5}*{$3$} &  $+$ & $a_{4} a_{-1} | 0 \rangle$ \\
&  &  $+$ & $a_{2} a_{1} a_{1} a_{-1}| 0 \rangle$ \\
&  &  $-$ & $a_{3} a_{1} a_{-1} | 0 \rangle$ \\
&  &  $-$ & $a_{2} a_{2} a_{-1} | 0 \rangle$ \\
&  &  $-$ & $a_{1} a_{1} a_{1} a_{1} a_{-1} | 0 \rangle$ \\\midrule
\multirow{7}*{$N_L=5,N_R=0$} & \multirow{7}*{$5$} &  $+$ & $a_{4} a_{1} | 0 \rangle$ \\
&  &  $+$ & $a_{3} a_{2} | 0 \rangle$ \\
&  &  $+$ & $a_{2} a_{1} a_{1} a_{1}| 0 \rangle$ \\
&  &  $-$ & $a_{5} | 0 \rangle$ \\
&  &  $-$ & $a_{3} a_{1} a_{1} | 0 \rangle$ \\
&  &  $-$ & $a_{2} a_{2} a_{1} | 0 \rangle$ \\
&  &  $-$ & $a_{1} a_{1} a_{1} a_{1} a_{1} | 0 \rangle$ \\\midrule
\multirow{9}*{$N_L=3,N_R=3$} & \multirow{9}*{$0$} &  $+$ & $a_{3} a_{-3} | 0 \rangle$ \\
&  &  $+$ & $a_{2} a_{1} a_{-2} a_{-1} | 0 \rangle$ \\
&  &  $+$ & $a_{1} a_{1} a_{1} a_{-1} a_{-1} a_{-1} | 0 \rangle$ \\
&  &  $+$ & $a_{1} a_{1} a_{1} a_{-3} | 0 \rangle$ \\
&  &  $+$ & $a_{3} a_{-1} a_{-1} a_{-1} | 0 \rangle$ \\
&  &  $-$ & $a_{3} a_{-2} a_{-1} | 0 \rangle$ \\
&  &  $-$ & $a_{2} a_{1} a_{-3} | 0 \rangle$ \\
&  &  $-$ & $a_{2} a_{1} a_{-1}  a_{-1}  a_{-1}| 0 \rangle$ \\
&  &  $-$ & $a_{1} a_{1} a_{1} a_{-2} a_{-1} | 0 \rangle$ \\\midrule
\multirow{10}*{$N_L=4,N_R=2$} & \multirow{10}*{$2$} &  $+$ & $a_{4} a_{-2} | 0 \rangle$ \\
&  &  $+$ & $a_{3} a_{1} a_{-1} a_{-1} | 0 \rangle$ \\
&  &  $+$ & $a_{2} a_{2} a_{-1} a_{-1} | 0 \rangle$ \\
&  &  $+$ & $a_{2} a_{1} a_{1} a_{-2} | 0 \rangle$ \\
&  &  $+$ & $a_{1} a_{1} a_{1} a_{1} a_{-1} a_{-1}| 0 \rangle$ \\
&  &  $-$ & $a_{4} a_{-1} a_{-1} | 0 \rangle$ \\
&  &  $-$ & $a_{3} a_{1} a_{-2} | 0 \rangle$ \\
&  &  $-$ & $a_{2} a_{2} a_{-2} | 0 \rangle$ \\
&  &  $-$ & $a_{2} a_{1} a_{1} a_{-1} a_{-1}| 0 \rangle$ \\
&  &  $-$ & $a_{1} a_{1} a_{1} a_{1} a_{-2} | 0 \rangle$ \\\midrule
\multirow{7}*{$N_L=5,N_R=1$} & \multirow{7}*{$4$} &  $+$ & $a_{5} a_{-1} | 0 \rangle$ \\
&  &  $+$ & $a_{3} a_{1} a_{1} a_{-1}| 0 \rangle$ \\
&  &  $+$ & $a_{2} a_{2} a_{1} a_{-1}| 0 \rangle$ \\
&  &  $+$ & $a_{1} a_{1} a_{1} a_{1} a_{1} a_{-1} | 0 \rangle$ \\
&  &  $-$ & $a_{4} a_{1} a_{-1} | 0 \rangle$ \\
&  &  $-$ & $a_{3} a_{2} a_{-1} | 0 \rangle$ \\
&  &  $-$ & $a_{2} a_{1} a_{1} a_{1} a_{-1}| 0 \rangle$ \\\midrule
\end{tabular}}}
\end{center}
\vspace{-0.5cm}
\caption{\label{table_NGstates}
Table with the states of the lowest Nambu-Goto levels with $q=0,1,2, \dots, 5$ and $N_L+N_R \le 6$.}
\end{table}

We note that if we take the square root of both sides of the energy in 
eqn(\ref{eqn_EnNG}), then the resulting expression can be expanded
as a series in $1/\sigma l^2$. Assuming $p=0$ for simplicity, one has
\begin{eqnarray}
E_n(l)
& = &
\sigma l \left(1 + 
\frac{8\pi}{\sigma l^2} 
\left(n -\frac{1}{24}\right)\right)^\frac{1}{2} \nonumber \\
& \stackrel{l\surd\sigma \to \infty}{=} &
\sigma l + \frac{4\pi}{l} 
\left(n -\frac{1}{24}\right) + O\left( \frac{1}{\sigma l^3} \right)
\label{eqn_EnNGexpansion}
\end{eqnarray}
where $n=(N_L+N_R)/2=N_R=N_L$. Here the second term is the universal 
L\"uscher correction
\cite{LSW}. 
We also see from eqn(\ref{eqn_EnNG}) that the ground state, $E_0(l)$, 
becomes tachyonic for $\sigma l^2 < \pi/3$ signalling a change of phase, 
which one might in the present context interpret as a deconfining 
Hagedorn transition. Of course, in the real world the large-$N$
deconfining transition is first order and occurs for $l_c^2\sigma > \pi/3$ 
so such a tachyonic transition does not appear 
for any physically realisable value of the flux tube length, $l$. 
(But see
\cite{BBMT_Hagedorn}.)

\subsection{Effective string action}
\label{subsection_strings}

In this Section we shall begin with a sketch of the current status of 
analytic attempts to determine the effective string action
for closed flux tubes. We shall focus on work directly related to
the subject of this paper. We shall also point to relevant numerical
work that has appeared over the last year or two. For earlier work
we refer the reader to the literature quoted in these papers
and in 
\cite{AABBMT_d4k1,MT_Krakow}. 
Finally we briefly comment on our earlier paper
\cite{AABBMT_d3k1} 
and specifically on those aspects that are superseded by the present
analysis.
 
Consider a flux tube that is wrapped around the $x$-torus and propagates
around the (Euclidean) time torus. It will sweep out a surface that is a simple 
2-torus, at least if we are in the large-$N$ limit where handles and higher 
genus surfaces are suppressed. If we have an effective string action for 
such surfaces, $S_{eff}[S]$, then we can calculate the path integral over 
all such  surfaces, $Z_{torus}(l,\tau)$, where $l$ and $\tau$ are the sizes 
of the $x$ and $t$ tori. This should equal the partition function of the
closed flux tubes in this large-$N$ gauge theory:
\begin{equation}
Z_{torus}(l,\tau) = \int_{T^2=l\times\tau} dS e^{-S_{eff}[S]}
=
\sum_{n,p} e^{-E_n(p,l)\tau}
\label{eqn_Ztorus}
\end{equation}
where $E_n(p,l)$ is the energy of the $n$'th flux tube state of length
$l$ and of momentum $p$ (which now also includes transverse momenta). 
Thus the effective string action predicts the spectrum of such closed flux 
tubes. On the other hand Lorentz invariance constrains the $p$-dependence
of $E_n(p,l)$ and this in turn will constrain the possible form of $S_{eff}$
\cite{LW04,HM06}.
More generally, the conformal invariance of the effective string action
\cite{PS}
can also be used to constrain its form
\cite{JD04}.

It was realised long ago that the leading $O(1/l)$ correction to the linear 
$\sigma l$ piece of $E_n(l)$ is in fact universal -- the `L\"uscher correction'
\cite{LSW}.
This corresponds to noting that if we write the effective string action
in `static gauge' and express it in a series of powers of the
derivative of the transverse fluctuation field $h(x)$, then the
leading Gaussian kinetic term for $h$ gives this universal
$O(1/l)$ contribution to $E_n(l)$. Much more recently it was found
\cite{LW04}
that the next term in the derivative expansion of $S_{eff}[h]$ is universal,
so that the next term in an expansion of $E_n(l)$, at $O(1/l^3)$, is also 
universal. This was also shown
\cite{JD04},
at much the same time, and with a stronger result in $D=3+1$,
using the Polchinski-Strominger conformal gauge approach
\cite{PS}.
More recently there has been further progress 
\cite{OA}
in both $D=2+1$ and $D=3+1$. (See also
\cite{HD}.)
In particular, in $D=2+1$ it is now known
that the $O(1/l^5)$ term is also universal. The physical constraints that
are used to derive this universality are satisfied by the Nambu-Goto
model, so that we can write
\begin{equation}
\frac{E_n(l)}{\surd\sigma} \stackrel{l\to\infty}{=} 
l\surd\sigma
+ \frac{c^{NG}_1}{l\surd\sigma}
+ \frac{c^{NG}_2}{(l\surd\sigma)^3}
+ \frac{c^{NG}_3}{(l\surd\sigma)^5}
+ O\left(\frac{1}{l^7}\right)
\label{eqn_Euniv}
\end{equation}
where the coefficients $c^{NG}_i$ are identical to those that arise
in the expansion of $E$ in powers of $1/l$ in the Nambu-Goto model, as in
eqns(\ref{eqn_EnNGexpansion}). 

An especially interesting result for us is the
demonstration that all the operators that appear in the derivative expansion
of the Nambu-Goto action appear with precisely the same coefficients 
in the general effective string action
\cite{OA}.
This provides a motivation for regarding  $S_{eff}[h]$ as being given,
in a non-trivial sense, 
by the full Nambu-Goto action plus a series of `corrections': in particular
at small $l$ where the expansion of the Nambu-Goto energy diverges
and needs to be resummed as in eqn(\ref{eqn_EnNGexpansion}). This
result is particularly significant in view of the numerical calculations
\cite{AABBMT_d3k1,AABBMT_d4k1}
that have shown that the spectrum of flux tubes of moderate $l$ is close
to the resummed Nambu-Goto prediction.

The above summarises the essential theoretical background for the 
analysis in this paper. There has of course been a great deal of
theoretical and, particularly, numerical work on this and related
problems, but most of that can be followed through the references
in the papers we have quoted and we do not repeat them here.
There are however a number of relevant papers that have appeared during 
the past year or so, which we would like to point the reader to.
Most directly relevant is 
\cite{MC_nonuniv}
where the static confining potential is calculated in the 3d Ising
model and the term corresponding to the $O(1/l^5)$ term in our
above discussion is found not to take the expected universal value.
The authors discuss possible reasons for this, but it is obviously
something that needs to be understood. Again in 
\cite{FG_nonuniv}
the corresponding term in the finite temperature expansion of the
string tension in a gauge dual of d3 random percolation is found not 
to take the universal Nambu-Goto value. (Note that this paper predates
\cite{OA}
and so does not comment on the expected universality of this term.)
Our expectation that there should be massive modes is closely linked 
to the idea that
the flux tube has an intrinsic width, and there have been papers
calculating that at both zero and non-zero $T$ in some confining field 
theories as well as ideas how to go about doing so
\cite{tube_width}.
There are interesting extensions to finite $T$
\cite{T_univ},
attempts to see to what scale the effective string action is valid
\cite{ESA_scale},
and a calculation of the excitations of the static potential in $D=2+1$
\cite{BB_Vd3}.
There have also been some interesting calculations from the gauge-gravity
side, on the flux tube intrinsic width
\cite{GGtube_width}
and on the Wilson line and Coulomb potential
\cite{UKJS}.

\section{Background}
\label{section_background}

\subsection{Lattice and continuum}
\label{subsection_lattice}

Our $D=2+1$ Euclidean space-time is discretised to a periodic cubic 
$L_x\times L_y\times L_t$ lattice with lattice spacing $a$. The degrees 
of freedom are SU($N$) matrices, $U_\mu(x,y,t)$ or more compactly $U_l$, 
assigned to the links $l$ of 
the lattice. Our action is the standard (Wilson) plaquette action,
so the partition function is 
\begin{equation}
Z(\beta)
=
\int \prod_l dU_l \,
e^{-\beta \sum_p\{ 1 - \frac{1}{N}{\mathrm{ReTr}}U_p\}}
\label{eqn_latticeZ} 
\end{equation} 
where $U_p$ is the ordered product of matrices around the
boundary of the elementary square (plaquette) labelled by $p$.
Taking the continuum limit of eqn(\ref{eqn_latticeZ}), and comparing
to the usual continuum path integral, one finds that
\begin{equation}
\beta\stackrel{a\to 0}{=}\frac{2N}{a g^2}
\label{eqn_beta} 
\end{equation} 
where $g^2$ is the coupling. In $D=2+1$  $g^2$ has dimensions of mass,
and so $a g^2$ is the dimensionless coupling on the length scale $a$. 
The continuum limit, $a\to 0$, is therefore approached by tuning 
$\beta = 2N/ag^2 \to \infty$.

If we calculate some physical masses (or energies) on the lattice, 
they will have
lattice corrections and they will be in lattice units, i.e. we will 
obtain them as $am_i(a)$. To obtain the continuum limit one can take 
ratios of masses, calculate these over some substantial range of $a$, and 
extrapolate to $a=0$, using the leading correction that is known
to be $O(a^2)$ for our plaquette action:
\begin{equation}
\frac{am_i(a)}{am_j(a)} 
= \frac{m_i(a)}{m_j(a)}
\stackrel{a\to 0}{=}
\frac{m_i(0)}{m_j(0)} + c (a\mu)^2.
\label{eqn_rcontinuum} 
\end{equation} 
Here we can use $a\mu = am_i(a)$ or any other calculated mass --
different choices correspond to different subleading $O(a^4)$
corrections in eqn(\ref{eqn_rcontinuum}), which we neglect. 
Obviously all this assumes that $a$ is sufficiently small for the 
leading $O(a^2)$ correction to dominate. If this is not the case, 
i.e. if the fit using  eqn(\ref{eqn_rcontinuum}) is found to be 
unacceptably poor, one can systematically drop the mass ratios coming 
from the largest values of $a$, i.e. the smallest values of $\beta$, 
until the fit becomes good. In practice one finds
\cite{MT98_d3}
that the approach to the continuum limit for typical dynamical
quantities is very rapid.

An alternative approach is to calculate the continuum value of 
$m_i/g^2$, using eqn(\ref{eqn_beta}) and
\begin{equation}
\frac{\beta}{2N} am_i(a) 
\stackrel{a\to 0}{=}
\frac{m_i(0)}{g^2} +  \frac{c}{\beta},
\label{eqn_mcontinuum} 
\end{equation} 
where again we have retained only the leading correction.
The lattice correction is $O(1/\beta) \propto O(a)$
rather than $O(a^2)$ because different lattice coupling definitions
will clearly differ at this order. In this way one can, for example, 
calculate the continuum string tension in units of $g^2$
\cite{MT98_d3,BBMT_d3k1}.

\subsection{Large-$N$ limit}
\label{subsection_largeNlimit}

\begin{figure}[htb]
\begin	{center}
\leavevmode
\input {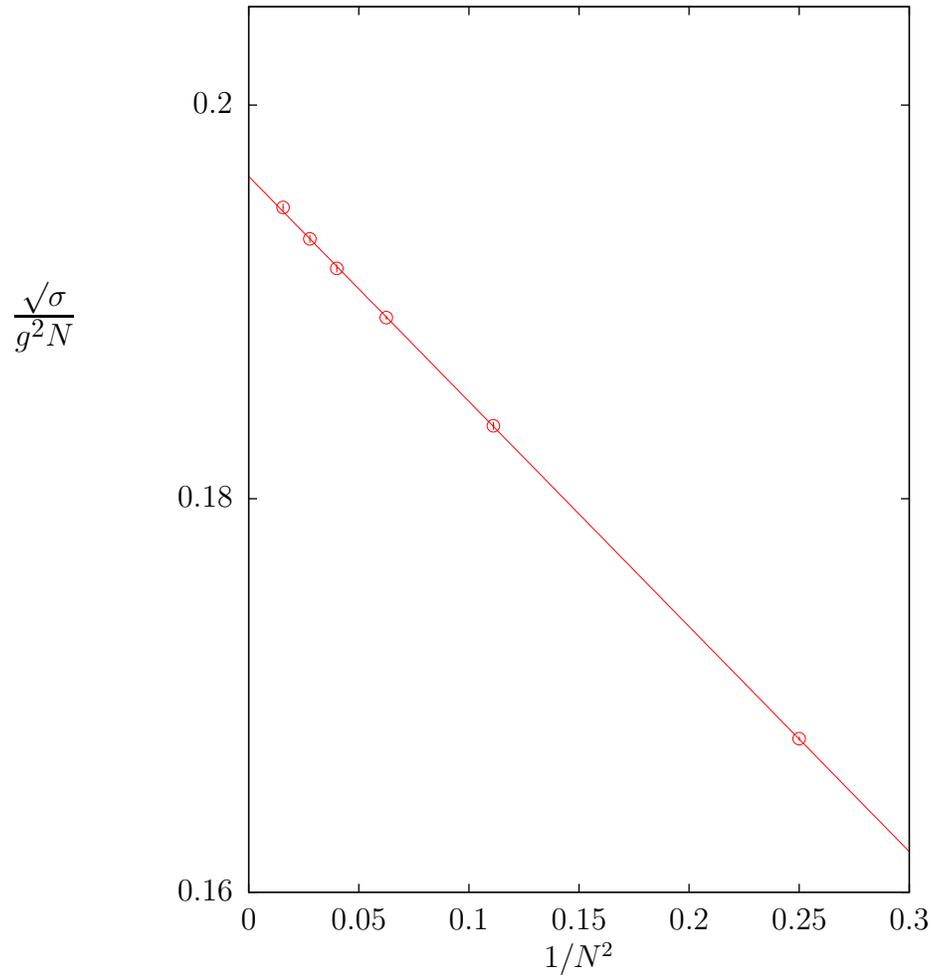}
\end	{center}
\caption{String tension in units of $g^2N$ for various continuum
 SU($N$) gauge theories. The curve is a best fit to $N\geq 2$ of
the conventional functional form: $\frac{\surd\sigma}{g^2N} =
0.19638  - \frac{0.1144}{N^2}$.}
\label{fig_g2ND3}
\end{figure}

One expects that at large $N$ physical masses will be proportional 
to the 't Hooft coupling $\lambda \equiv g^2 N$ with a leading
correction that is $O(1/N^2)$ 
\cite{largeN}, 
i.e.
\begin{equation}
\frac{m_i}{g^2N} 
=
\lim_{N\to\infty} \frac{m_i}{g^2N}  + \frac{c}{N^2}
\label{eqn_mgN} 
\end{equation} 
to leading order.
So if we vary $\beta \propto N^2$ we will be keeping the lattice 
spacing $a$  fixed in physical units, to leading order in $N$.
These expectations are largely based on an analysis of all-orders
perturbation theory, so it is interesting to ask how precisely they
are confirmed by  non-perturbative lattice calculations.
This question has been addressed in the past
\cite{MT98_d3,N_counting},
but here we can go somewhat further using the very precise string 
tensions calculated for $N\in [2,8]$ in
\cite{BBMT_d3k1}.
We display in Fig.~\ref{fig_g2ND3} the continuum values of 
$\surd\sigma/g^2N$ taken from the first row of Table 2 in
\cite{BBMT_d3k1}.
(Using the values in the other rows would produce slightly larger errors 
but would lead to the same conclusions.) We also show 
in Fig.~\ref{fig_g2ND3} the best fit of the conventional form, i.e. 
eqn(\ref{eqn_mgN}) with $\surd\sigma$ replacing $m_i$:
\begin{equation}
\frac{\surd\sigma}{g^2N} 
=
0.19638(9)  - \frac{0.1144(8)}{N^2}.
\quad ; \quad 
\chi^2/ndf \sim 0.4
\label{eqn_KNfit} 
\end{equation} 
Eqn(\ref{eqn_KNfit}) provides a very good fit to all our values of 
$N$, including SU(2). This is perhaps surprising given that higher order 
corrections in $1/N^2$ are surely present. To investigate this 
we can include an extra $c^\prime/N^4$ term in eqn(\ref{eqn_KNfit}) 
and we then find $c^\prime = 0.008(27)$, with little change
in the first two terms. This indicates that in the $1/N^2$ expansion 
of $\surd\sigma/g^2N$  the coefficients decrease rapidly, so that the 
large-$N$ limit is unexpectedly precocious.

If we now allow the power of the correction term in
eqn(\ref{eqn_mgN}) to vary we find
\begin{equation}
\frac{\surd\sigma}{g^2 N} 
=
c_0 + \frac{c_1}{N^\gamma}
\qquad \longrightarrow \qquad
\gamma = 1.97 \pm 0.10.
\label{eqn_g2ND3corr}
\end{equation} 
So if we assume that $\gamma$ has to be an integer, we can 
unambiguously conclude that the leading correction is in fact $O(1/N^2)$,
just as predicted by 't Hooft's diagrammatic analysis
\cite{largeN}. 
Let us now allow the leading power of $N$ to vary, i.e. 
$g^2 N \to g^2 N^\alpha$, then we find
\begin{equation}
\frac{\surd\sigma}{g^2 N^\alpha} 
=
c_0 + \frac{c_1}{N^2}
\qquad \longrightarrow \qquad
\alpha = 1.002 \pm 0.004.
\label{eqn_g2ND3corr2}
\end{equation} 
Thus if we assume a $O(1/N^2)$ correction, the lattice values of
the string tension tell us that $g^2 \propto 1/N^{1.002(4)}$
i.e. the conventional expection of  $g^2 \propto 1/N$ is confirmed
very accurately. Finally, if we allow both powers to vary, then 
\begin{equation}
\frac{\surd\sigma}{g^2 N^\alpha} 
=
c_0 + \frac{c_1}{N^\gamma}
\quad \longrightarrow \quad
\alpha = 1.008 \pm 0.015 \ , \  \gamma = 2.18 \pm 0.40.
\label{eqn_g2NcorrD3corr}
\end{equation} 
The constraint on the power of the correction is now significantly
looser, but the evidence for $g^2 \propto 1/N$ is still very convincing.
Altogether, we can conclude that these lattice calculations provide strong 
support for the non-perturbative validity of the usual large-$N$ counting.

\subsection{Calculating the spectrum}
\label{subsection_calculation}

To calculate the spectrum, we calculate the correlation functions
of some suitable (see below) set of lattice operators $\{\phi_i\}$.
Expanding the correlators in terms of the energy eigenstates,
$H | n \rangle = E_n | n \rangle$ and expressing $t=an_t$ in lattice
units, we have
\begin{equation}
C_{ij}(t) 
=
\langle \phi_i^\dagger(t)\phi_j(0) \rangle
=
\langle \phi_i^\dagger e^{-Han_t} \phi_j \rangle
=
\sum_k c_{ik} c^\star_{jk} e^{-aE_k n_t} 
\label{eqn_cortomass}
\end{equation}
where $c_{ik} = \langle vac | \phi^\dagger_i | k \rangle$.
We can now perform a variational calculation of the spectrum as follows.
Suppose that  $\phi = \psi_0$ maximises 
$\langle \phi^\dagger(t^\prime)\phi(0) \rangle
/\langle \phi^\dagger(0)\phi(0) \rangle$
over the vector space spanned by the $\{\phi_i\}$. (Obviously we
can restrict the $\{\phi_i\}$ to a desired set of quantum numbers.)
Here $t^\prime$ is some convenient small value of $t$, where all our 
$C_{ij}(t)$ are known quite precisely, and which we shall typically 
choose to be $t=a$. Then $\psi_0$ is our best estimate of the
wave-functional of the ground state. Repeating this calculation over
the basis of operators orthogonal to $\psi_0$ gives us  $\psi_1$,
our best estimate for the first excited state. And so on for the
higher excited states. If our basis is large enough for $\psi_i$
to be close to the true wave-functional, $\Psi_i$, then its correlator
should be dominated by the corresponding state,
$\langle \psi_i^\dagger(t)\psi_i(0) \rangle \propto \exp(-E_i t)$,
even for small values of $t$, where the signal to noise ratio is large
and where we are able to extract an accurate value for the energy, $a E_i$.

Here the states that we are interested in are loops of flux closed around
the $x$-torus. Thus our operators will also wind around the $x$-torus.
The simplest such operator is the Polyakov loop
\begin{equation}
l_p(n_y,n_t) = \mathrm{Tr} 
\left\{\prod^{L_x}_{n_x=1} U_x(n_x,n_y,n_t)\right\} 
\label{eqn_poly}
\end{equation}
where $l=aL_x$ and we have taken the product of the link matrices in 
the $x$-direction, around the $x$-torus. (We recall a standard
argument that uses the fact that the gauge potentials are only
periodic up to an element of the  centre of the SU($N$) group, to show
that in the confining phase $ \langle \phi_c \, l_p\rangle = 0$ 
for any contractible loop $\phi_c$, thus showing that such a winding 
operator has zero projection onto glueball states.)
The operator in  eqn(\ref{eqn_poly}) is localised in $n_y$ and so has 
transverse momentum $p_\perp \neq 0$. If we sum over $n_y$, to get
$l_p(n_t) = \sum_{n_y} l_p(n_y,n_t)$, then we obtain an operator with 
$p_\perp = 0$, and from now on we assume this has been done.  
This operator is manifestly invariant under longitudinal translations,
so $p=0$. It is also invariant under parity $P$. So in order to have 
$p\neq 0$ or $P\neq +$ we must introduce a deformation into the operator 
defined in eqn(\ref{eqn_poly}). For this purpose we choose the deformations 
displayed in Table~\ref{Operators}. Now, if we translate an operator 
by $\Delta x$ in the $x$ direction, multiply it by the phase factor 
$\exp\{i 2 \pi q \Delta x/l\}$ where $q$ is an integer,
and then add all such translations, we obtain an operator with
longitudinal momentum $p=2\pi q/l$. If we had done so with $p\neq 0$
to the simple Polyakov loop in eqn(\ref{eqn_poly}), we would have
obtained a null operator. But for the other operators in 
Table~\ref{Operators} this will not, in general, be the case.
%
%
\begin{table}[htb]
\centering{
\begin{tabular}[c]{||c||c||c||c||c||} \hline \hline
\includegraphics[height=2.5cm]{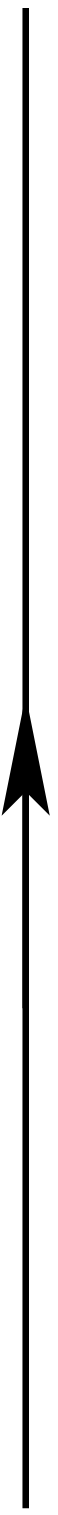}
&
\includegraphics[height=2.5cm]{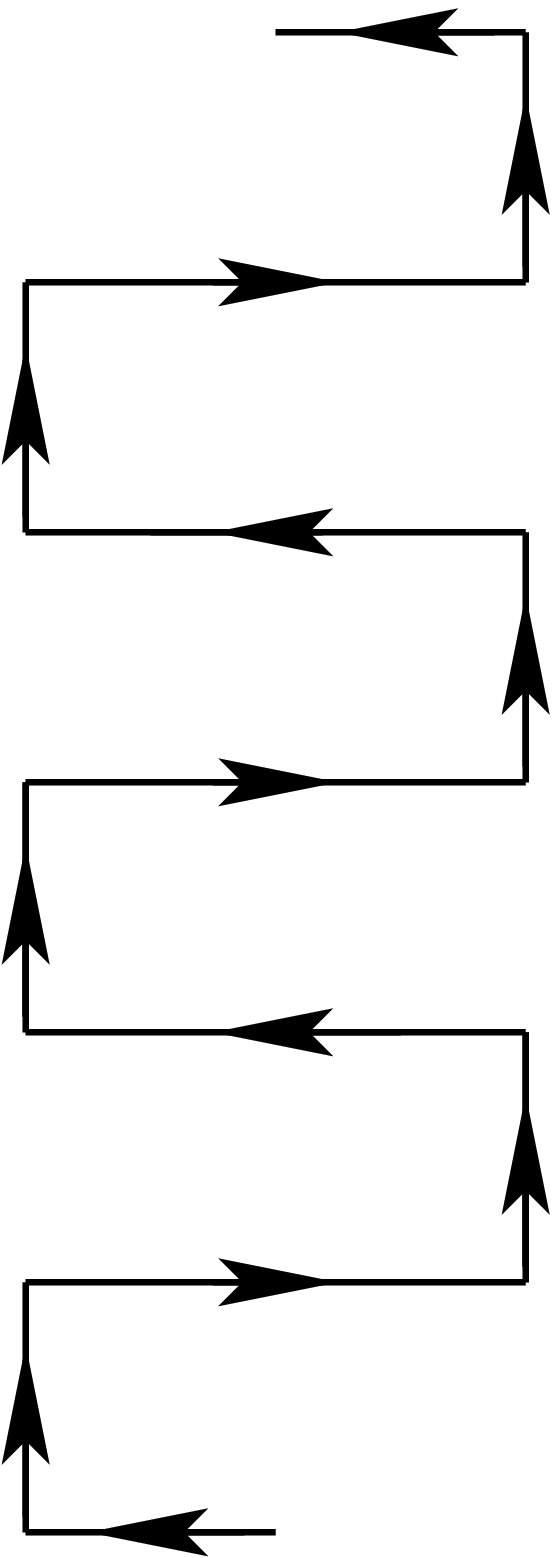}
&
\includegraphics[height=2.5cm]{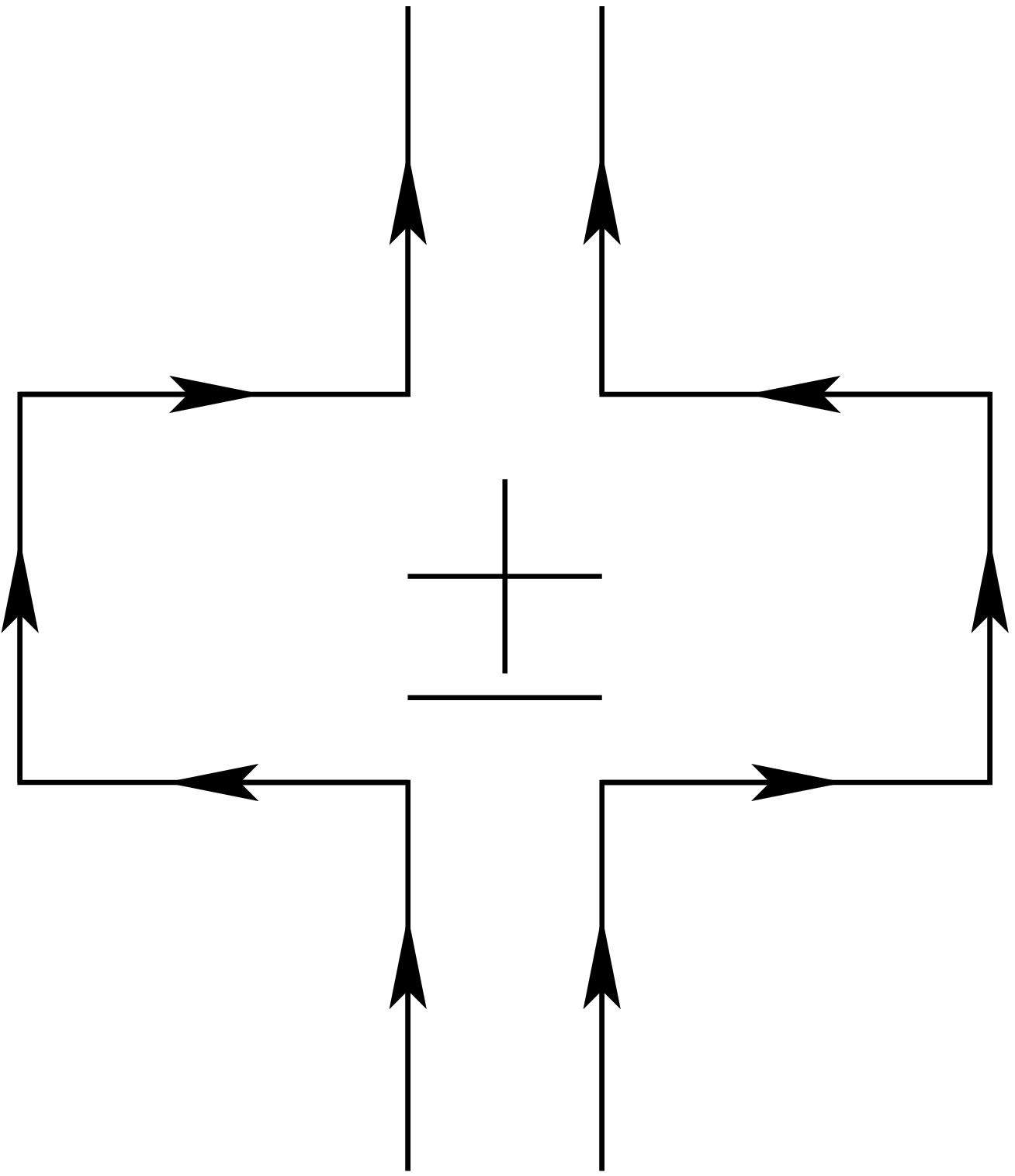}
&
\includegraphics[height=2.5cm]{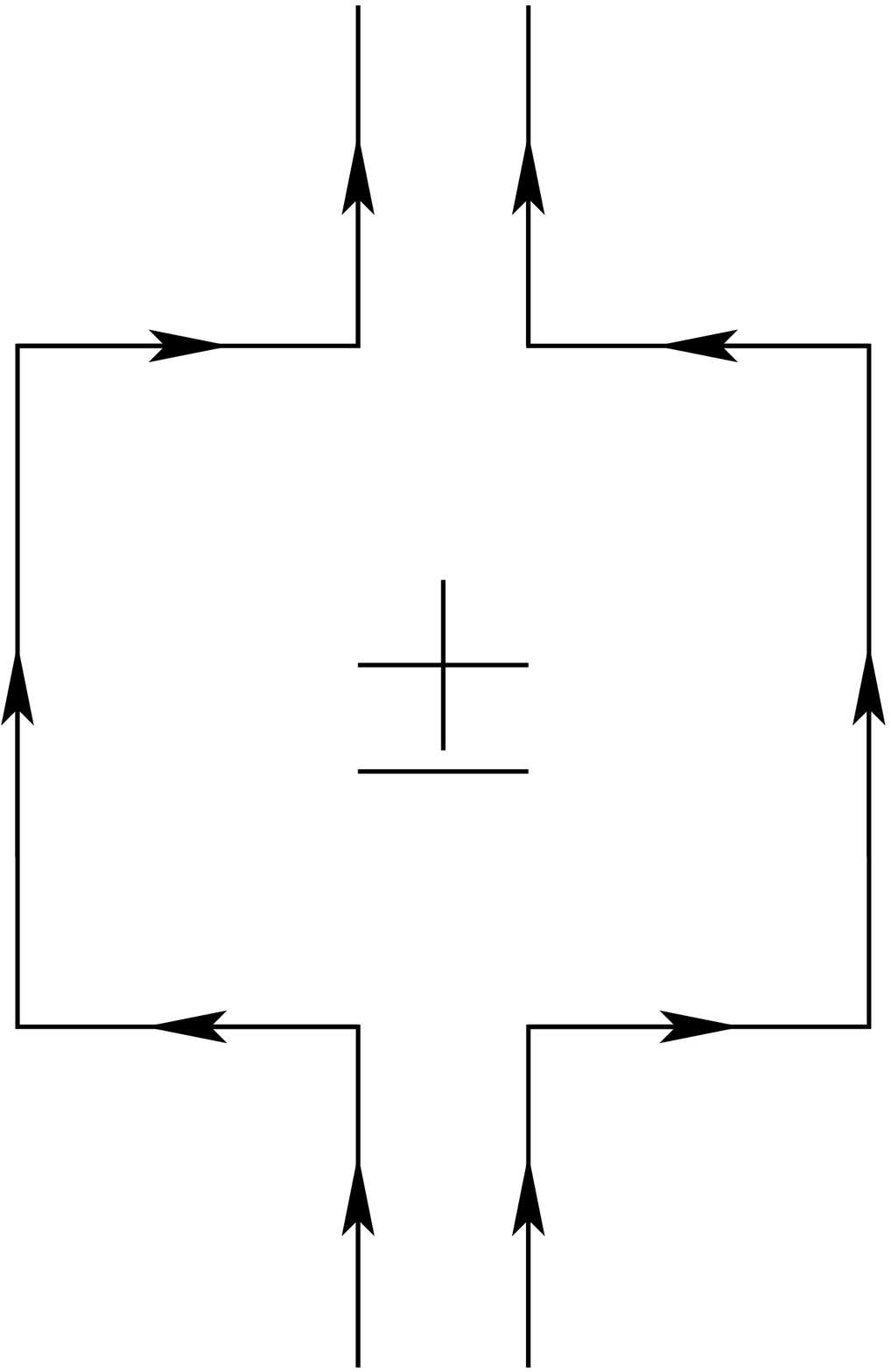}
&
\includegraphics[height=2.5cm]{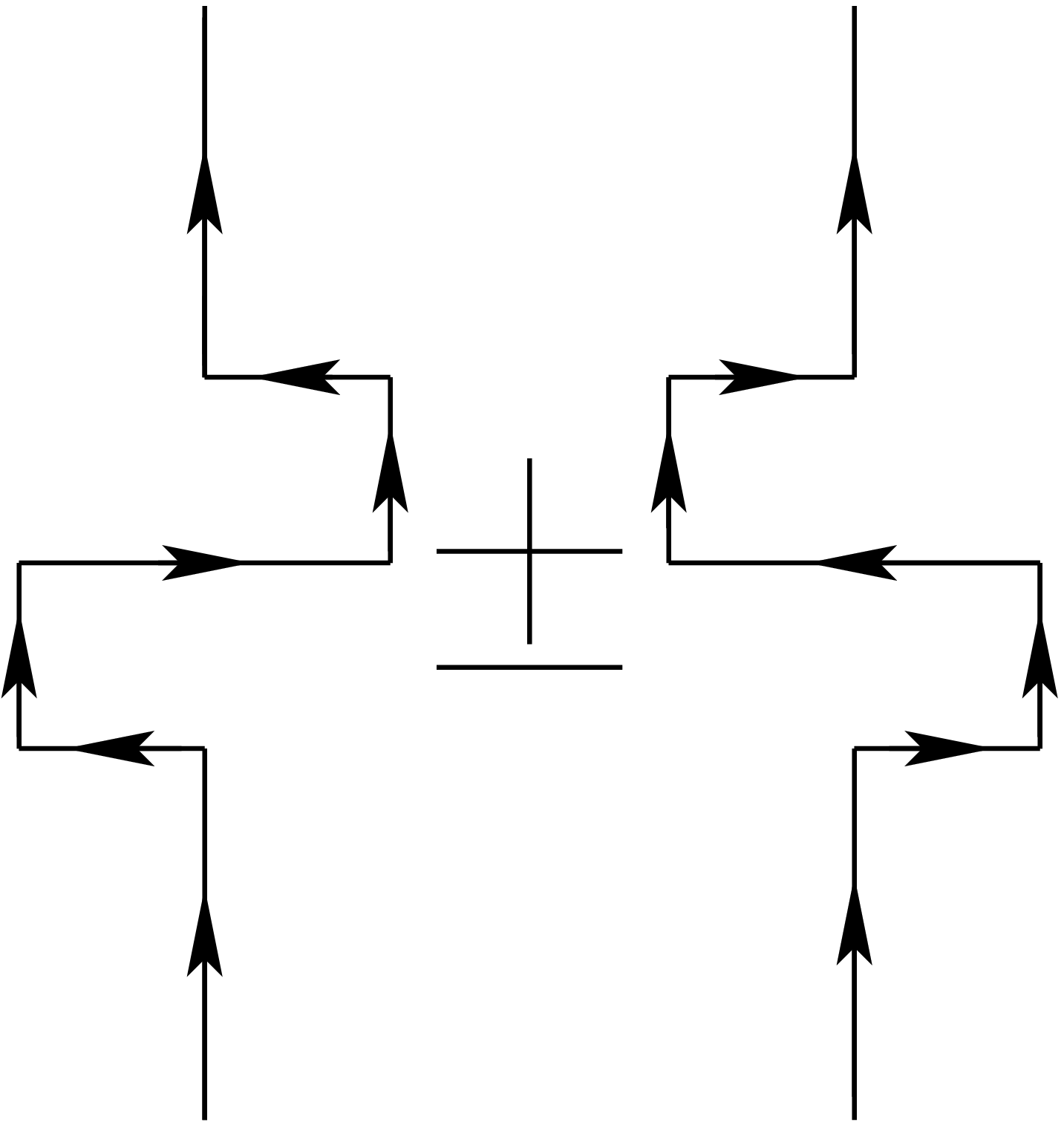} \\ \hline \hline

$1$ & $2$ & $3$ & $4$ &$5$ \\ \hline \hline

\includegraphics[height=2.5cm]{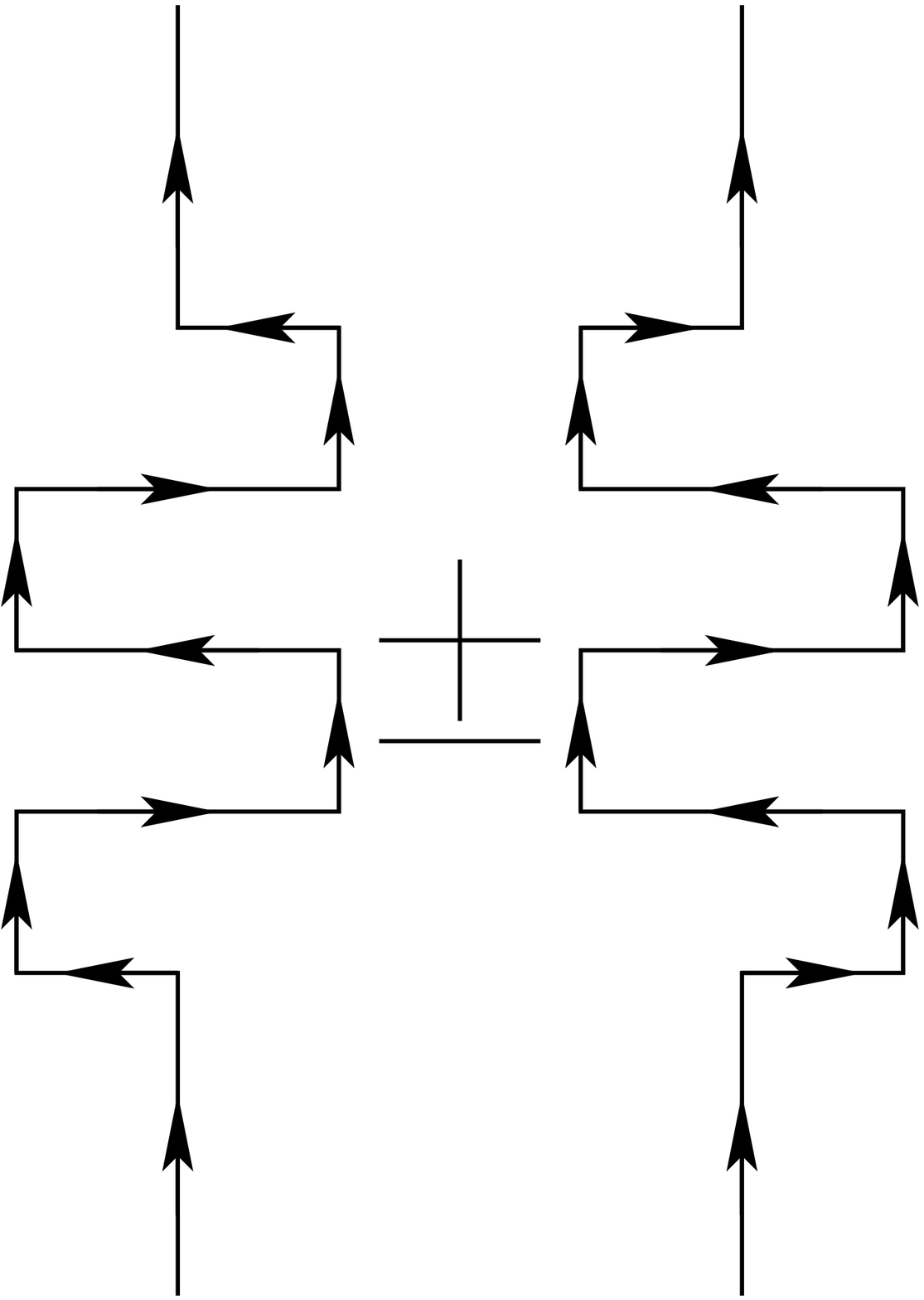}
&
\includegraphics[height=2.5cm]{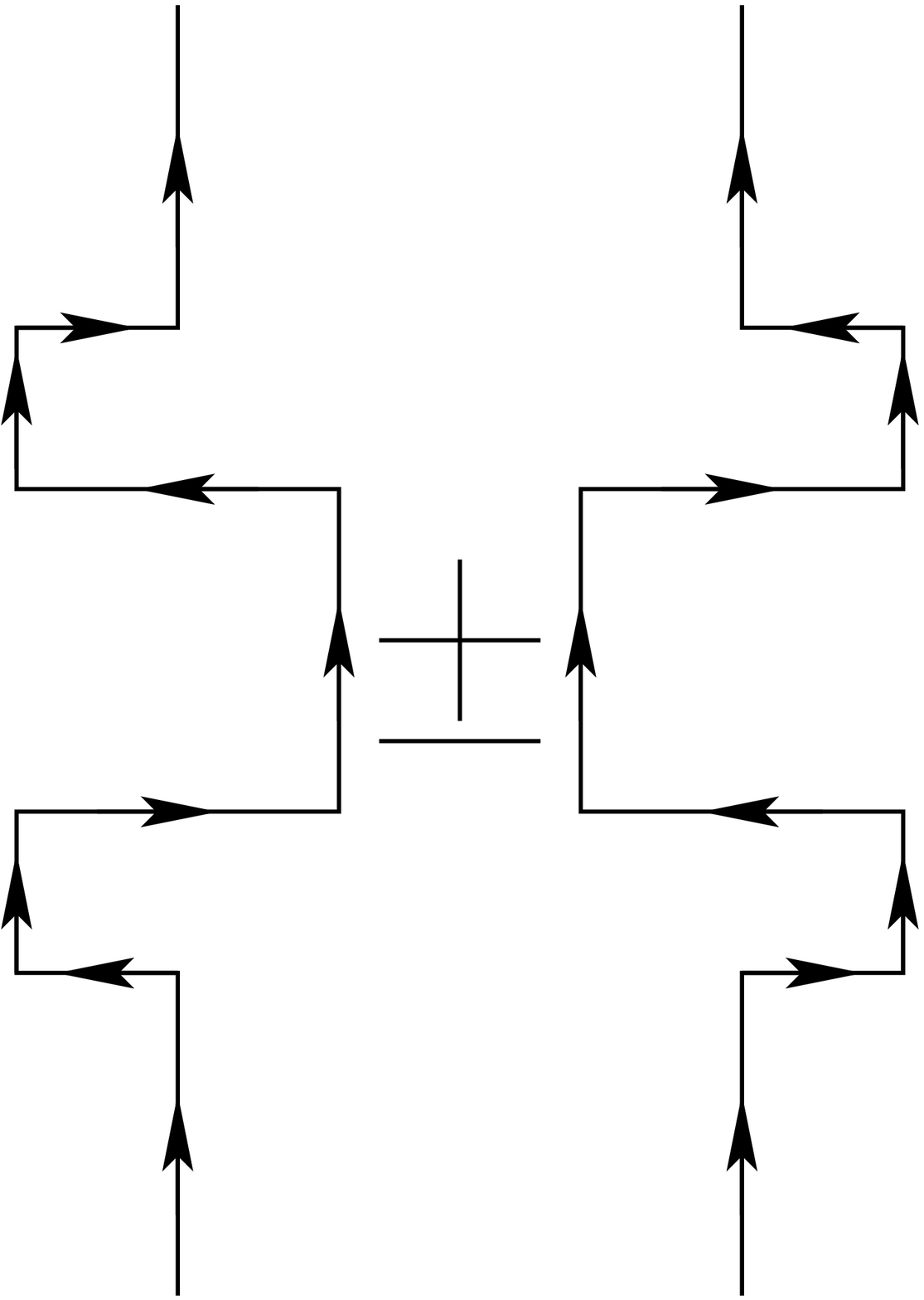}
&
\includegraphics[height=2.5cm]{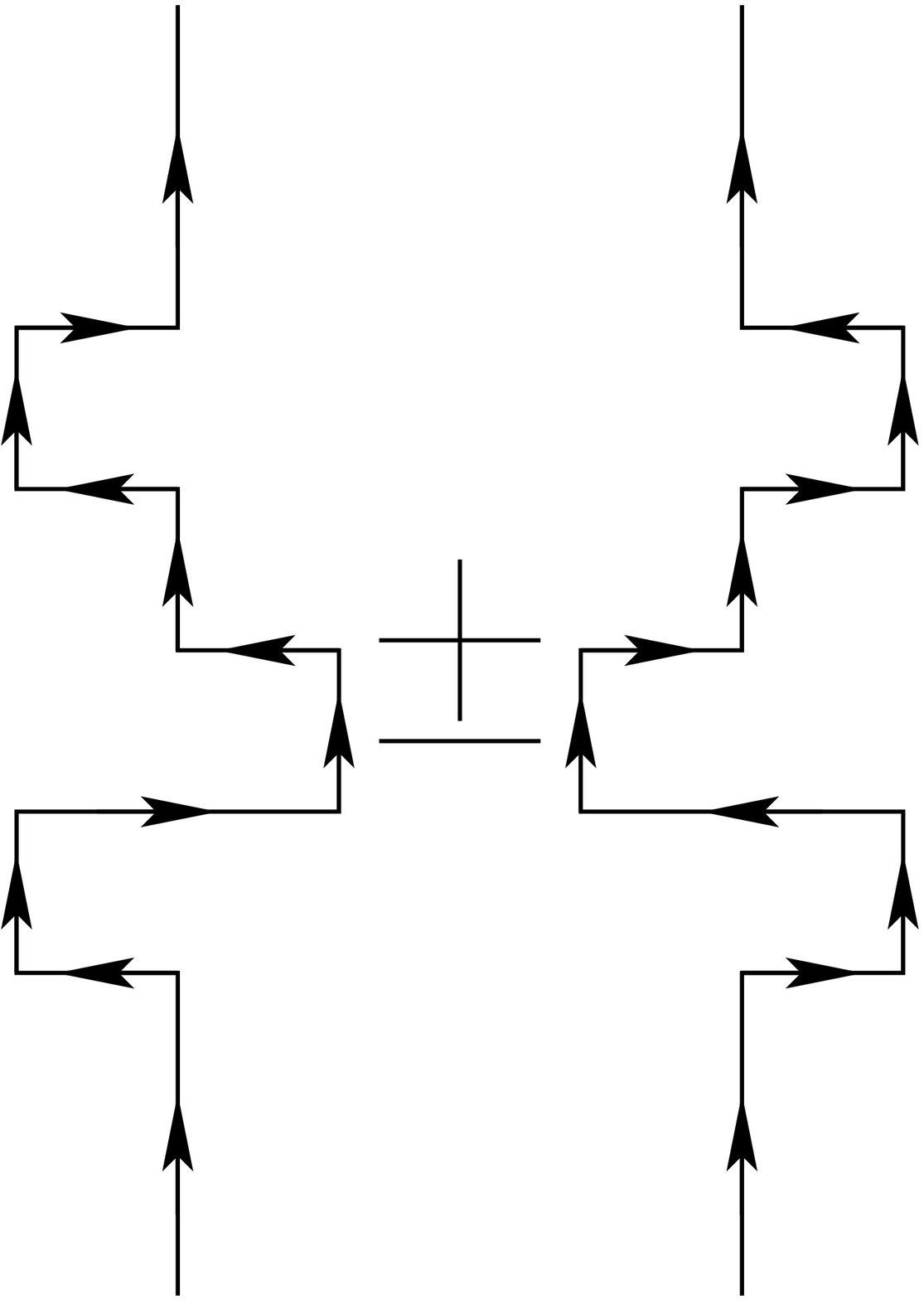}
&
\includegraphics[height=2.5cm]{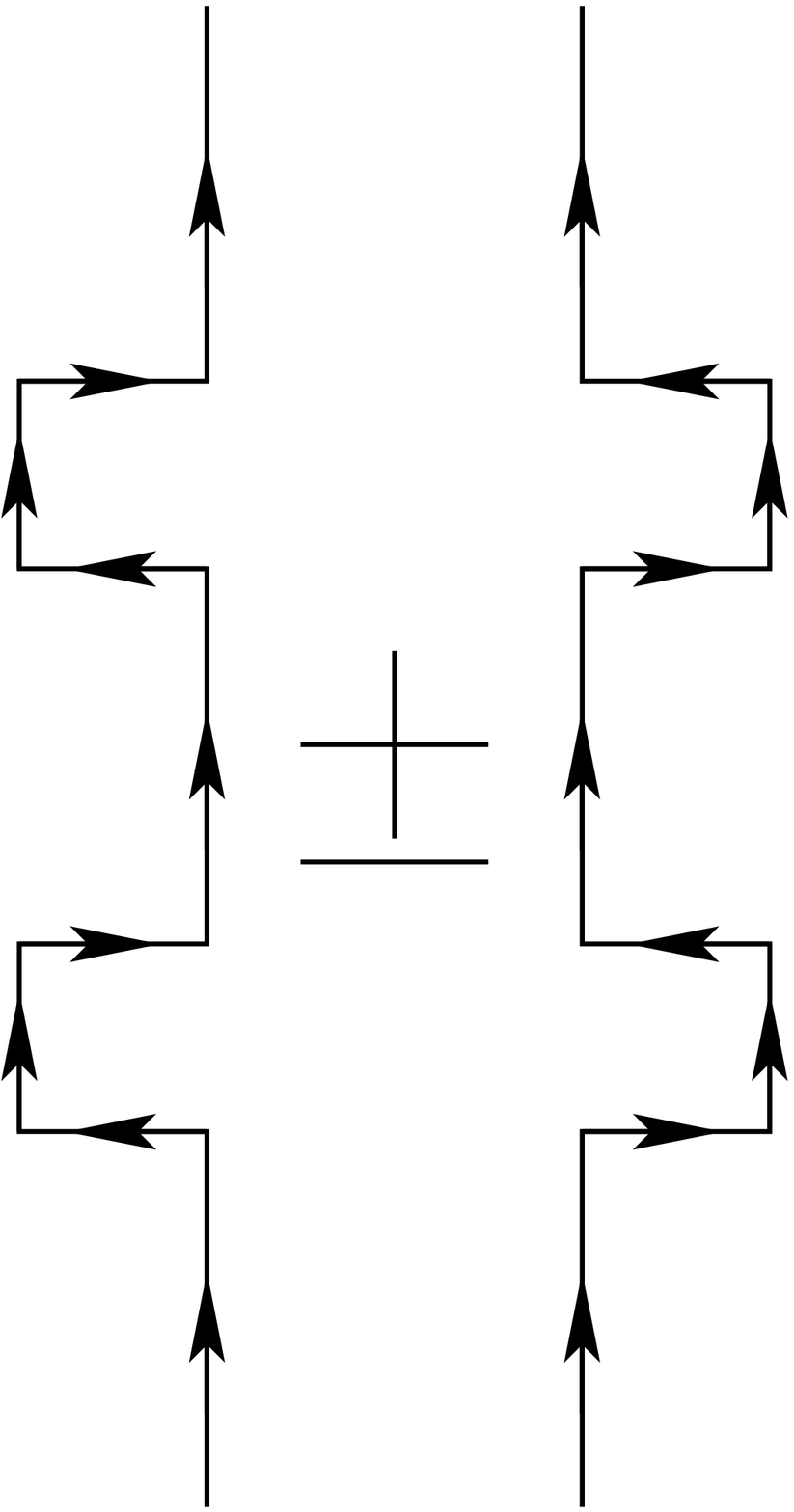}
&
\includegraphics[height=2.5cm]{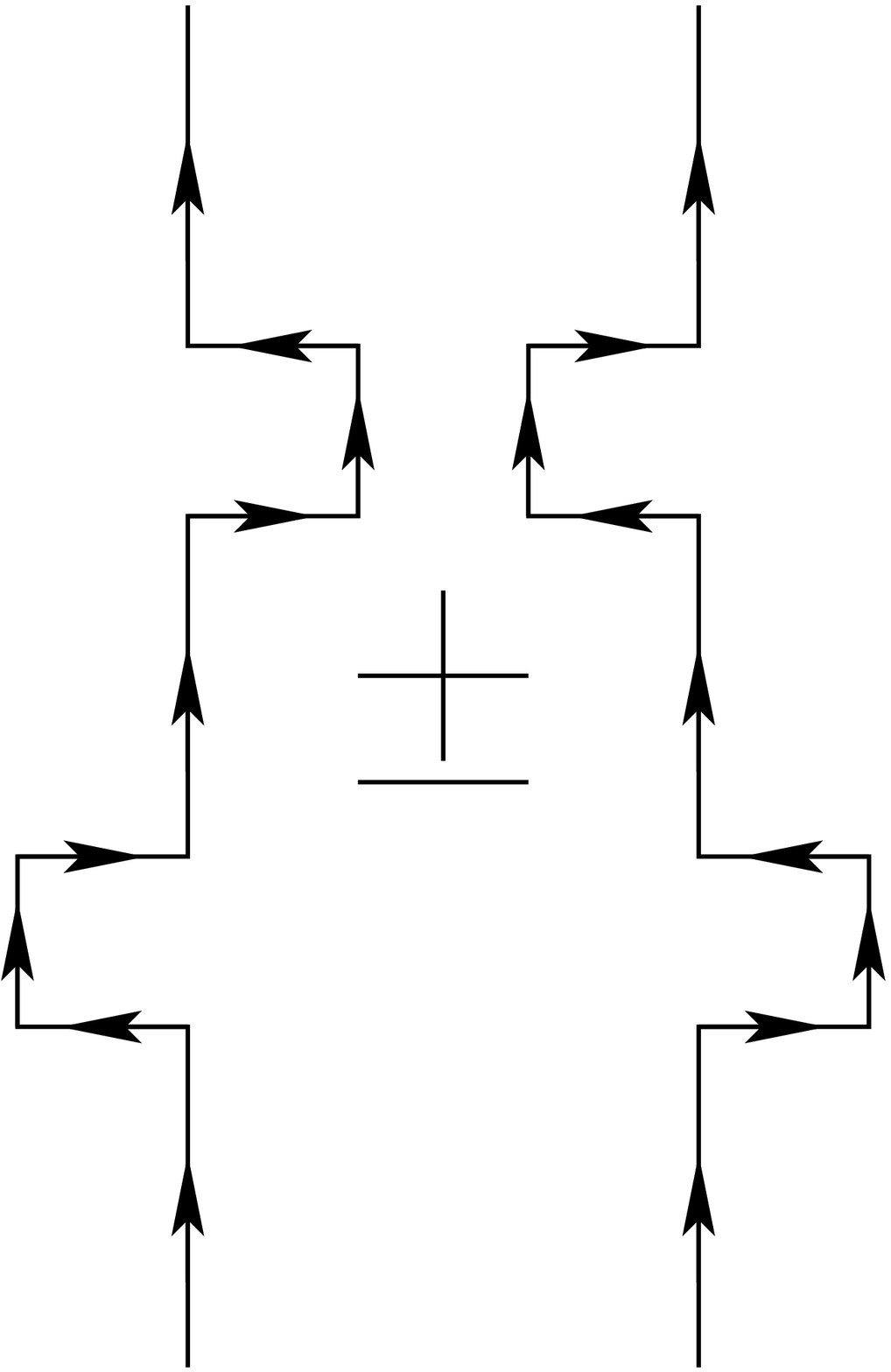} \\ \hline \hline

$6$ & $7$ & $8$ & $9$ & $10$ \\ \hline \hline

\includegraphics[height=1.75cm]{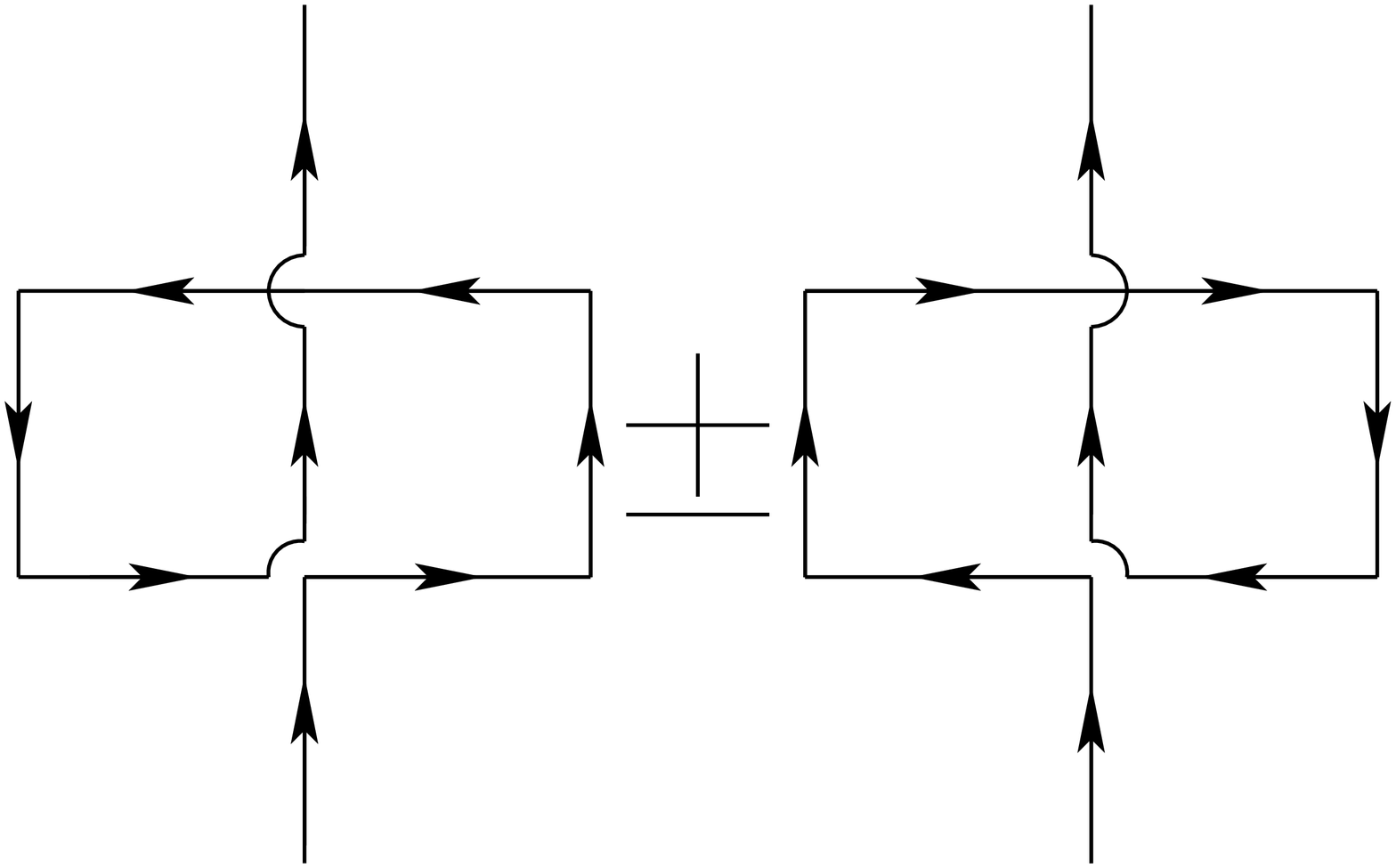}
&
\includegraphics[height=2cm]{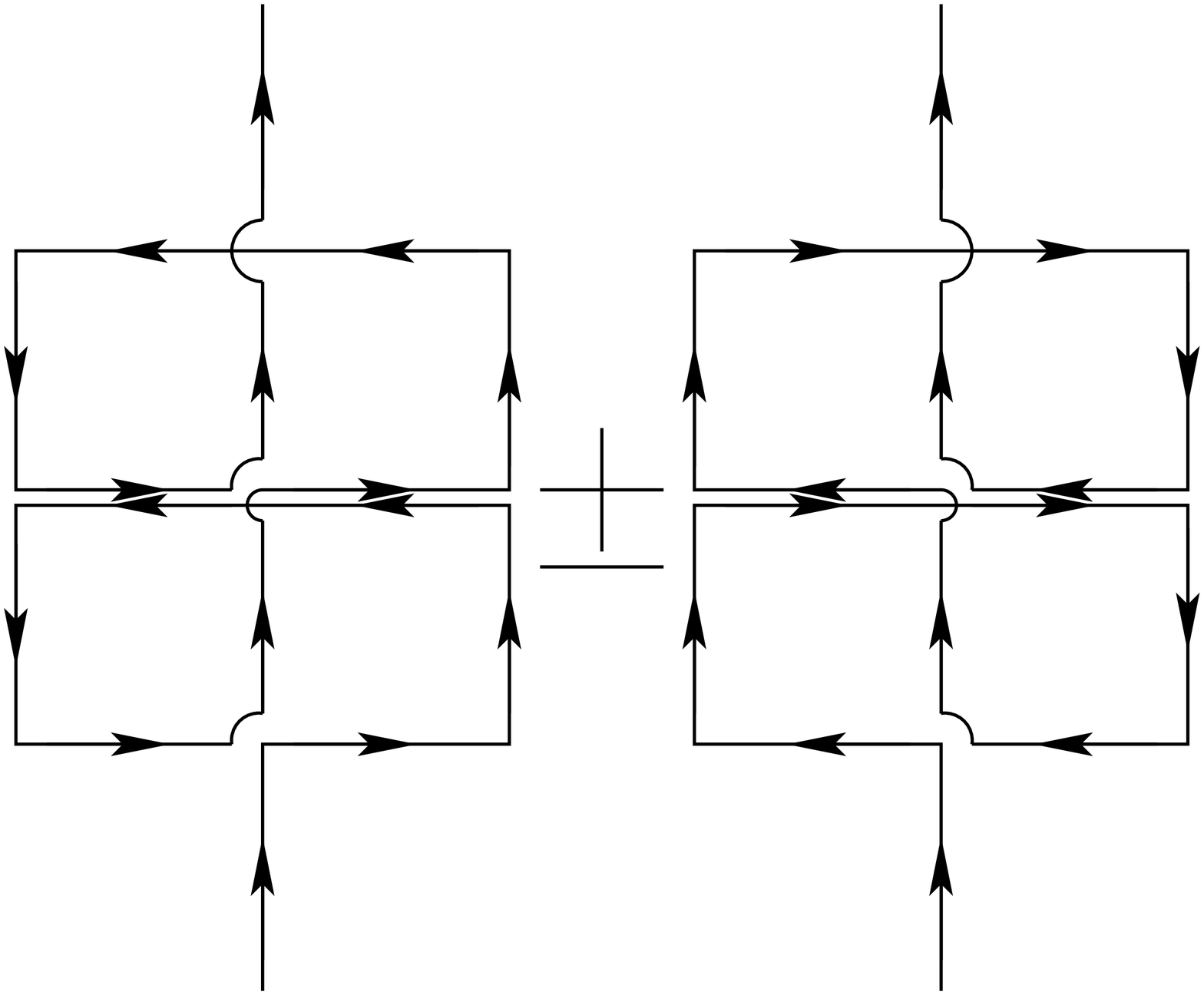}
&
\includegraphics[height=2cm]{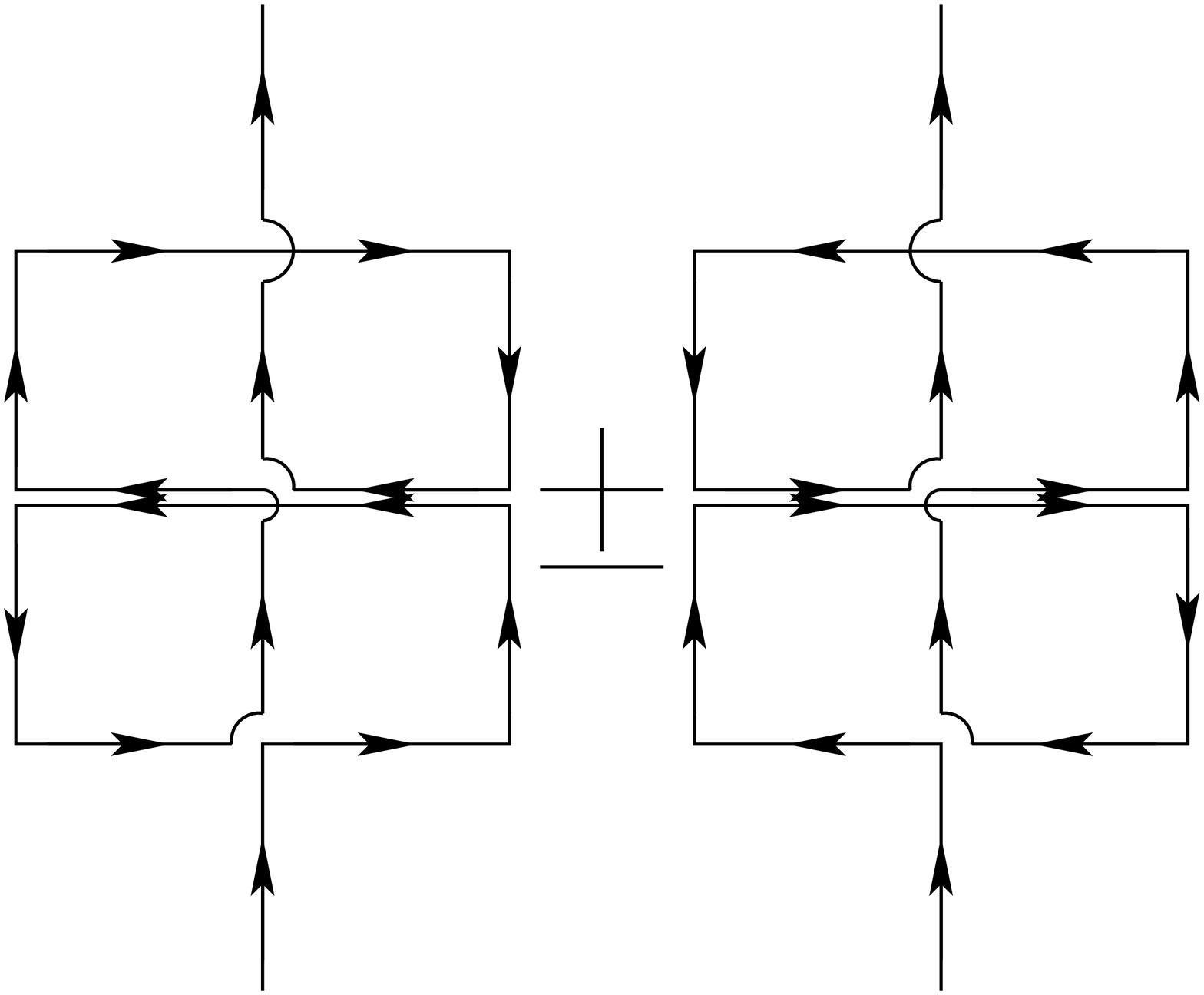}
&
\includegraphics[height=2cm]{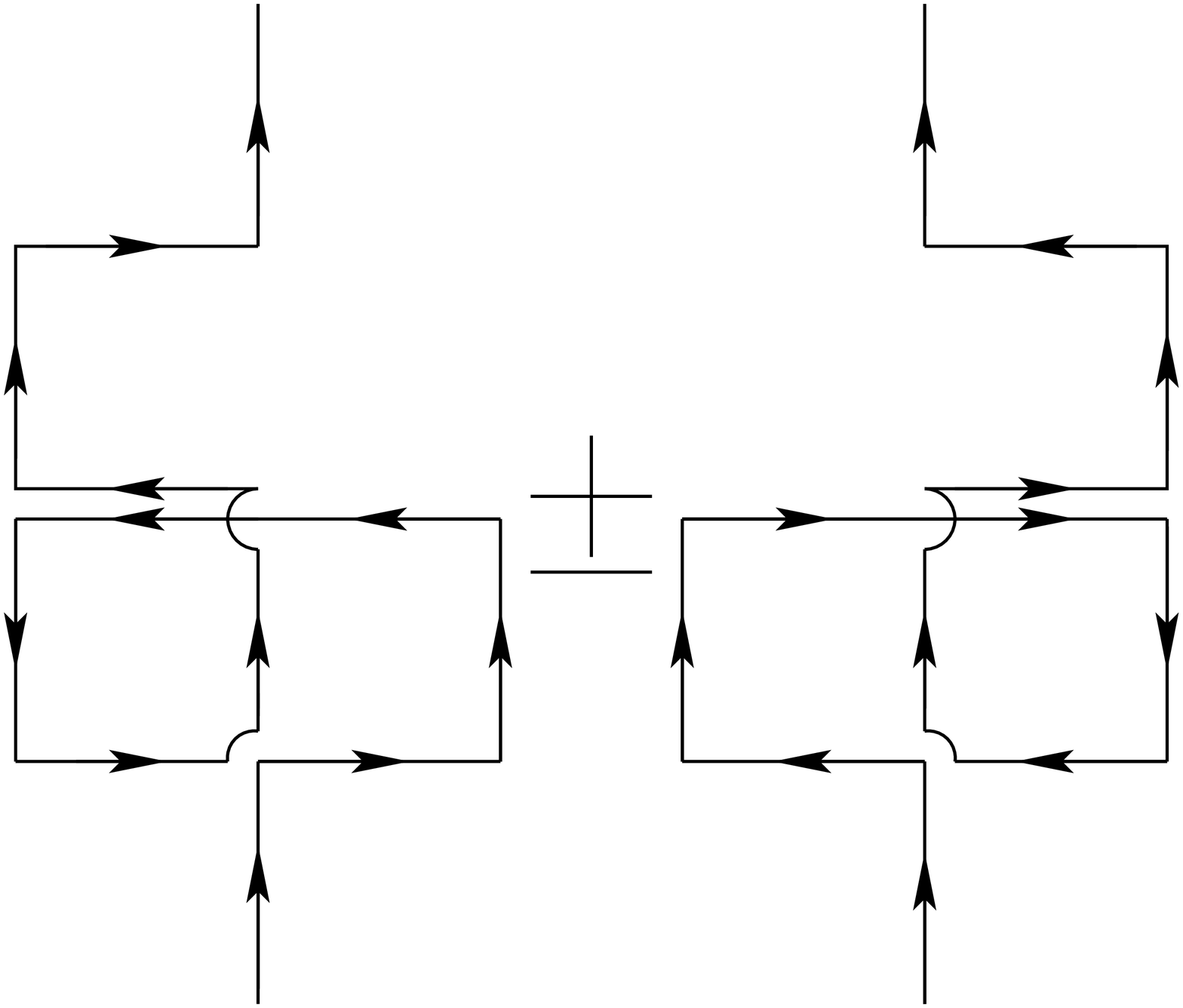}
&
\includegraphics[height=2cm]{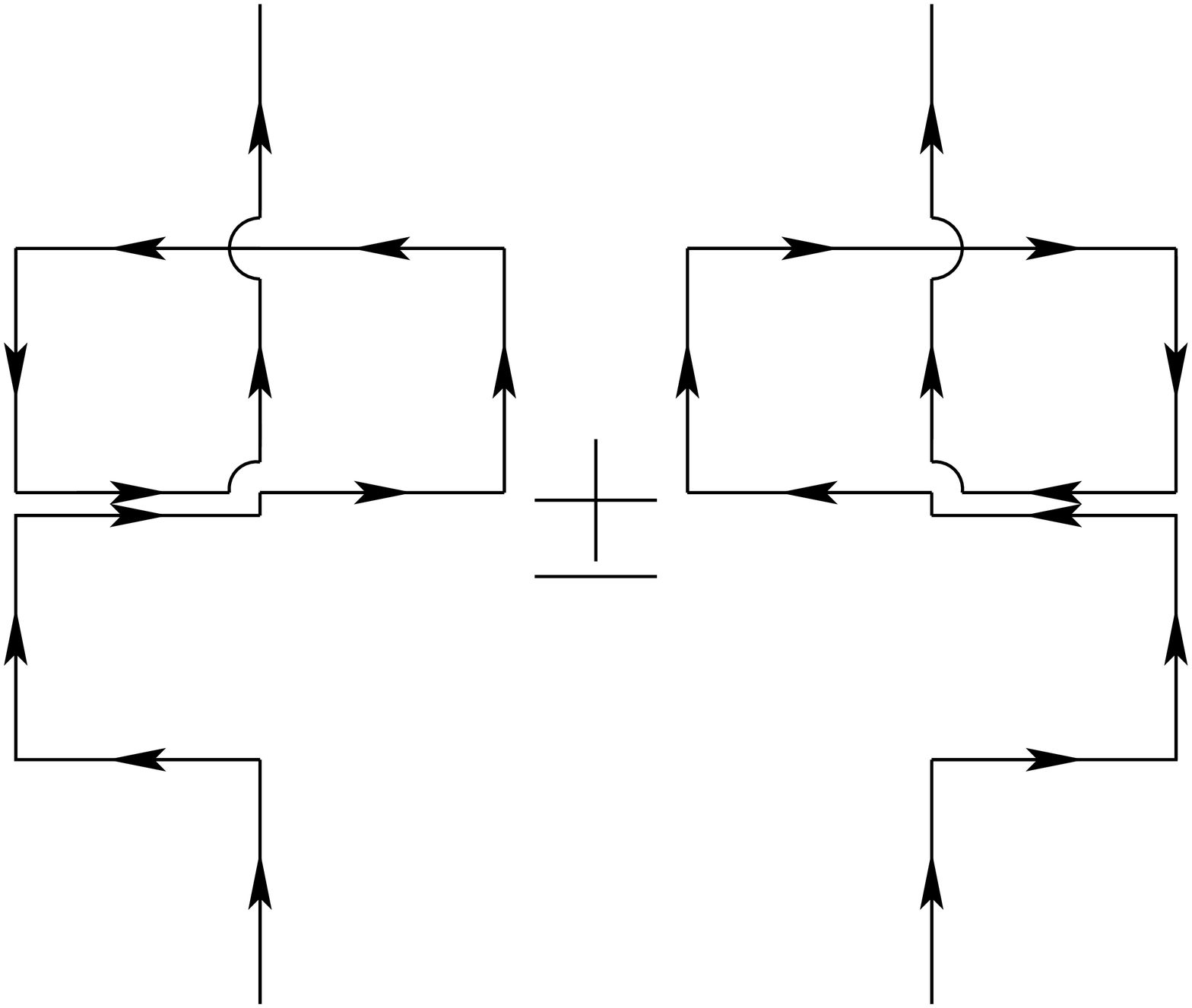} \\ \hline \hline

$11$& $12$ & $13$ & $14$ & $15$ \\ \hline \hline

\end{tabular}
}
\caption{The lattice paths used in the construction of Polyakov loops in this work. 
Our set of operators can be divided into three subsets: (a) the
simple line operator (1) in several smearing/blocking
levels; (b) the wave-like operator (2) whose number depends
upon $L_x$, $L_{\perp}$, and the smearing/blocking level; (c) the
pulse-like operators (3-15) in several different
smearing/blocking levels. In addition the extent of the transverse 
deformations is varied.  The $\pm$ combinations correspond
to $P=\pm$. The operators in (1) and (2) are intrinsically $P=+$.} 
\label{Operators}
\end{table}

In practice, to obtain good overlaps onto any states at all, one needs 
to smear
\cite{smear}
and/or block 
\cite{block}
the `link matrices' that appear in the operators in Table~\ref{Operators}.
Taking into account the various blocking levels, our typical basis
has $\sim 80$ operators for each set of quantum numbers. (In our newer
calculations we have not included the wavelike operators shown in box 2
of Table~\ref{Operators} since we found in our earlier calculations
that they have $\sim 100\%$ overlap onto our simple blocked line
operators in box 1 and therefore bring nothing new to the calculation.)

\subsection{Control of systematic errors}
\label{subsection_errors}

The systematic errors in  $D=2+1$ are much the same as in $D=3+1$
and the latter have been discussed in some detail in our recent 
companion paper
\cite{AABBMT_d4k1}.
In $D=2+1$ our operator basis has a much better overlap onto the
light flux tube states of interest, and so many of the systematic
errors will be much smaller. We will not repeat here the full 
discussion in 
\cite{AABBMT_d4k1},
some of which has been covered in our earlier $D=2+1$ papers
\cite{AABBMT_d3k1,BBMT_d3k1},
but will comment on three particular issues. 

\subsubsection{effective energies}
\label{subsubsection_Eeff}

We calculate energies by identifying the asymptotic exponential fall-off
of correlation functions $\langle \psi^\dagger_i(t)\psi_i(0) \rangle$, 
as described above. Typically the statistical error is roughly constant
in $t$, so the error/`signal' ratio grows exponentially with $t$. This
means that we need to extract the energy at small $t$. So one requirement
is that our best variational wavefunction $\psi_i$ should have
a high overlap onto the state $|i\rangle$, so that the corresponding 
exponential, $|c_i|^2\exp\{-aE_i n_t\}$, already dominates the sum in
\begin{equation}
\langle \psi_i^\dagger(t)\psi_i(0) \rangle
=
\sum_k |c_{ik}|^2 e^{-aE_k n_t} 
\label{eqn_cortomass2}
\end{equation}
at small $t=an_t$. An additional requirement
is that $aE_i$  should be small enough that we can accurately identify  
such an exponential fall-off over a sufficient range of $t=an_t$ for
us to be able to estimate  $aE_i$, and indeed the overlap. That is to say,
as the energy of interest becomes larger, both the statistical and systematic
errors become larger.

To illustrate this systematic error we define an effective energy
obtained by doing a local exponential fit to neighbouring values 
of the correlation function:
\begin{equation}
e^{-aE_{i,eff}(n_t)}
=
\frac{\langle \psi_i^\dagger(n_t)\psi_i(0) 
\rangle}{\langle \psi_i^\dagger(n_t-1)\psi_i(0) \rangle}
\label{eqn_Eeff}
\end{equation}
It is apparent from eqn(\ref{eqn_cortomass2}) that if $|i\rangle$ is
the lightest state in some quantum number sector, then as $n_t$ grows
$E_{i,eff}(n_t)$ decreases and approaches $E_i$. 
So we can identify  $E_i$ when the values of  $E_{i,eff}(n_t)$ form a plateau
in $n_t$. If $\psi_i$ is not a ground state it may contain some small
component of a lower energy state, and
then at larger $n_t$ it will decrease to the corresponding lower plateau.
This may create ambiguities which we note are absent for the lowest energy 
state of any given quantum numbers.

\begin{figure}[htb]
\begin	{center}
\leavevmode
\input {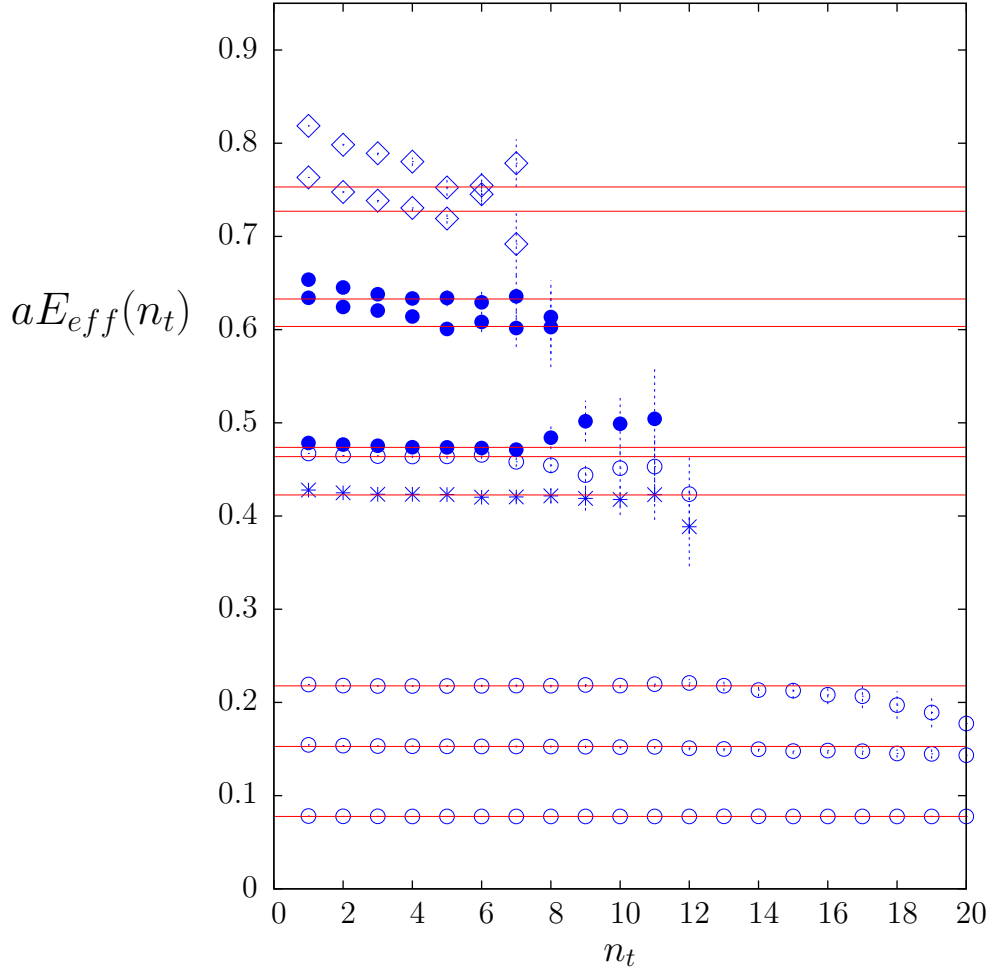}
\end	{center}
\caption{Effective energies extracted from the correlator $C(t=an_t)$ 
using eqn(\ref{eqn_Eeff}). For the absolute ground state
of a flux tube of length $l/a=16,24,32,64$, $\circ$ in ascending order.
Also for the $l=32a$ flux tube: the first, second and third excitations
with $p=0,\, P=+$, $\bullet$; the ground state with $p=2\pi/l$ 
and $P=-$, $\star$; the ground and first excited states
with $p=0,\, P=-$, $\diamond$, shifted upwards by $\Delta E = 0.1$
for clarity. All from SU(6) at $\beta=171$.}
\label{fig_Eeff}
\end{figure}

In Fig.~\ref{fig_Eeff} we display the values of $aE_{eff}(n_t)$ for a number
of states from our SU(6) calculation at $\beta=171$. The open circles 
are for the absolute $p=0, P=+$ ground state, for flux tube lengths
$l/a=16,24,32,64$. For all but the largest $E$, the statistical errors are 
invisible on this plot except at large $n_t$.
The horizontal red lines are the extracted energies. We note how once
the errors become large, at larger $n_t$, the points have a tendency to 
drift away from the plateau value. Nonetheless, even for $l=64a$ where
the plateau is shorter, it is clear that the calculation of $aE_0(l)$
is unambiguous and under good control. This is aided by the fact that
the plateau begins at very small $n_t$: the overlaps are close to $100\%$.

The solid circles in Fig.~\ref{fig_Eeff} represent the 1st, 2nd and 3rd 
excited states of a flux tube with $p=0, P=+$ and with a length $l/a=32$. 
The lightest of these is still well determined, but the two higher
excited states begin to demonstrate the joint problem of a less
good overlap and larger energy making the identification of a plateau
less clear-cut. In fact the normalised overlap of the second state
is  $|c_i|^2\sim 0.9$. This problem becomes more pronounced for the 
lightest two states with  $p=0, P=-$ which are represented by the open 
diamonds. Here the identification of an energy plateau is still 
plausible, but we are clearly leaving the area of certainty. We note
that the upper of the two states has an overlap  $|c_i|^2\sim 0.8$.

As we can see from the latter cases, if the overlap is smaller, there is a 
greater contribution from higher excited states at smaller $n_t$, so that 
the effective energy at those $n_t$ appears larger. If the overlap is
very small then the `signal' will disappear into the statistical errors
long before we reach large enough $n_t$ to see an energy
plateau, and we are then left with what appears to be an ill-defined but
highly excited state. Roughly speaking, it is very hard to identify
states with an overlap of less than $|c_i|^2 \sim 0.5$, and the
energy calculation typically becomes difficult for $|c_i|^2 \leq 0.75$.
As a good example of this, we expect any state that involves multi-trace 
operators to have a much smaller overlap onto our single trace operators
than this, and to be completely invisible in our variational calculation. 
So a state consisting of the ground state flux tube accompanied
by the lightest scalar glueball, although it is certainly present 
and although it is well within the range of the energies we study,
at least at smaller $l$,  
does not appear in the spectrum we calculate. That is to say, 
for larger $N$ all our states are composed of single closed
flux tubes, that sweep out surfaces of the lowest genus.

In summary: the examples in Fig.~\ref{fig_Eeff} show that while our 
results in this SU(6) calculation
are mostly under very good control, this control begins to slip for
the states with highest energies, particularly when such a state is not
the ground state of some quantum numbers. The reader should 
bear this caveat in mind, although the detailed fits from which 
we attempt to draw quantitative conclusions will involve those
states over which we believe we do have good control.

\subsubsection{finite volume corrections}
\label{subsubsection_V}

When we perform spectrum calculations of flux tubes of length $l$ 
on $l\times l_\perp \times l_t$ lattices, it is important to make
sure that corrections due to the finite transverse spatial size,
$l_\perp$, and the finite temporal extent, $l_t$, are negligible.
In our previous papers we have described tests of such corrections
in some detail, and the volumes used in this paper have been chosen
accordingly. However most of those tests were done with a small basis 
of operators, which allowed us to calculate the absolute ground state
but did not allow an accurate determination of excited states.
Since (some) excited states will have a larger total `width' than the
ground state, and hence might be more sensitive to the transverse
boundaries (the temporal extent is not a problem here), 
we have performed a small selection of calculations with
our full operator basis and with our usual statistics, so that we 
can test for finite volume effects at a level of accuracy appropriate
to most of our calculations.

The test we do is in SU(3) at $\beta=21$. Since many of the finite
volume effects are suppressed with increasing $N$, by looking at $N=3$
we are being deliberately conservative. Moreover as we reduce $l$ 
towards $l_c$ we expect the flux tube `width' to diverge since,
for $N\leq 3$, this is a critical point where the correlation length 
diverges. In our SU(6) calculation, the transition is robustly first 
order, and the finite volume corrections at small values of $l$
should be much smaller than for SU(3).

We have performed calculations for two values of $l$, one moderately
short, $l=12a$, and one moderately long, $l=20a$. In physical units 
these lengths correspond to $l\surd\sigma \simeq 2.0, \, 3.5$
respectively. We have not performed calculations for very small
values of $l_\perp$ where the corrections will undoubtedly be large,
but rather have compared results obtained with our `standard' value of
$l_\perp$ with those obtained with significantly larger $l_\perp$.
We calculate the effective energy $E_{eff}(n_t)$ of a particular state 
using eqn(\ref{eqn_Eeff})
where the operator $\psi_i$ is chosen by our variational calculation 
as the `best' operator over our basis for this state. 
In practice our overlaps are good enough that the contribution 
of excited states to $aE_{eff}(n_t)$ is already very small for $t=a$, 
and often negligible for  $t=2a$. The calculations at such small 
values of $t$ are very accurate and so even small finite volume corrections 
should be visible.

\begin{table}[htb]
\begin{center}
\begin{tabular}{|c|c|c|c|c|c|}\hline
\multicolumn{6}{|c|}{ $aE_{eff}(t,l=12a) \ , \ p=0$} \\ \hline
$P$ & state & $t$  & $l_\perp=28a$ &  $l_\perp=22a$ & $l_\perp=18a$ \\ \hline
+   &  1  & $a$  & 0.3177(8)   & 0.3168(6)   & 0.3162(8)     \\
    &     & $2a$ & 0.3169(11)  & 0.3161(9)   & 0.3151(9)     \\ \hline
+   &  2  & $a$  & 0.9319(10)  & 0.9263(9)   & 0.9191(12)    \\
    &     & $2a$ & 0.9142(26)  & 0.9073(21)  & 0.9064(30)    \\ \hline
+   &  3  & $a$  & 1.2333(17)  & 1.2234(13)  & 1.2347(16)    \\
    &     & $2a$ & 1.1374(42)  & 1.1520(35)  & 1.1702(46)    \\ \hline
+   &  4  & $a$  & 1.3404(20)  & 1.3356(15)  & 1.3174(17)    \\
    &     & $2a$ & 1.3059(82)  & 1.2961(58)  & 1.2768(65)    \\ \hline
-   &  1  & $a$  & 1.3537(17)  & 1.3632(16)  & 1.3698(19)    \\
    &     & $2a$ & 1.2978(68)  & 1.3254(60)  & 1.3111(54)    \\ \hline
-   &  2  & $a$  & 1.4603(19)  & 1.4638(18)  & 1.4671(19)    \\
    &     & $2a$ & 1.3776(76)  & 1.3934(62)  & 1.3990(77)    \\ \hline
\end{tabular}
\caption{Effective energies extracted at $t=a$ and $t=2a$ for the
low-lying $p=0$ and $P=\pm$ spectrum. For a short flux tube of length 
$l=12a$, i.e. $l\surd\sigma \sim 2$, on lattices of transverse size 
$l_\perp$ (and temporal extent $l_t=24a$).}
\label{table_Vl12}
\end{center}
\end{table}

\begin{table}[htb]
\begin{center}
\begin{tabular}{|c|c|c|c|c|}\hline
\multicolumn{5}{|c|}{ $aE_{eff}(t,l=20a) \ , \ p=0$} \\ \hline
$P$ & state & $t$  & $l_\perp=22a$ &  $l_\perp=28a$  \\ \hline
+   &  1  & $a$  & 0.5813(8)   & 0.5808(8)  \\
    &     & $2a$ & 0.5770(15)  & 0.5798(14)  \\ \hline
+   &  2  & $a$  & 1.0539(11)  & 1.0557(15)     \\
    &     & $2a$ & 1.0415(40)  & 1.0517(36)    \\ \hline
+   &  3  & $a$  & 1.3618(20)  & 1.3704(18)    \\
    &     & $2a$ & 1.3264(75)  & 1.3532(65)    \\ \hline
+   &  4  & $a$  & 1.3744(19)  & 1.3801(19)    \\
    &     & $2a$ & 1.3571(63)  & 1.3601(76)    \\ \hline
-   &  1  & $a$  & 1.4071(20)  & 1.4050(23)    \\
    &     & $2a$ & 1.3793(81)  & 1.3700(69)    \\ \hline
-   &  2  & $a$  & 1.4267(19)  & 1.4274(20)    \\
    &     & $2a$ & 1.3771(76)  & 1.3898(71)    \\ \hline
\end{tabular}
\caption{As in Table~\ref{table_Vl12} but for a longer flux tube,
$l=20a$, i.e. $l\surd\sigma \sim 3.5$.}
\label{table_Vl20}
\end{center}`
\end{table}

In Tables~\ref{table_Vl12} and ~\ref{table_Vl20} we display 
the values of $E_{eff}(t=a)$ and $E_{eff}(t=2a)$ for flux tubes of
length $l=12a$ and $l=20a$ respectively. We do so for the lightest four 
states with $P=+$ and the lightest two with $P=-$. All this in SU(3)
at $\beta=21$ where $a\surd\sigma \simeq 0.174$. We show how the 
energies change when the transverse size is increased from $l_\perp = 18a$
to $22a$ to $28a$. 

A preliminary aside is that in almost all cases the decrease in 
$E_{eff}(t)$ when we extract it from $t=2a$ rather than $t=a$ is
very small, at the $O(1\%)$ level. This confirms that our variationally
selected operators are in fact very good wavefunctionals for these 
states. 

Comparing the values of $E_{eff}$ for different values of $l_\perp$,
we see from Tables ~\ref{table_Vl12} and ~\ref{table_Vl20}
that the change as we go from the smaller to the largest values of
the transverse lattice size, is often invisible within errors (which 
are typically at the level of a fraction of a percent) and where
there might be some variation, it is almost always $ < 1\%$.
This check therefore provides us with important and convincing evidence 
that the finite volume corrections to our results in this paper are not
significant.

\subsubsection{approaching the critical point in SU(2)}
\label{subsubsection_TcN2}

When the `deconfining' finite volume transition at $l=l_c$
is robustly first order, as it is for $N\geq 5$, it makes sense
to compare the spectrum of closed flux tubes to the predictions 
of an effective string theory all the way down to $l=l_c$. However 
when the transition is second order 
one expects the behaviour of the spectrum as $l\to l_c$ to be governed 
by the critical exponents of the critical point. (Which might also
influence a weakly first-order transition such as in SU(4).) For an SU($N$)
gauge theory in $D=2+1$ these will be in the universality class of a $Z_N$
spin model in two dimensions. That is to say, the behaviour of
$E(l)$ will be governed by these critical exponents as $l\to l_c$
and we do not expect to obtain useful information about the generic
effective string theory for SU($N$) gauge theories by studying this 
limit in such a case.

That the ground state energy does indeed decrease as
\begin{equation}
E_0(l)
\stackrel{l\to l_c^+}{\propto}
(l-l_c)^\gamma
\quad , \qquad
\gamma = 
\begin{cases} 
1  &  \quad SU(2) \\
\frac{5}{6 } &  \quad SU(3) 
\end{cases}
\label{eqn_ltolc}
\end{equation}
was shown numerically a long time ago; see for example Fig 1 in
\cite{N2d3_MT}
for the case of SU(2) and Fig~\ref{fig_E0n3t2} for an example in SU(3)
\cite{unpub_MT}.
It is interesting to see over what range of $l$ the transition from 
eqn(\ref{eqn_ltolc}) to something like the Nambu-Goto behaviour
\begin{equation}
E_0(l) \simeq E^{NG}_0(l)
=
\sigma l
\left( 1 - \frac{\pi}{3} \frac{1}{\sigma l^2} \right)^{\frac{1}{2}}
\label{eqn_E0NG}
\end{equation}
actually takes place. We analyse this for SU(2) where the location 
of the phase transition at $l_c\surd\sigma \simeq 0.9$ is significantly
smaller than the value of $l$ at which eqn(\ref{eqn_E0NG}) would 
imply that the state becomes tachyonic. The calculation
\cite{INI07_MT}
is with $l=4a$ and this is varied by varying $\beta$ (and hence $a$) in 
small increments. At each value of $\beta$ the string tension is calculated
in a separate calculation. As $aE_0(l)$ decreases, the other lattice 
dimensions are increased (ultimately up to $4\times 72\times 144$) 
so as to avoid 
finite volume corrections. The resulting values of $E_0/\surd\sigma$ 
are plotted against $1/l\surd\sigma$ in Fig~\ref{fig_EkTk_n2t4}.
We also plot there the Nambu-Goto prediction in eqn(\ref{eqn_E0NG})
and the linear behaviour in  eqn(\ref{eqn_ltolc}) that is predicted
by universality. We see from  Fig~\ref{fig_EkTk_n2t4} that the transition
between the critical and Nambu-Goto behaviours is very smooth and occurs 
at $l\surd\sigma \sim 1.2$, which is quite far from the critical point
at $l_c\surd\sigma \sim 0.9$. It is interesting to note that
if we expand the Nambu-Goto square root in  eqn(\ref{eqn_E0NG})
and keep only the terms up to $O(1/l^5)$, which are the terms that
have been shown to be universal for any effective string action
\cite{OA},
then we get the curve shown in Fig~\ref{fig_EkTk_n2t4}, which is quite 
close to the numerical values over the whole range of $l$. 
Finally we note that calculations like these have also been made for SU(3)
\cite{TtoTcN3d3}
and for a percolation model
\cite{TtoTcpercd3}.

\section{Results}
\label{section_results}

There are two main features of this paper that are new as compared to our 
earlier work
\cite{AABBMT_d3k1}.\\
(1) We have performed SU(6) calculations at $\beta=171$, which 
corresponds to a small lattice spacing, comparable to that of our older 
SU(3) calculation at $\beta=40$. In contrast to the latter, we cover 
a much wider range of flux tube lengths, $1.2 \leq l\surd\sigma \leq 5.5$.
In addition, we cover a wider range of momenta. Altogether this is
by far our `best' calculation. And the fact that it is at larger $N$
makes it of particular relevance, since the phase transition at $l=l_c$
is robustly first order, so that a simple effective string action might
be applicable all the way down to $l_c$. (And indeed even somewhat below 
$l_c$ if the metastability of the confined phase is strong enough
\cite{BBMT_Hagedorn}.)  Moreover mixings, decays, and 
higher genus contributions should be strongly suppressed. \\
(2) Our comparison with what one expects from an effective string action
will take into account the important recent progress 
\cite{OA}
described in Section~\ref{subsection_strings}.

We have also made some calculations in SU(4) and SU(5) at values of
$a$ that are intermediate between our large and small lattice spacings
in SU(3) and SU(6). (These calculations were primarily performed
to obtain higher representation $k=2$ flux tube spectra
\cite{AABBMT_d3k}.)
We will  occasionally comment upon these, but they will play a role
that is very much secondary to our SU(6) analysis. 
Finally, we have some very high statistics calculations of the
absolute ground state in SU(2) and SU(4) performed with a small
basis of operators (and so not designed for extracting excited states).

In Table~\ref{table_physics} we provide the values of some basic 
physical quantities, for each of the calculations in which we
calculate the closed flux tube spectrum. In each case the string 
tension comes from fitting the ground state energy to the Nambu-Goto 
expression with a $O(1/l^7)$ correction, as expected from the most
recent analytic analyses.
(In actual fact the correction is so small that its particular form
is not important to the extracted value of $a^2\sigma$.) 
The mass gap comes from
\cite{MT98_d3,N_counting}
and the critical length from calculations of the $D=2+1$ deconfining 
temperature in
\cite{Tcd3}.
In Table~\ref{table_physics2} we do the same for the SU(2) and SU(4)
calculations that are dedicated to calculating the ground state.

\begin{table}[htb]
\begin{center}
\begin{tabular}{|c|c|c|c|c|c|}\hline
 & $\beta$ & $l/a\in$  &  $a\surd\sigma$ & $l_c/a$ & $a m_G$ \\ \hline
SU(3)  & 21.0  & [8,32]  &  0.17392(11) &  5.89(2) & 0.760(7) \\
SU(3)  & 40.0  & [16,48] &  0.08712(10) & 11.65(4) & 0.381(3) \\
SU(4)  & 50.0  & [12,24] &  0.13084(21) &  8.09(3) & 0.563(2) \\
SU(5)  & 80.0  & [12,32] &  0.12976(11) &  8.31(2) & 0.548(3) \\
SU(6)  & 90.0  & [8,24]  &  0.17184(12) &  6.37(3) & 0.738(4) \\  
SU(6)  & 171.0 & [14,64] &  0.08582(4)  & 12.47(5) & 0.367(2) \\  \hline
\end{tabular}
\caption{Parameters of our flux tube spectrum calculations: the SU($N$) 
group, the value of the inverse bare coupling, $\beta = 2N/ag^2$, and
the range of flux tube lengths, $l$. Also listed  are some of the 
corresponding physical properties: the string tension, $\sigma$, the
deconfining length, $l_c$, and the mass gap, $m_G$, all in lattice units.}
\label{table_physics}
\end{center}`
\end{table}

\begin{table}[htb]
\begin{center}
\begin{tabular}{|c|c|c|c|c|c|}\hline
 & $\beta$ & $l/a\in$  &  $a\surd\sigma$ & $l_c/a$ & $a m_G$ \\ \hline
SU(2)  & 5.6   & [4,16]  & 0.27316(4)  & 3.43(3) & 1.285(5) \\
SU(4)  & 32.0  & [6,32]  & 0.21523(5)  & 4.98(1) & 0.911(4) \\ \hline
\end{tabular}
\caption{As in Table~\ref{table_physics} but for the two
high statistics calculations dedicated to the ground state 
of the flux tube.}
\label{table_physics2}
\end{center}`
\end{table}

Our earlier work
\cite{AABBMT_d3k1},
comparing the then available SU(3) and SU(6) spectra, provided good evidence 
that any $a$ and $N$ dependence was small. We will therefore initially
assume this in our discussion of the ground state.
That same work demonstrated that the simple
Nambu-Goto free string spectrum is a remarkably good first approximation
to the numerically determined spectrum, and we shall therefore focus
upon that as our initial point of comparison.

We begin with an analysis of the absolute ground state. We then check
for lattice corrections to the continuum limit, and for $O(1/N^2)$
corrections to the $N=\infty$ limit. We then move on to
an overview of our results for the low-lying spectrum and follow
that with a more detailed comparison with current theoretical expectations.

\subsection{Absolute ground state}
\label{subsection_ground state}

The energy $E_0(l)$ of the absolute ground state is our most easily and
accurately calculated energy. However, because the string corrections 
to the linear piece, $\sigma l$, come from the zero-point energies of the 
string excitation modes, they are very small and it is not clear how well 
they can be pinned down. We can see this if we write down what has been 
established for $E_0(l)$ from the universal properties of the effective 
string action,
\cite{OA}
\begin{eqnarray}
E_0(l) = E^{NG}_0(l) + O\left( \frac{1}{l^7} \right)
& = &
\sigma l \left( 1 - \frac{\pi}{3} \frac{1}{\sigma l^2} \right)^{\frac{1}{2}}
+ O\left( \frac{1}{l^7} \right)
\nonumber \\
& = &
\sigma l - \frac{\pi}{6} \frac{1}{l} 
- \frac{\pi^2}{72} \frac{1}{\sigma l^3} 
- \frac{\pi^3}{432} \frac{1}{\sigma^2 l^5}
+ O\left( \frac{1}{l^7} \right) .
\label{eqn_E0univ}
\end{eqnarray}
The second line shows explicitly all the known universal terms. These
are identical to the Nambu-Goto energy in the first line, when that
is expanded in powers of $1/\sigma l^2$ to that order. Since the higher 
order terms in the expansion are of $O(1/l^7)$ the equality between
the two lines in eqn(\ref{eqn_E0univ}) is formally automatic. However
we also know
\cite{OA}
that the operators that arise from the expansion of the Nambu-Goto
action are universal to all orders, and in that sense the resummed
Nambu-Goto term in the top line may be regarded as universal.
Of course this expression becomes tachyonic for 
$l\surd\sigma \leq \sqrt{\pi/3}$, but such values of $l$ are unphysical
when $N$ is large enough for the deconfining transition to be first
order since $l_c\surd\sigma > \sqrt{\pi/3}$ in those cases. (And when the 
transition is second order, $E_0(l)$ is determined by the critical 
behaviour in this range of $l$, as we have seen in Fig~\ref{fig_EkTk_n2t4}.)

We shall begin with our high statistics SU(2) calculation. In
Fig~\ref{fig_EgsQ0_n2b5.6} we plot the ground state energy, normalised
to $\sigma l$. (The variation in the value of $\sigma$ as extracted
from different fits to $E_0(l)$ is negligible in this context.) 
The deviation of $E_0(l)/\sigma l$ from unity exposes 
the $O(1/\sigma l^2)$ corrections to the leading linear term. 
We show the best fit with the Nambu-goto expression, and also the best 
fit using just the leading $O(1/l)$ L\"uscher correction in 
eqn(\ref{eqn_E0univ}). We see that while the former is very close to 
the numerical values, except at the very smallest value of $l$, the 
latter fit, while accounting for much of the deviation from the dominant 
linear $\sigma l$ piece, visibly misses all except the largest $l$ 
values. This tells us that our calculations of $E_0$ are indeed 
accurate enough to be very sensitive to corrections that are of higher 
order than the L\"uscher term.

To analyse this in more detail, we subtract from $E_0(l)$ the Nambu-Goto 
expression $E_0^{NG}(l)$ and plot the difference, against $l\surd\sigma$, 
in Fig.~\ref{fig_DEOAgsQ0_n2b5.6}. On this highly expanded scale we 
can now see a visible difference at the smallest values of $l$,
starting at around  $l\surd\sigma \sim 2$. Even there this is very 
small, $\leq 0.2\%$, until the very smallest value of $l$. Here the
change of sign of the deviation is a clear signal that we are now in
the basin of attraction of the critical point lying just below unity
(see Fig.~\ref{fig_EkTk_n2t4}). It is interesting to see what happens
if we subtract from  $E_0(l)$ a model that includes only the universal 
terms up to $O(1/l)$, $O(1/l^3)$, and $O(1/l^5)$ respectively. This 
is shown in Fig.~\ref{fig_DEOAgsQ0_n2b5.6}. If we (rather arbitrarily)
decide to focus on the values with $l\surd\sigma \geq 1.5$ as perhaps
being outside the influence of the $l=l_c$ critical point, then we clearly 
see that the fits that exclude the known universal term at $O(1/l^5)$ 
have larger deviations from the calculated values and so are
disfavoured. However our numerical values cannot really tell us if 
the full Nambu-Goto expression is better or worse than if we just
include all the known universal terms, i.e. up to and including $O(1/l^5)$. 
It is interesting to note from Fig.~\ref{fig_DEOAgsQ0_n2b5.6} that 
including a non-universal $O(1/l^7)$ correction to $E_0^{NG}$ does not 
really help except in suggesting that all the values below
$l\surd\sigma = 2$ are probably influenced by the deconfining
critical point and hence not reflecting the behaviour of $E_0(l)$
at large $N$.

Since the higher powers of $1/l$ only become significant at smaller $l$,
it is clear from the above SU(2) analysis that it is important
not to be under the influence of a nearby critical 
point at small $l$. For this reason we now turn to SU(4) where the transition
is (weakly) first order. We show in Fig.~\ref{fig_DEOAgsQ0_n4b32}
the analogue of  Fig.~\ref{fig_DEOAgsQ0_n2b5.6}. We now see a monotonic
increase of the deviation from Nambu-Goto as $l$ becomes very small. As we 
can see in  Fig.~\ref{fig_DEOAgsQ0_n4b32} this deviation can in fact be
accounted for by a leading non-universal  $O(1/l^7)$ correction. (How
constraining this fit is, given that the deviations relevant to it are 
only from the two or three lowest values of $l$, is something
we shall address more quantitatively below.)
We also see from Fig.~\ref{fig_DEOAgsQ0_n4b32}, that all this is 
also true if we take as our model all the known universal terms, up to and 
including $O(1/l^5)$. Here the  deviation at the smallest value of $l$ 
is significantly larger than for Nambu-Goto, although, as we can see,
it can also be accommodated by a (larger) non-universal $O(1/l^7)$ 
correction. In this sense, our results slightly favour Nambu-Goto as being
the better model of the two.

Irrespective of these details, the most striking feature of these 
comparisons is how well the known universal part of the effective 
string action describes very short flux tubes. As we see from 
Fig.~\ref{fig_DEOAgsQ0_n4b32} even at $l\surd\sigma \simeq 2$ any 
mis-match with $E_0(l)$ is at most at the $\sim 0.1\%$ level. 
To emphasise the significance of this we show in 
Fig.~\ref{fig_DEOAgsQ0_n4b32} the difference between $E_0(l)$ and
the linear $\sigma l$ piece, with $\sigma$ being determined at the 
largest $l$ value. At $l\surd\sigma \simeq 2$ this difference would 
be $\sim 0.1$, on this plot. So the fact that Nambu-Goto or just the
universal pieces account for this difference, tells us that the 
zero-point energies from the excitations of an ideal thin string
account for  at least $\sim 99\%$ of the string excitation energy at
this $l$. The fact that this should be so for a flux tube that, at 
$l\surd\sigma \simeq 2$, is not much longer than its expected
$\sim 1/\surd\sigma$ intrinsic width, is quite 
counterintuitive. Why should what is essentially a short fat periodic
blob behave so accurately like a thin string? Where are the contributions
of the zero-modes of the massive modes of a flux tube? Clearly our
calculations are telling us that the actual dynamics is somehow very 
much simpler than our naive intuition.

We can be more systematic about this flux tube/string agreement as follows.
Consider Fig.~\ref{fig_DEOAgsQ0_n4b32}. For $l\surd\sigma \simeq 3.5$, we 
see that the $O(1/l)$ L\"uscher correction very nearly equals the difference
$E_0(l)-\sigma l$, and the higher order contributions in $1/l$ contribute 
a negligible amount here. So this confirms the universal  $O(1/l)$ correction 
to a corresponding high level of accuracy. Moving on to 
$l\surd\sigma \simeq 2.5$ and  $l\surd\sigma \simeq 2.1$ we see that
here the next universal $O(1/l^3)$ term accurately accounts for the
gap between the linear plus L\"uscher term contributions and the 
calculated energy, $E_0(l)$, while the higher order terms still
contribute a total amount that is negligible. Thus we confirm 
the universal  $O(1/l^3)$ correction to a corresponding level of accuracy.
Going to smaller $l$ we find that simultaneously with the $O(1/l^5)$  
becoming important, unknown higher order contributions also become important,
so we cannot claim to have evidence for the  $O(1/l^5)$ contribution
with any precision.

It is clear that to do better we need more values of $E_0(l)$ at
the small values of $l$ where the deviations from Nambu-Goto become visible.
To achieve a much higher resolution in $l$ at small $l$, one clearly needs 
a much smaller value of $a$. We make a small step towards this with
our SU(6) calculation at $\beta=171$. Of course, every time we increase
$N$, our accuracy decreases significantly, since the cost of a basic matrix
multiply increases $\propto N^3$. And decreasing $a$ while
keeping the volume fixed in physical units means that the cost grows 
$\propto 1/a^3$. Moreover this SU(6) calculation is designed to calculate 
excited states and so has a correspondingly large basis of operators,
which is computationally expensive. So, despite some compensating factors, 
our SU(6) calculations are, statistically, far less accurate than
the $N=2,4$ calculations at the small values of $l$ relevant here;
e.g. by a factor  $O(10)$ at $l\surd\sigma \sim 2$. At large $l$,
on the other hand, the benefits of the larger overlaps due to the
larger operator basis make the SU(6) calculations much more accurate,
and so they are the best place to confirm the asymptotic linearity of the
ground state flux tube energy. We show the corresponding SU(6) plot in 
Fig.~\ref{fig_DEOAgsQ0_n6f}. Here the first order deconfining
transition is robustly first order and the fact that the shape at small
$l$ is very similar to that in SU(4) reassures us that the weakness
of the first order transition in the latter case plays no significant
role. Apart from that the conclusions are much as for SU(4) except that
the much larger statistical errors lessen the accuracy with which one 
confirms the universal terms, although the smaller systematic errors
(in particular considerably smaller $O(a^2)$ lattice corrections) mean
that one is much more confident about the relevance of the fits to the
continuum limit. In particular we observe that the fitted coefficient 
of the non-Nambu-Goto $1/(l\surd\sigma)^7$ contribution to $E_0(l)$ is
$\sim 0.09$ which is comparable to the coefficient, $\sim 0.05$, of the
corresponding term in the series expansion of Nambu-Goto, and so may be
regarded as taking a `natural' value. 

We have seen that one can describe the deviations from the Nambu-Goto 
expression, $E^{NG}_0(l)$, with a  $O(1/l^7)$ correction. This is a
natural choice since all lower powers are known to be universal.
It is interesting to ask how well this $1/l^7$  power is determined
by our calculations since it might be that this term will also turn
out to be universal so that the first non-universal
term actually starts at a higher power. If it is universal then it is
plausible that it equals the corresponding term of Nambu-Goto, because
of the universality of the scaling-0 operators in terms of which 
the Nambu-Goto action can be expanded
\cite{OA}.
(Although the $D=3+1$ case shows us that a universal term can differ
from Nambu-Goto
\cite{OA}.) 
Accordingly we fit our above SU(4) and SU(6) calculations to the form
\begin{equation}
E_0(l) = E^{NG}_0(l) + \frac{c}{l^\gamma}
\label{eqn_E0corrn}
\end{equation}
and display the $\chi^2$ per degree of freedom in Fig.\ref{fig_NGgamma}.
To expose the importance of this correction, we include in the 
displayed value of $\chi^2$ just the lowest 3 values of $l$, since only 
these are affected significantly by this correction term, and we take the
number of relevant fit parameters to be 2, i.e. $c$ and $\gamma$
in eqn(\ref{eqn_E0corrn}). (The value of $\sigma$ is determined by 
the larger values of $l$.) Although we know on theoretical grounds
that $\gamma \geq 7$ we show the  $\chi^2$ values for fits with  
$\gamma < 7$ as well. We observe 
that $\gamma = 11$ is excluded, $\gamma = 9$ is disfavoured (although not
impossible), while  $\gamma = 7$ is quite strongly favoured. The value
$\gamma = 5$ would be equally plausible, and this is consistent with
our above analysis where we saw that we could not claim significant 
numerical evidence for the universal  $O(1/l^5)$ term. However 
$\gamma = 3$ is strongly excluded, as one would expect from the fact
that we had good numerical evidence for the universal  $O(1/l^3)$ term.
So if we exclude $\gamma=5$ on theoretical grounds, 
we see that the numerical evidence points to the leading
non-Nambu-Goto like contribution being either $O(1/l^7)$ or, less
probably, $O(1/l^9)$.

As we remarked earlier, the contributions of the string excitation modes 
to $E_0(l)$ are very small, because only the zero-point energies 
contribute, and this has an obvious practical disadvantage. But it
also brings an important advantage:  it is plausible to expect that 
the expansion of $E_0(l)$ in powers of $1/\sigma l^2$ is convergent 
throughout the range $l>l_c$ (or nearly all of it) just like $E^{NG}_0(l)$. 
This means that we can analyse the corrections order by order 
at smaller $l$, just as we have done above. Indeed it 
is plausible that an SU(4) calculation with $a$ reduced by a factor 
of two and with the same statistical accuracy as the calculation presented 
here, would be able to confirm the universal $O(1/l^5)$ term quite
accurately, and would simultaneously be able to determine unambiguously 
the power of the leading non-Nambu-Goto-like term.

\subsection{Continuum limit}
\label{subsection_continuum}

In 
\cite{AABBMT_d3k1}
we compared our SU(3) flux tube spectra at the two lattice spacings
listed in Table~\ref{table_physics}. We looked at the lightest few
$p=0$ states and found no significant differences between the two values
of $a$, providing evidence that the $O(a^2)$ lattice corrections were 
negligible in these calculations. However in that paper our SU(3) 
calculations at the smaller value of $a$ did not go to very small values 
of $l$ and, in addition, we did not compare calculations with 
$p=2\pi q/l \neq 0$. 

Our new SU(6) calculation allows us to make a much more complete 
comparison. As we see from Table~\ref{table_physics} the lattice spacings 
of our two SU(6) calculations differ by about a factor of two. So the 
$O(a^2)$ lattice corrections at $\beta=171$ should be about $\sim 1/3$ of
any observed difference between  $\beta=90$ and $\beta=171$.

In Fig~\ref{fig_DENGgsQ0_n6n6f} we compare the ground states in the
two calculations. In each case we fit the Nambu-Goto expression
$E^{NG}_0(l)$ to obtain the corresponding value of the string tension 
$a^2\sigma$. (This is primarily determined by the values at large $l$.)
We see that at both values of $\beta$, one has $E_0(l)=E^{NG}_0(l)$ 
within errors (which are similar for both calculations) down to
$l\surd\sigma \simeq 1.7$, i.e. down to these very short flux tubes 
there are no visible lattice corrections to the continuum Nambu-Goto
expression. At the smallest common value of $l$, $l\surd\sigma \simeq 1.37$, 
we do appear to see a difference, although it is only at the 2 standard 
deviation level of significance. While it is amusing that a
naive continuum extrapolation would bring the $a\to 0$ value at this $l$
closer to the Nambu-Goto prediction, the errors are too large 
for this to be a significant observation. All this suggests that if there
are any significant $O(a^2)$ corrections to our values of $E_0(l)$, they
are confined to $l\surd\sigma \leq 1.4$. We note that any such 
uncertainty would only affect 
our determination in Section~\ref{subsection_ground state} of the 
non-universal correction to Nambu-Goto, and not the evidence for the 
lower-order universal terms.

We now compare some of the excited states in the two calculations.
In Fig.~\ref{fig_EQ0_n6n6f} we plot the energies of the lightest 
three $P=+$ states and the lightest two with $P=-$, and compare
these to the Nambu-Goto predictions, which for the excited states are
completely parameter-free. As we can see, and have observed 
in our earlier calculations
\cite{AABBMT_d3k1},
these states coalesce to the first two excited energy levels of the
Nambu-Goto model with the expected degeneracies and quantum numbers, as 
we increase $l$. (See eqn(\ref{eqn_EnNG}) and Table~\ref{table_NGstates}.)
We see that to a good approximation the first excited state shows 
no $a$-dependence until we are down to $l\surd\sigma \simeq 1.37$.
This is just like the ground state, except that here the deviations of 
$E_n(l)$ from $E^{NG}_n(l)$ already become visible at $l\surd\sigma \sim 3$. 
That is to say, here we can claim that our analysis of the
corrections to Nambu-Goto will be largely unaffected by lattice 
corrections. For the four states belonging to the second  energy level,
it is again true that the discrepancy between the two calculations
only becomes large once $l$ is decreased to $l\surd\sigma \simeq 1.37$.
All this suggests that we do not need to be concerned about $O(a^2)$
corrections in our analysis of the excited states, as long as we
avoid placing too much weight on the energies for $l\surd\sigma < 1.4$.

We turn now to the states with non-zero momentum $p=2\pi q/l$ along the 
flux tube axis. The momentum is carried by the `phonons' running along 
the flux tube, and this leads to the energy-momentum dispersion 
relation in eqns(\ref{eqn_NLR},\ref{eqn_NLRmom},\ref{eqn_EnNG}) for
the Nambu-Goto model in the continuum limit. We observe that
as we decrease $l$ at a fixed $q$, $p^2$ grows and can become
the dominant component of the energy. Thus any $O(a^2)$ corrections
to the dispersion relation could lead to a significant shift in 
the total energy. If such corrections affect the individual phonon 
contributions to $p^2$, then we would expect the effect to be
largest for the states where the whole of the momentum is carried by a 
single phonon. These are states with $P=-$. 
(See eqn(\ref{eqn_Pd3}).) For $q=1,2$ this is the only 
state in the lowest $P=-$ energy level (see Table~\ref{table_NGstates})
and so it is easy to identify it. For higher $q$ the degeneracy of the 
lowest $P=-$ energy level grows, but we shall assume that it is the 
lightest state that is relevant and plot that, since we shall see
that the lattice spacing corrections to the dispersion relation
typically act to decrease the energy.

So we plot in Fig~\ref{fig_EgsQall_n6} the energies of the lightest 
$P=-$ states with $q=1$ and $q=2$, for SU(6) at $\beta=90$ (together 
with the $q=0$ $P=+$ ground state for comparison).
The continuous curves are the continuum Nambu-Goto 
predictions: recall that there is no free parameter since 
$a^2\sigma$ is obtained from a fit to the $P=+$ ground state. We see that
the agreement at larger $l$ is excellent, but that as we reduce $l$ there 
are increasing deviations, especially at $q=2$. Here the lattice momentum
at the smallest value of $l$, i.e. $l=8a$, is $ap=\pi/2$. This is half the
maximum momentum which is at $ap=\pi$, since on a lattice $ap=0$ 
and $ap=2\pi$ are the same. At $ap=\pi$ we expect $E(p^2)$ to have
a maximum and to be far from its continuum value, so a significant
lattice correction at  $ap=\pi/2$ would be no surprise. We do not of course
know the correct form of this correction, but what we can do is to
see what happens if we replace $p^2$ by the form that enters the
lattice dispersion relation for $E^2(p)$ for a free scalar field, 
i.e.
\begin{equation}
(ap)^2 \quad \rightarrow \quad  2 - 2 \cos(ap),
\label{eqn_momlat}
\end{equation}
if we use the most local lattice discretisation of the derivative.
Making this replacement in eqn(\ref{eqn_EnNG}) we obtain the dashed curves 
in Fig~\ref{fig_EgsQall_n6}. We observe that the calculated values are
remarkably close to these `lattice' Nambu-Goto predictions.
Of course, to confirm that this is not just an accident, we need to
look at a different $a$, and this we do in  Fig~\ref{fig_EgsQall_n6f}
for our SU(6) calculation at $\beta=171$. Since $a$ is smaller by a
factor of $\sim 2$ here, we would expect lattice spacing corrections
in Fig~\ref{fig_EgsQall_n6f}
to be the same for $q$ as they were for $q/2$ in Fig~\ref{fig_EgsQall_n6}, 
at the same value of $l\surd\sigma$. This is indeed so and, as we see, the
deviations from the continuum NG predictions are well accounted for by
using eqn(\ref{eqn_momlat}). 

So we see that there are significant lattice corrections to $E(p)$
at large $ap$ through its dispersion relation, at least for those states
where all the momentum is carried by a single phonon. Presumably this will 
also be significant in states composed of several phonons, as long as
some of these have large enough momenta. It would be useful to
find a plausible way to
estimate these effects for the general Nambu-Goto state on the lattice.

\subsection{Large-$N$}
\label{subsection_largeN}

Our earlier work provided evidence that the spectrum of 
flux tubes behaves like Nambu-Goto for values of $l$ that are not 
very small, and since this is true for $N=3$ and $N=6$, one
can safely assume that this is also true at $N=\infty$. 
Here we strengthen this observation using our new $N=4,5,6$ calculations. 
And we address in some detail the remaining interesting question: 
can we assume that the deviations from Nambu-Goto 
that we observe at smaller $l$ in SU(6), are similar to those that
one would see at $N=\infty$? 

We begin by comparing the deviations of the ground state energies 
from their best Nambu-Goto fits. We do so in 
Fig.~\ref{fig_DENGgsQ0_n3fn4n5n6f} for the small-$a$ SU(3) and SU(6)
calculations in  Table~\ref{table_physics} and for the SU(4) and SU(5)
calculations listed there. We observe in Fig.~\ref{fig_DENGgsQ0_n3fn4n5n6f} 
no evidence for any $N$ dependence, even on this expanded scale. 
Since all of these calculations are at small enough lattice 
spacings that we can expect any $O(a^2)$ corrections to be small,
we can conclude that this $N$-independence also holds in the 
continuum limit.

In 
\cite{AABBMT_d3k1}
we compared the first excited $q=0, P=+$ state in our $\beta=21$ SU(3) 
and $\beta=90$ SU(6) calculations, and found no significant differences. 
Since we have just seen that $O(a^2)$ corrections are small at these $a$
(except where high momenta are involved), this provides some evidence that
the $O(1/N^2)$ corrections are already small for $N=3$. In
Fig.~\ref{fig_EQ0_n3n6} we repeat the comparison, but now include
the next two excited states in this channel, as well as the lightest
two $q=0, P=-$ states. As we can see, as we increase $l$ these extra 
four states converge rapidly to the second excited Nambu-Goto energy level 
And for $l\surd\sigma > 3$ there
appears to be no significant difference between the SU(3) and SU(6)
values. However for  $l\surd\sigma \lesssim 3$ the discrepancy
between the two calculations rapidly grows, especially for the lighter 
of the two $P=+$ states. So here, at smaller $l$, the $O(1/N^2)$ 
corrections appear to be large, at least for SU(3).

To see if they are also significant for SU(6), we
compare the spectrum of our new SU(6) calculation at $\beta=171$
with the SU(4) and SU(5) calculations listed in Table~\ref{table_physics}
as well as with SU(3) at $\beta=40$. These are all at smaller $a$ than the
calculations in Fig.~\ref{fig_EQ0_n3n6}. We begin with the three lightest
$q=0, P=+$ excited states, which we show in Fig.~\ref{fig_EP+Q0_n3fn4n5n6f}.
We see that the first excited state indeed shows no significant
$N$ dependence. For the higher excited states it appears that the
only large and significant $N$ dependence comes from SU(3),
as long as we remain with $l\surd\sigma \gtrsim 2$. 
In  Fig.~\ref{fig_EP-Q0_n3fn4n5n6f} we see that the same appears to 
be the case for the lightest two $q=0, P=-$ states.

It may be that the anomalously large deviations seen at small $l$
for the SU(3) excited states have to do with the fact that the small-$l$ 
deconfining transition is second order for SU(3). Irrespective of that, the
evidence is that any $O(1/N^2)$ corrections will be very small
for SU(6), certainly as long as we remain at $l\surd\sigma \gtrsim 2$,
and probably significantly below that. So in the case of SU(6) this leaves 
a substantial range of $l$ where the observed  deviations from 
Nambu-Goto are both significant and representative of the $N=\infty$ theory.

\subsection{Excited states : an overview}
\label{subsection_ex_overview}

If we consider an excited state of a long string in the Nambu-Goto 
model with $p=0$, its energy can be expanded in powers of $1/\sigma l^2$ as in 
eqn(\ref{eqn_EnNGexpansion}). We also know
\cite{OA}
that the terms up to and including $O(1/l^5)$ are universal and equal to the
corresponding terms in the Nambu-Goto expansion with any non-universal terms 
starting at $O(1/l^7)$ or later. 
As we see from  eqn(\ref{eqn_EnNGexpansion}), this expansion 
only converges for
\begin{equation}
l\surd\sigma \, > \, l_c^{NG}(n)\surd\sigma 
= \left\{8\pi \left(n-\frac{1}{24}\right)\right\}^{\frac{1}{2}}
\simeq \begin{cases} 4.91 & \quad  n=1 \\
7.02 & \quad n=2 \\
8.62 & \quad n=3 \\
\cdots &
\end{cases} 
\label{eqn_NGcvgce}
\end{equation}
where $n=1,2,3$ correspond to the first, second and third excited energy
levels in the $p=0$ sector. Nonetheless, as we see from Fig~\ref{fig_EQ0_n6f}
where we plot our lattice values of $E(l)$ for the lightest $p=0$ 
eigenstates in SU(6) at $\beta=171$, these values do in fact remain very 
close to the full Nambu-Goto prediction for values of $l$ that are
well below the point at which the power expansion ceases to converge -- 
and where we need to use the full, resummed (square root) formula in 
eqn(\ref{eqn_EnNGexpansion}). So it is no surprise that if we plot the 
sum of the known universal terms, as in Fig~\ref{fig_EQ0_n6f}, we find
that they are unable to account for this precocious onset of free string 
behaviour.

We see the same phenomenon with the $p \neq 0$ eigenstates whose energies 
we calculate. To better expose any (dis)agreement between Nambu-Goto and
our calculated values for these states, we construct the
`excitation energy (squared)'
\cite{AABBMT_d3k},
\begin{equation}
\Delta E^2(q,l)
=
E^2(q;l) - E_{0}^2(l) 
- \left ( \frac{2\pi q}{l}\right )^2
\stackrel{NG}{=} 4\pi\sigma (N_L+N_R),
\label{eqn_exq} 
\end{equation} 
where  $E_0(l)$ is the calculated energy of the (absolute) ground 
state (with $p=0$), and where we show the Nambu-Goto prediction
that follows from eqn(\ref{eqn_EnNG}). (We choose to subtract the 
calculated value of $E_{0}(l)$ rather than the Nambu-Goto prediction 
for it. In the present context the difference is insignificant.)
Note that while eqn(\ref{eqn_exq}) assumes a continuum dispersion 
relation, we shall occasionally modify it as in eqn(\ref{eqn_momlat}). 

As our first example, consider the lightest $P=-$ state at each nonzero
momentum $p=2\pi q/l$, with $q=1,2,3,4,5$. In the Nambu-Goto model we
can expect these to be single phonon states $a_q|0\rangle$. (Certainly for 
$q=1,2$ and plausibly for $q>2$.) In Fig~\ref{fig_DlatEgsQall_n6f}
we plot the excitation energies from our SU(6) calculation at $\beta=171$, 
using eqn(\ref{eqn_exq}) modified by the lattice dispersion relation in
eqn(\ref{eqn_momlat}). This is a replotting of Fig~\ref{fig_EgsQall_n6f} 
designed to render more precise the comparison with Nambu-Goto. Where our 
errors are reasonably small, i.e. for $q=1,2,3$, we observe very good 
agreement with Nambu-Goto for $l\surd\sigma > 1.5$. For $q=4,5$ there
appears to be some systematic upward deviation from Nambu-Goto, but we note
that the errors are much larger here, because these states have much larger 
energies, and for the same reason the systematic errors are also larger.
In particular, as emphasised in Section~\ref{subsection_errors} we cannot 
go to higher $t$ in the correlation functions
so as to check that there is no  contamination from the presence of higher 
excited states, and this creates a systematic error that grows with $q$
and which leads to an overestimate of $E(l)$. It is plausible that this
explains the overshoot visible in Fig~\ref{fig_DlatEgsQall_n6f}.

In Figs~\ref{fig_DEQ1_n6f},~\ref{fig_DEQ2_n6f} and ~\ref{fig_DEQ3_n6f}
we show the excitation energies of the lightest few states 
with longitudinal momenta $p=2\pi/l,\, 4\pi/l,\, 6\pi/l$ respectively.
Again, where the errors are small, we observe excellent agreement
with the Nambu-Goto prediction: roughly speaking, for 
$l\surd\sigma \geq 3$ for all our states, and for $l\surd\sigma \geq 2$ 
for the lighter ones with the smallest errors. 

In summary, we have seen that all our calculated values of the flux tube
energies are remarkably close to the Nambu-Goto prediction, down to
values of $l$ well below the point at which the series expansion
in powers of $1/\sigma l^2$ has ceased to converge and all orders
of Nambu-Goto are important to the resummed expression. We see this for
the lightest $q=0$ states in Fig.~\ref{fig_EQ0_n6f}, for the $P=-$
ground states with $q=1,2,3,4,5$ in Fig.~\ref{fig_DlatEgsQall_n6f},
and for the lightest few states in each of the $q=1,2,3$ sectors in
Figs~\ref{fig_DEQ1_n6f},~\ref{fig_DEQ2_n6f} and ~\ref{fig_DEQ3_n6f}.
This provides much more evidence for our earlier claim
\cite{AABBMT_d3k1}
that a good first approximation to the effective string action
must be the full resummed Nambu-Goto action, and that the corrections 
to that must be small even down to small values of $l$.  
This observation leads to the question we address in the next subsection: 
what do our results tell us about the specific nature of these corrections?

A final more general aside on all these results is that there 
appears to be no room for any 
states in addition to those that converge at larger $l$ to the
Nambu-Goto predictions. The fact that the corrections to Nambu-Goto remain 
small even at small $l$, means that even here, where we are looking 
over a range of energies $E - E_0(l) \gg \surd\sigma$, there is no 
sign of the extra states one might expect to arise from the excitation
of massive $\sim O(\surd\sigma)$ modes. 

\subsection{Excited states: fits}
\label{subsection_ex_fits}

The correction terms to Nambu-Goto are known 
\cite{OA} 
to begin with a power that is no less then seven, so we
assume we can write
\begin{eqnarray}
\frac{1}{\surd\sigma}E_n(l)
& = &
\frac{1}{\surd\sigma} E^{NG}_n(l) 
+ \frac{1}{\surd\sigma} \Delta E_n(l) \\ \nonumber
& \stackrel{l\to\infty}{=} &
\frac{1}{\surd\sigma} E^{NG}_n(l) 
+ \frac{c}{(l\surd\sigma)^7}
\left\{1 + \frac{c_1}{l^2\sigma} + \frac{c_2}{(l^2\sigma)^2} + \cdots
\right\}
\label{eqn_NGcor} 
\end{eqnarray} 
where the coefficients $c,c_1,...$ are unknown. For large $l$
we can also expand  $E^{NG}_n(l)$ in powers of $1/l^2\sigma$
as in eqn(\ref{eqn_EnNGexpansion}). However as we decrease $l$ this
latter expansion diverges when $l=l_c^{NG}(n)$ in eqn(\ref{eqn_NGcvgce}).
Nonetheless we have seen that even well below this value of $l$, 
at which all the terms in the expansion become important, the values 
of $E_n(l)$ remain very close to the resummed Nambu-Goto expression.
It is possible that the correction $\Delta E_n(l)$, 
regarded as a series in $1/l$, 
also diverges at some finite $l=l_d(n)$ that is within or above the 
range  of our calculations, i.e. $l_d \gtrsim l_c$. In that case, 
the fact that our calculated energies differ from Nambu-Goto by a finite
amount for all $l \gtrsim l_c$ tells us that the series of correction terms 
in eqn(\ref{eqn_NGcor}) can be resummed for  $l_c \lesssim l < l_d$ just like 
Nambu-Goto can be for $l < l_c^{NG}$. If on the other hand 
$l_d < l_c$ then we can expect
the leading $(1/l^7)$ term to dominate  $\Delta E_n(l)$ most of the
way down to $l_c$. We attempt to capture these
expectations with the heuristic parameterisation,
\begin{equation}
\frac{1}{\surd\sigma} \Delta E_n(l) 
=
\frac{c}{(l\surd\sigma)^7} 
\left(1 + \frac{c^\prime}{l^2\sigma} \right)^{-\gamma}
\simeq
\begin{cases} 
\frac{c}{(l\surd\sigma)^7}   &  \quad l \gg l_{d} \\
\frac{c c^{\prime-\gamma}}{(l\surd\sigma)^{7-2\gamma}}   &  \quad  l \ll l_{d}
\end{cases}
\label{eqn_NGcor_resum} 
\end{equation} 
where $l^2_d \sigma = c^\prime$. The expression for $\Delta E_n(l)$
could certainly be more complicated, with more parameters. However, as
we shall soon see, our calculated values of  $E_n(l)$ are not good enough
to justify an analysis with more parameters, and in any case
eqn(\ref{eqn_NGcor_resum}) embodies the essential feature of a
resummed formula: the correct $\propto 1/l^7$ behaviour at large $l$,
with the possibility of a quite different effective power behaviour at 
small $l$.  

As an aside we remark that one exception to this is the absolute ground 
state, where the Nambu-Goto series expansion converges for all $l>l_c$,
at least for $N\geq 4$ where the phase transition at $l=l_c$ is first
order. So here it is particularly plausible (although not guaranteed)
that the leading $O(1/l^7)$ correction to Nambu-Goto continues to dominate 
the total correction, $\Delta E_0(l)$, down to our smallest values of $l$
i.e. that $l_d(n=0) < l_c$. This is of course precisely what our
analysis of the ground state in Section~\ref{subsection_ground state}  
suggested. 
  
To analyse the correction to the Nambu-Goto prediction for the excited 
states, we need to look at those states where $E(l)$ is accurately 
determined at small $l$, and where there are simultaneously substantial 
deviations from Nambu-Goto. Looking at Fig.~\ref{fig_EQ0_n6f} and
Fig.~\ref{fig_DEQ1_n6f}, some obvious candidates are the first and 
second excited states in the $p=0,\, P=+$ sector and the lightest
state in the  $p=2\pi/l,\, P=+$ sector. We consider these in turn.

We begin with the first excited state in the $p=0,\, P=+$ sector, plotting 
in Fig.~\ref{fig_DENGex1Q0_n6f} the deviation of its energy 
from the Nambu-Goto prediction. We include several fits. First, we show
a `fit' to $c/l^7$: this (or a higher power) is expected to be the 
leading correction term as $l\to\infty$. Its complete failure
to describe the observed deviation (and a higher power would evidently 
perform even worse) emphasises that we must here
be in a region of $l$ where the correction to Nambu-Goto can
no longer be expressed as a convergent series but must be resummed.
We also show two fits that are variations on the form given in 
eqn(\ref{eqn_NGcor_resum}):
\begin{equation}
\frac{1}{\surd\sigma} \Delta E_n(l) 
=
\begin{cases} 
\frac{-1.0}{(l\surd\sigma)^7} 
\left(\frac{1}{25.0}+\frac{1}{l^2\sigma} \right)^{-2.75} & \\
\frac{-1.0}{(l\surd\sigma)^7} 
\left(\frac{1}{31.1}+\frac{1}{l^2\sigma} \right)^{-2.61} & 
\end{cases}
\label{eqn_corfit_ex1q0} 
\end{equation}
The first is the lower solid curve in  Fig.~\ref{fig_DENGex1Q0_n6f},
while the second is the one slightly higher. The latter
has a slightly better overall $\chi^2$ but the former fits the points 
in the middle somewhat better. In any case we see that:\\
(1) in both cases the radius of convergence of the correction term,
$\sim 25-31$, is not far from that of the Nambu-Goto expression, 
i.e. $\sim 24$; \\
(2) the coefficient of the leading $1/(l\surd\sigma)^7$ term at large 
$l$ is in the range $0.7-0.8 \times 10^3$ which is of the same order
as the coefficient of the $1/(l\surd\sigma)^7$ term in the Nambu-Goto 
expansion, which is $ \sim 1.3 \times 10^3$. \\
This tells us that the correction terms we are seeing have coefficients
of the same order as those of the corresponding Nambu-Goto terms, as
one would expect if they were to arise from `natural' correction terms 
in the effective string action.

It is interesting to ask how well the calculations shown in
Fig.~\ref{fig_DENGex1Q0_n6f} constrain the parameters of the fit
in eqn(\ref{eqn_corfit_ex1q0}). We perform separate fits that 
include (case A) and exclude (case B) the very smallest value of $l$, 
and find: 
\begin{eqnarray}
\gamma & = & 2.61\binom{12}{17}, \quad c^\prime= 31\binom{26}{9}   
\qquad : \quad \text{fit \ A} \\ \nonumber
\gamma & = & 2.88\binom{28}{33}, \quad c^\prime= 20\binom{41}{7}   
\qquad : \quad \text{fit \ B} 
\label{eqn_corpar_ex1q0} 
\end{eqnarray}
This confirms that while the power $\gamma$ is quite well constrained,
the value of $c^\prime$ is much more weakly constrained. Indeed within
two or three standard deviations, one can set $c^\prime \to \infty$.
That is to say, the values of $E(l)$ plotted in Fig.~\ref{fig_DENGex1Q0_n6f}
provide little evidence for an asymptotic $\propto 1/l^7$ behaviour.
We demonstrate this in Fig.~\ref{fig_DENGex1Q0_n6f} with the black
dashed curve which is simply given by $-1.12/(l\surd\sigma)^{2.51}$
and does not incorporate this asymptotic $1/l^7$ correction. This fit is 
visibly worse, but not very much worse. The reason is obvious:
the onset of the $1/l^7$ dependence only occurs at larger $l$ where
the deviations from Nambu-Goto have become very small. Thus our use
of a $1/l^7$ prefactor in eqn(\ref{eqn_NGcor_resum}) must be primarily
motivated by the theoretical analysis, with some significant support from our
earlier analysis of the absolute ground state. Assuming such a prefactor,
the location of the transition region, around $l^2\sigma \sim c^\prime$, 
and the corresponding coefficient of the asymptotic $1/l^7$ correction are,
as we  have noted already, not far from the corresponding Nambu-Goto values,
and hence quite `natural'.

It is interesting to see whether our other calculations
support this analysis. In Fig.~\ref{fig_DENGex1Q0_n4n5n6} we plot the
same quantity as in Fig.~\ref{fig_DENGex1Q0_n6f}, but this time for
the SU(4) and SU(5) calculations at a coarser value of $a$ and 
for SU(6) at the much coarser $a$ corresponding to $\beta=90$.   
(We exclude SU(3) from consideration because of its second order
phase transition at $l=l_c$.) We plot just one curve from 
Fig.~\ref{fig_DENGex1Q0_n6f}, the upper fit listed in 
eqn(\ref{eqn_corpar_ex1q0}). We see that this curve is a good
fit to the SU(6) values (except at the smallest value of $l$)
and adequate for SU(5), and for the larger $l$ values of SU(4).
Bearing in mind possible $O(a^2)$ lattice corrections to this
SU(6) calculation, and possible effects from the weak first order nature
of the transition in SU(4), the level of agreement we see with this
curve is reassuring. On the other hand, these calculations are
visibly rougher than the new SU(6) calculations shown in
Fig.~\ref{fig_DENGex1Q0_n6f}, and it is clear that they would not
add much to any quantitative analysis.

In Fig.~\ref{fig_DENGex2Q0_n6f} we show a corresponding plot 
for the second excited state, taken from our new SU(6) calculation.
We also show the next higher excited state, which at large $l$ converges
to the same energy level, since the combined plot highlights
a potential ambiguity: the two states might actually `cross' at 
$l\surd\sigma \simeq 3$. It is obviously important, when we try to fit 
the energy of the state with a correction to Nambu-Goto, to be sure that it
is the same state at all $l$. In principle this ambiguity can be
resolved by a careful examination of the operators that contribute
to the wavefunctionals of the two energy eigenstates, as a function 
of $l$. We do not attempt to do so here, but instead will assume that 
the lightest values of $E$ do indeed belong to the same state at all $l$.
Returning to Fig.~\ref{fig_DENGex2Q0_n6f}, it is clear that these 
calculations are much less precise than those in Fig.~\ref{fig_DENGex1Q0_n6f}, 
so we cannot expect to draw very detailed conclusions. Nonetheless
it is clear that the corrections are larger and probably increase
more slowly with decreasing $l$. We show some fits of the form in
eqn(\ref{eqn_NGcor_resum}). The fits with values of $\gamma$ similar to 
those of the fits in Fig.~\ref{fig_DENGex1Q0_n6f}, have much larger values 
of $c^\prime$: the transition region to the asymptotic $\propto 1/l^7$
behaviour occurs at $l\surd\sigma \sim \surd c^\prime \sim 15 - 30$ rather 
than the $\sim 4 - 8$ in Fig.~\ref{fig_DENGex1Q0_n6f}. (From the 
Nambu-Goto model one might expect a factor of only $\sim \surd 2$.)
In addition such fits do not seem to capture well the overall trend. In
particular, the alternative curve which rises at very small $l$, 
because the power $\gamma = 4.16$ leads to the $(l^2\sigma)^\gamma$
factor overwhelming the $1/l^7$ factor at small $l$ , looks `better',
and also has a modest  $c^\prime$. However the quality of these calculations
is not good enough to justify anything more than such impressionistic remarks.

We now turn to the lightest state with non-zero momentum, $p=2\pi/l$, 
and with $P=+$. In  Fig.\ref{fig_DENGgsQ1P+_n6f} we plot the energy
minus the value predicted by Nambu-Goto, using our SU(6), $\beta=171$
calculation. We also plot a curve that is not a fit, but is exactly
the same as the $\gamma=2.75$ fit in eqn(\ref{eqn_corfit_ex1q0}) which
was displayed in Fig.~\ref{fig_DENGex1Q0_n6f}. In 
Fig.~\ref{fig_DENGgsQ1P+_n4n5n6} we show the energy of the same
state from some of our other calculations, together with exactly
the same curve. Again this calculation is not so precise that 
this comparison can be considered unambiguous, but what we do learn is 
that the correction to Nambu-Goto is certainly consistent with a resummed 
series with the correct large $l$ behaviour and with natural coefficients.

In the  $p=2\pi/l$ sector, the above $P=+$ state is not the simplest,
containing as it does two phonons $\sim a_2a_{-1}|0\rangle$. The
simplest state has  $P=-$ and has just one phonon   $\sim a_1|0\rangle$.
In Fig.~\ref{fig_DENGgsQ1P-_n6f} we plot the energy of the latter minus
the Nambu-Goto prediction. The values of $E(l)$ are very accurate,
and the reason we left it out till now is that it was already clear from
Fig.~\ref{fig_DEQ1_n6f} that the corrections were very small and only
occur at very small $l$. At such small $l$ even for the lowest non-zero 
momentum, i.e. $p=2\pi/l$, we
need to worry about lattice corrections to the dispersion relation,
and so we plot values using both the lattice free-field and continuum
dispersion relations, as shown. Although we can see in 
Fig.~\ref{fig_DENGgsQ1P-_n6f} that this does create a visible shift
in the values of $E-E_{NG}$,  this shift does not affect our conclusions.
The first is that the deviations only begin at very small $l$,
just as for the absolute ground state in  Fig.~\ref{fig_DENGgsQ0_n6n6f}.
The second is that the deviation then grows rapidly with decreasing $l$ 
consistent, just as in Fig.~\ref{fig_DENGgsQ0_n6n6f}, with the leading 
asymptotic $\propto 1/l^7$ dependence. By contrast, 
as shown in Fig.~\ref{fig_DENGgsQ1P-_n6f}, it is certainly not 
consistent with a much smoother resummed $l$ dependence of the kind we 
saw working well in Fig.~\ref{fig_DENGex1Q0_n6f} and 
Fig.\ref{fig_DENGgsQ1P+_n6f}. This naturally raises the question whether
the same might not apply to other states containing a single phonon.
We therefore repeat the exercise for the lightest $p=4\pi/l$, $P=-$
state, which should be just $a_2|0\rangle$. The results, in 
Fig.~\ref{fig_DENGgsQ2P-_n6f}, are both good and bad. The good is
that we can confirm that, irrespective of the dispersion relation
employed, here too the deviations are negligible except at very small $l$.
The bad is that at the smallest $l$, where the deviations become 
significant, they depend so strongly on the dispersion relation used
that it is hard to draw any useful conclusion about the functional
form of the correction.

In summary, most of the excited states that are accurately calculated
unambiguously demand a correction to Nambu-Goto that varies 
much more slowly with $1/l$ than the $\sim 1/l^7$ leading asymptotic
behaviour that is expected theoretically and for which we have evidence
from our analysis of the ground state. This gross discrepancy implies 
that in the range of $l$ relevant to our fits, the correction to
Nambu-Goto can no longer be expressed as a convergent series in $1/l$
but has to be resummed, just like the Nambu-Goto series itself. 
However, and unexpectedly, the $p \neq 0$ ground states 
with $P=-$, i.e. those with a single phonon, are consistent with just 
a leading $O(1/l^7)$ correction term just like the absolute ground state.

\subsection{Reflection parity, $P_r$}
\label{subsection_Pr}

Our choice of operators in Table~\ref{Operators}, was not 
originally made with a view to labelling the states by their reflection
parity, $P_r$. That is to say, for the majority of operators in the
Table we have not included the corresponding $x$-reflected operators.
Moreover, some of the operators are intrinsically $P_r = +$ after we
sum over translations in $x$ to produce $p=0$. So our
overlap onto $P_r=-$ states is likely to be smaller than onto the
$P_r=+$ states. Since one of our goals is to search for states 
that manifest the excitation of massive modes and are additional to 
the stringy states that rapidly converge to Nambu-Goto, it is important
that we have a good overlap onto sectors of all quantum numbers,
including $P_r=-$, since otherwise we trivially risk not observing such 
extra states even if they are present.

Since the operators in Table~\ref{Operators} are mostly far from being 
orthogonal, it is quite possible that even if for most operators 
we do not have their exact $x$-reflections (where different), 
we may well have operators that are approximate $x$-reflections.
This is of course hard to 
know just by staring at the operators. However if we calculate the 
lightest two $p=0$ states with $P=-$, which in Nambu-Goto should be
the $ \sim \{ a_2a_{-1}a_{-1} \pm a_1a_1a_{-2} \}|0\rangle$ combinations
with $P_r=\pm$, we see  from Fig.~\ref{fig_EQ0_n6f} that these two states 
do indeed converge rapidly to the appropriate Nambu-Goto energy level.
So we certainly have enough overlap to identify the lightest $P_r=-$
Nambu-Goto-like state. We see this in more detail in our effective energy
plot in Fig.~\ref{fig_Eeff}, where we can estimate that in this
particular case we have an overlap of $\sim 80\%$ onto the $P_r=-$
state. While this is not as good as the $>90\%$  overlap onto the
associated $P_r=+$ state, it  reassures us that our overlap onto the 
$P_r=-$ sector is large enough that there is no special reason
to worry about missing states in this particular sector.  

We complete this section with an analysis of the $l$-dependence of the 
lightest of the two $p=0$, $P=-$ states discussed above. The first question
concerns their $P_r$ quantum numbers. Due to the blocking choices, it
turns out that for $l/a=32,48,64$ we have enough pairs of 
operators that are exact $P_r$ transforms of each other that we are able
to cleanly identify the $P_r$ quantum numbers. This tells us that
the lightest of the the two states is the one with $P_r=+$. This
is what one would naively expect: any splitting between the $P_r=\pm$ 
states leaves the one that is antisymmetric in $x$ as the heavier one.
Turning then to the lighter $P_r=+$ state, we plot its 
deviations from Nambu-Goto in Fig.~\ref{fig_DENGgsQ0P-_n6f}.
What we see is quite interesting. While the lowest $l$
values are consistent with a steep fall-off, it is difficult not
to ignore the values below Nambu-Goto around $l\surd\sigma \sim 3$
or those  above, around $l\surd\sigma \sim 4.5$. In fact it is
hard not to see here a clear hint of an oscillating behaviour around
something like the top fit in eqn(\ref{eqn_corfit_ex1q0}), which we
have also plotted in  Fig.~\ref{fig_DENGgsQ0P-_n6f}. Although we have 
not remarked upon it earler in this paper, hints of oscillations can be 
equally found in Fig.~\ref{fig_DENGgsQ1P+_n6f}, in the third excited 
state in Fig.~\ref{fig_DENGex2Q0_n6f} and elsewhere. And where we have 
more than one state converging on an energy level, with both
oscillating, then states may also `intertwine'. Our fitting functions 
are of course only heuristic and chosen for simplicity; it is entirely
possible that the real variation can also incorporate something like 
a Bessel-function oscillation. A relatively minor improvement in 
the quality of the calculation would unambiguously clarify this
issue of possible oscillations.

\section{Summary and conclusions}
\label{section_conclusion}

We have calculated the low-lying energy spectrum of closed flux tubes 
with lengths ranging from the moderately long, $l\surd\sigma \simeq 5.5$,
down to the very short, close to the `deconfining' phase transition 
at $l=l_c\simeq 1.1/\surd\sigma$. By contrast
analytic investigations of the effective string action 
\cite{OA},
which make powerful predictions for the first few terms of the
expansion of $E_n(l)$ in powers of  $1/l^2\sigma$, focus on
large values of  $l$ where such a series 
converges and where the energy gaps are small, i.e.  
$E_n(l)-E_0(l) \ll \surd\sigma$. Thus our calculations
are mostly complementary to the analytic ones, although there is a 
substantial overlap for the absolute ground state and some overlap, 
at our largest values of $l$, for the very lightest excited states. 

We checked that for our main SU(6) calculation the $O(a^2)$ 
lattice corrections are small, except possibly for the very smallest 
values of $l$ and for large momenta where deviations from the continuum 
energy-momentum dispersion relation can be significant (as shown in 
Fig.~\ref{fig_EgsQall_n6f}). We also saw from comparisons such as those 
in Figs~\ref{fig_DENGgsQ0_n3fn4n5n6f}-\ref{fig_EP-Q0_n3fn4n5n6f},
that $N=6$ is very close to $N=\infty$ (with any significant correction
once again limited to the smallest $l$). Thus we can assume that the states 
whose energies we calculate using our basis of single-trace operators 
contain only a single flux tube, so that the partition function is over 
surfaces of lowest genus that wrap around the torus. Such a partition 
function can be calculated using what we have recently learned  about 
the universal terms of the effective string action
\cite{OA},
with corresponding predictions for the energy spectrum. 
Comparing this to our numerical spectrum is one of the main motivations 
of this work. However equally interesting is to see how much our more 
accurate calculations confirm and quantify our earlier observation
\cite{AABBMT_d3k1}
that the free string Nambu-Goto model provides a very good description
of the low-lying energy spectrum, even at very small $l$, and to 
attempt to find some states that reflect the excitation
of the additional massive modes of the flux tube.
 
Our most accurate calculation is that of the absolute ground state. 
In this case the stringy corrections to the dominant linear $\sigma l$
term are small because they come from the zero-point energies of the modes 
of the string. Thus it is plausible that the expansion of $E_0(l)$ in
powers of $1/l^2$ converges all the way down to $l_c$, just like the
Nambu-Goto energy, and that we can attempt to identify the leading
correction. We were able to show, in Section~\ref{subsection_ground state},
that both Nambu-Goto and the sum of known universal terms 
work very well down to small $l$. In the process we found that we
could find good numerical evidence for the universality of the $O(1/l)$ and 
$O(1/l^3)$ terms (which are the same as Nambu-Goto) but were not
able to test the known universality of the $O(1/l^5)$ term.
Assuming the latter, we can then predict that the first term 
that differs from Nambu-Goto is most likely to be $O(1/l^7)$ or, less 
likely, $O(1/l^9)$, but not a higher power. (See Fig.~\ref{fig_NGgamma} 
and also Figs~\ref{fig_DEOAgsQ0_n2b5.6}-\ref{fig_DEOAgsQ0_n6f}.)
This is consistent with the known universality results
\cite{OA}
that predict a power $\geq 7$.

For the excited states the analysis is very different.
As we see in Figs~\ref{fig_EQ0_n6f}-\ref{fig_DEQ3_n6f}, the calculated 
energies are very close to Nambu-Goto well below the value of $l$ at which
the Nambu-Goto power series no longer converges. Here all orders are 
important, and we have to use the well-known resummation. It is thus
no surprise that merely using the known universal terms of the effective
string action cannot capture this striking agreement, as we see in 
Fig.~\ref{fig_EQ0_n6f}. Moreover it is also no surprise that the
deviations from Nambu-Goto at smaller $l$ cannot be fitted with some
leading $1/l^{\gamma \geq 7}$ correction, as we see from 
Fig.~\ref{fig_DENGex1Q0_n6f} and Fig.~\ref{fig_DENGex1Q0_n4n5n6}. Indeed
one important conclusion of this study is that one requires a resummation 
of the correction terms, which we heuristically parameterised by 
$\frac{c}{(l\surd\sigma)^7} \left(1 + \frac{c^\prime}{l^2\sigma} 
\right)^{-\gamma}$. A good example
is provided in Fig.~\ref{fig_DENGex1Q0_n6f}, suggesting a power
$\gamma \sim 2.7$ and a radius of convergence similar to that of the
corresponding Nambu-Goto expression. So the effective power of the 
correction for smaller $l$ is $\sim 1/l^{7-2\gamma}\sim 1/l^{\sim 2}$ 
rather than $\sim 1/l^7$, and in fact we cannot 
claim any significant evidence for the latter power from this 
excited state's calculation. A similar conclusion follows from an
analysis of the lightest $P=+$ state with momentum $p=2\pi/l$ as shown 
in Fig.~\ref{fig_DENGgsQ1P+_n6f} and Fig.~\ref{fig_DENGgsQ1P+_n4n5n6}. 
By contrast the lightest $P=-$ state with $p=2\pi/l$, shown in 
Fig.~\ref{fig_DENGgsQ1P-_n6f}, behaves like the absolute ground state,
displaying a correction that only becomes significant at very small $l$ 
and varies rapidly in a manner  consistent with $\sim 1/l^7$,
but is definitely not consistent with the softer resummed behaviour
discussed above. This state is special in that it contains a single
phonon, $\sim a_1|0\rangle$, in the Nambu-Goto model. The lightest $P=-$ 
state with $p=4\pi/l$ also has one phonon in Nambu-Goto, and displays
a similar behaviour. 

In summary, we have confirmed in some detail our earlier observation
\cite{AABBMT_d3k1}
that typical low-lying 
energy eigenstates of a closed flux tube remain close to the free-string 
Nambu-Goto prediction well below the values of $l$ where a series
expansion of $E^{NG}_n(l)$ in $1/l^2\sigma$ diverges, and where all
powers become important. Our analysis of the (absolute) ground state
suggests that the leading correction to Nambu-Goto is most likely to be 
$O(1/l^7)$, consistent with recent analytic studies of the effective action
\cite{OA}.
By contrast, at modest values of $l$ our lightest excited states show 
corrections to the Nambu-Goto predictions which clearly demand a 
resummation of the series of correction terms, so as to give a 
behaviour closer to $\sim 1/l^2$ than  $\sim 1/l^7$. The exceptions
appear to be the lightest states with a single `phonon' which are very 
much like the absolute ground state: the corrections are small, only
becoming visible at very small $l$, and are consistent with 
a $\sim 1/l^7$ behaviour. Interestingly, very recent analytic studies 
have shown that the `scaling 0' operators that arise in the expansion
of the square-root Nambu-Goto action are all universal
\cite{OA}.
If for some reason the series of correction terms displays the
desired resummation properties, then one may be most of the way
to a theoretical understanding of most of these remarkably simple 
numerical results. Of course, one needs to understand the
special behaviour of the single phonon states and, most importantly, why
there is no sign of excitations of massive modes, even at small $l$,
(unlike the case of $D=3+1$
\cite{AABBMT_d4k1}).
The answer to the latter might explain how even at $l\surd\sigma \sim 2$,
where the flux tube surely `looks like' a fat periodic blob rather than
a thin string, at least $99\%$ of the difference $E_0(l) - \sigma l$
is given by the zero-point energies of the excitations of an ideal thin 
string.

\section*{Acknowledgements}

During the course of this work, MT participated in the {\it Confining 
Flux Tubes and Strings} Workshop at the ECT*, Trento, where 
there were many talks relevant to the present work (available on 
the ECT web-site): MT is grateful to the participants for useful 
discussions and to the ECT for its support in part of this research
under the European Community - Research Infrastructure Action under 
the FP7 `Capacities' Specific Programme, project `HadronPhysics2'.
The computations were carried out on EPSRC and Oxford funded computers 
in Oxford Theoretical Physics. 

\clearpage

\begin{appendix}
%
%
%
\section{Compilation of energy spectra}
\label{section_appendix_results}

In this Appendix we list the energies that we have calculated in
our SU(6) calculation at $\beta=171.0$. These are our best results
for the flux tube spectrum, both in terms of closeness to the
continuum limit, $a=0$, and closeness to $N=\infty$, and indeed
in terms of accuracy. We present the spectrum in enough detail so that
interested readers are able to make their own analyses.

In Table~\ref{table_n6b171Q0P+} we list the lightest 5 states in the 
with longitudinal momentum $p=2\pi q/l =0$ and parity $P=+$.
We also show the lattice sizes used (in lattice units). 
Table~\ref{table_n6b171Q0P-} lists the 3 lightest $P=-$ states
with $p=0$.  
Table~\ref{table_n6b171Q1} lists the lightest states with   
$p=2\pi q/l=2\pi/l$, in both $P=\pm$ sectors. 
Table~\ref{table_n6b171Q2} and Table~\ref{table_n6b171Q3} do
the same for $p=4\pi/l$ and  $p=6\pi/l$ respectively. And
Table~\ref{table_n6b171Q456} lists the ground states with
momenta  $p=8\pi/l,\, 10\pi/l,\, 12\pi/l$ and parities $P=\pm$.

\begin{table}[h]
\begin{center}
\begin{tabular}{|cc|ccccc|}\hline
$l/a$ & $l_\perp\times l_t$ & \multicolumn{5}{|c|}{ $aE(l;q=0) \ ; \ P=+$ } \\ \hline
14  & $100\times 200$  & 0.0519(4)  & 0.3620(99) &  &  &  \\
16  & $100\times 200$  & 0.0777(3)  & 0.3963(25) & 0.4954(93) & 0.5768(222) &  \\
20  & $70\times 120$  & 0.1176(5)  & 0.4219(19) & 0.5809(41) & 0.6300(63) & 0.6787(155) \\
24  & $48\times 60$  & 0.1528(9)  & 0.4369(30) & 0.5986(55) & 0.6284(86) & 0.6964(119) \\
28  & $48\times 60$  & 0.1842(8)  & 0.4550(36) & 0.6038(72) & 0.6393(76) & 0.7375(123) \\
32  & $40\times 48$  & 0.2177(10) & 0.4736(36) & 0.6033(83) & 0.6328(63) & 0.7292(171) \\
36  & $40\times 48$  & 0.2490(12) & 0.4903(27) & 0.6303(77) & 0.6428(97) & 0.7406(89) \\
40  & $48\times 48$  & 0.2817(14) & 0.5065(27) & 0.6464(93) & 0.6649(60) & 0.7789(76) \\
44  & $48\times 48$  & 0.3113(14) & 0.5281(19) & 0.6535(56) & 0.6777(71) & 0.7959(92) \\
48  & $48\times 48$  & 0.3425(13) & 0.5481(23) & 0.6806(53) & 0.6985(60) & 0.8035(110) \\
52  & $52\times 52$  & 0.3723(10) & 0.5652(19) & 0.7011(45) & 0.7225(57) & 0.8097(115) \\
56  & $56\times 56$  & 0.4056(9)  & 0.5886(19) & 0.7178(68) & 0.7371(78) & 0.8441(106) \\
60  & $60\times 60$  & 0.4340(11) & 0.6134(12) & 0.7328(50) & 0.7533(42) & 0.8484(106) \\
64  & $64\times 64$  & 0.4637(17) & 0.6343(45) & 0.7630(39) & 0.7626(59) & 0.8694(108) \\ \hline
\end{tabular}
\caption{The energies, $E(q,l)$, of the lightest five flux tube states with 
length $l$, parity $P=+$ and longitudinal momentum $p=2\pi q/l=0$. 
For SU(6) at $\beta=171.0$.}
\label{table_n6b171Q0P+}
\end{center}
\end{table}

\clearpage

\begin{table}[h]
\begin{center}
\begin{tabular}{|c|ccc|}\hline
$l/a$ & \multicolumn{3}{|c|}{ $aE(l;q=0) \ ; \ P=-$ } \\ \hline
16  &  0.5243(71)  & 0.6182(117) & 0.6794(101)  \\ 
20  &  0.5808(74)  & 0.6718(127) & 0.7237(159)  \\
24  &   0.6151(70) & 0.6739(102) & 0.7399(107)  \\
28  &   0.6377(80) & 0.6624(75)  & 0.7586(79)  \\
32  &   0.6270(66) & 0.6531(99)  & 0.7754(82)  \\
36  &   0.6384(81) & 0.6699(115) & 0.7641(92)  \\
40  &   0.6572(93) & 0.6717(99)  & 0.7931(107)  \\
44  &   0.6810(98) & 0.6900(107) & 0.7993(82)   \\
48  &   0.6966(62) & 0.7129(56)  & 0.8350(108)  \\
52  &   0.7174(72) & 0.7262(63)  & 0.8409(99)   \\
56  &   0.7394(57) & 0.7395(76)  & 0.8443(111)  \\
60  &   0.7462(74) & 0.7555(57)  & 0.8902(102)  \\
64  &   0.7697(69) & 0.7713(80)  & 0.8748(111)  \\ \hline
\end{tabular}
\caption{The energies, $E(q,l)$, of the lightest three flux tube states with 
length $l$, parity $P=-$ and longitudinal momentum $p=2\pi q/l=0$. 
For SU(6) at $\beta=171.0$.}
\label{table_n6b171Q0P-}
\end{center}
\end{table}

\begin{table}[h]
\begin{center}
\begin{tabular}{|c|ccc|cc|}\hline
\multicolumn{6}{|c|}{ $aE(l;q=1)$ } \\ \hline
$l/a$ & \multicolumn{3}{|c|}{ $P=-$ } & \multicolumn{2}{|c|}{ $P=+$ } \\ \hline
14  & 0.5222(32)  & 0.7331(109) &             & 0.6457(152) &  \\
16  & 0.4927(41)  & 0.6553(63)  & 0.7666(169) & 0.6207(128)  & 0.7979(203)  \\
20  & 0.4524(11)  & 0.6255(31)  & 0.7704(118) & 0.5853(123)  & 0.7385(205)  \\
24  & 0.4297(19)  & 0.6005(41)  & 0.7415(71)  & 0.5805(148)  & 0.7240(133)  \\
28  & 0.4206(20)  & 0.5898(56)  & 0.7413(74)  & 0.5916(116)  & 0.7205(124)  \\
32  & 0.4226(14)  & 0.5980(54)  & 0.7244(60)  & 0.5987(65)  & 0.7155(113)  \\
36  & 0.4288(16)  & 0.6091(38)  & 0.7279(71)  & 0.5947(70)  & 0.7247(125)  \\
40  & 0.4393(22)  & 0.6185(44)  & 0.7504(78)  & 0.6049(58)  & 0.7362(83)  \\
44  & 0.4595(26)  & 0.6295(52)  & 0.7499(81)  & 0.6216(43)  & 0.7452(85)  \\
48  & 0.4769(24)  & 0.6362(48)  & 0.7651(76)  & 0.6414(23)  & 0.7503(97)  \\
52  & 0.4997(20)  & 0.6581(58)  & 0.7883(99)  & 0.6600(24)  & 0.7637(70)  \\
56  & 0.5163(19)  & 0.6790(64)  & 0.8098(84)  & 0.6745(25)  & 0.7868(77)  \\
60  & 0.5423(22)  & 0.6974(51)  & 0.8298(98)  & 0.6915(29)  & 0.7950(87)  \\ 
64  & 0.5645(21)  & 0.7107(67)  & 0.8445(102) & 0.7108(24)  & 0.8314(93) \\ \hline
\end{tabular}
\caption{The energies, $E(q,l)$, of the some of the lightest flux tube states with 
length $l$, parity $P=\pm$ and longitudinal momentum $p=2\pi q/l=2\pi/l$. 
For SU(6) at $\beta=171.0$.}
\label{table_n6b171Q1}
\end{center}
\end{table}

\begin{table}[h]
\begin{center}
\begin{tabular}{|c|ccc|cccc|}\hline
\multicolumn{8}{|c|}{ $aE(l;q=2)$ } \\ \hline
$l/a$ & \multicolumn{3}{|c|}{ $P=-$ } & \multicolumn{4}{|c|}{ $P=+$ } \\ \hline
14  & 0.9636(55)  &           &           & 1.011(15) &  &  &    \\
16  & 0.8758(115) & 1.046(11) &           & 0.913(13) &  &  &    \\
20  & 0.7519(93) &  0.902(10) & 1.006(6)  & 0.759(9) & 0.876(6) & 0.916(11) &    \\
24  & 0.6941(42) & 0.834(9)   & 0.909(16) & 0.691(9) & 0.820(7) & 0.845(9)  & 0.950(14)   \\
28  & 0.6485(45) & 0.800(16)  & 0.915(15) & 0.654(10)& 0.788(9) & 0.796(11) & 0.927(14)   \\
32  & 0.6266(49) & 0.792(18)  & 0.859(12) & 0.612(9) & 0.748(14) & 0.747(9) & 0.888(12)   \\
36  & 0.6042(47) & 0.741(8)   & 0.857(12) & 0.612(6) & 0.755(8) & 0.744(6) &  0.847(12)  \\
40  & 0.5999(44) & 0.754(7)   & 0.855(11) & 0.610(5) & 0.734(8) & 0.742(9) &  0.838(13)  \\
44  & 0.6010(43) & 0.745(8)   & 0.854(12) & 0.606(5) & 0.737(8) & 0.743(7) &  0.850(11)  \\
48  & 0.6164(52) & 0.746(8)   & 0.880(12) & 0.612(5) & 0.758(4) & 0.768(7) &  0.859(10)  \\
52  & 0.6190(48) & 0.756(9)   & 0.877(12) & 0.622(4) & 0.761(7) & 0.774(9) &  0.874(11)  \\
56  & 0.6341(50) & 0.775(7)   & 0.893(12) & 0.633(5) & 0.758(9) & 0.772(8) &  0.857(11)  \\
60  & 0.6488(97) & 0.778(9)   & 0.904(11) & 0.656(8) & 0.772(7) & 0.785(8) &  0.873(13)  \\
64  & 0.6615(87) & 0.792(7)   & 0.911(14) & 0.656(6) & 0.802(8) & 0.826(8) &  0.900(14)  \\ \hline
\end{tabular}
\caption{The energies, $E(q,l)$, of the some of the lightest flux tube states with 
length $l$, parity $P=\pm$ and longitudinal momentum $p=2\pi q/l=4\pi/l$. 
For SU(6) at $\beta=171.0$.}
\label{table_n6b171Q2}
\end{center}
\end{table}

\begin{table}[h]
\begin{center}
\begin{tabular}{|c|ccc|cccc|}\hline
\multicolumn{8}{|c|}{ $aE(l;q=3)$ } \\ \hline
$l/a$ & \multicolumn{3}{|c|}{ $P=-$ } & \multicolumn{4}{|c|}{ $P=+$ } \\ \hline
14  & 1.340(7)  & 1.518(10) & 1.587(14)  & 1.459(48)  & 1.481(42)  &   &   \\
16  & 1.190(16) & 1.342(28) & 1.448(11)  & 1.284(24)  & 1.382(29)  &   &   \\ 
20  & 1.051(12) & 1.115(7)  & 1.228(10)  & 1.091(23)  & 1.184(25)  &   &   \\ 
24  & 0.945(12) & 0.991(7)  & 1.110(10)  & 0.992(20)  & 1.056(23)  &   &   \\ 
28  & 0.886(10) & 0.907(11) & 1.003(23)  & 0.875(14)  & 0.960(19)  &   &   \\ 
32  & 0.823(11) & 0.816(25) & 0.931(42)  & 0.844(12)  & 0.903(35)  &   &   \\ 
36  & 0.796(9)  & 0.782(19) & 0.885(35)  & 0.790(10)  & 0.886(13)  &   &   \\ 
40  & 0.770(8)  & 0.772(10) & 0.890(34)  & 0.771(9)   & 0.885(15)  & 0.928(28)  &   \\ 
44  & 0.746(8)  & 0.745(14) & 0.843(30)  & 0.737(17)  & 0.862(14)  & 0.867(29)  & 0.999(59)  \\ 
48  & 0.745(8)  & 0.737(18) & 0.880(27)  & 0.751(19)  & 0.876(29)  & 0.852(27)  & 0.986(20)  \\ 
52  & 0.749(6)  & 0.749(14) & 0.864(27)  & 0.754(10)  & 0.874(12)  & 0.872(20)  & 0.960(13)  \\ 
56  & 0.734(15) & 0.743(17) & 0.875(31)  & 0.757(7)   & 0.868(13)  & 0.875(28)  & 0.957(18)  \\ 
60  & 0.764(7)  & 0.784(8)  & 0.892(33)  & 0.772(7)   & 0.861(12)  & 0.887(28)  & 0.989(20)  \\ 
64  & 0.774(8)  & 0.788(8)  & 0.898(23)  & 0.770(14)  & 0.871(19)  & 0.913(27)  & 0.999(19)  \\  \hline
\end{tabular}
\caption{The energies, $E(q,l)$, of the some of the lightest flux tube states with 
length $l$, parity $P=\pm$ and longitudinal momentum $p=2\pi q/l=6\pi/l$. 
For SU(6) at $\beta=171.0$.}
\label{table_n6b171Q3}
\end{center}
\end{table}

\begin{table}[h]
\begin{center}
\begin{tabular}{|c|cc|cc|cc|}\hline
\multicolumn{7}{|c|}{ $aE(l;q)$ } \\ \hline
      & \multicolumn{2}{|c|}{ $q=4$ } & \multicolumn{2}{|c|}{ $q=5$ } & \multicolumn{2}{|c|}{ $q=6$ } \\ \hline
$l/a$ & $P=-$ & $P=+$ & $P=-$ & $P=+$ & $P=-$ & $P=+$ \\ \hline
14  & 1.523(60)  &            &   &   &   &     \\
16  & 1.505(11)  &            &   &   &   &     \\ 
20  & 1.339(15)  & 1.389(51)  & 1.553(29)  & 1.684(40)  &   &     \\ 
24  & 1.202(12)  & 1.259(11)  & 1.389(23)  & 1.497(32)  &   &     \\ 
28  & 1.065(23)  & 1.139(9)   & 1.240(49)  & 1.377(21)  &   &     \\ 
32  & 1.019(7)   & 1.035(8)   & 1.156(41)  & 1.286(21)  &   &     \\ 
36  & 0.984(8)   & 0.953(16)  & 1.141(31)  & 1.184(40)  & 1.295(66)  & 1.339(73)  \\ 
40  & 0.935(15)  & 0.908(15)  & 1.104(29)  & 1.140(37)  & 1.182(47)  & 1.174(61)  \\ 
44  & 0.917(14)  & 0.926(14)  & 1.058(28)  & 1.066(29)  & 1.164(38)  & 1.234(48)  \\ 
48  & 0.889(15)  & 0.892(14)  & 1.006(23)  & 1.050(25)  & 1.116(35)  & 1.176(47)  \\ 
52  & 0.877(10)  & 0.855(29)  & 0.997(22)  & 1.007(16)  & 1.147(31)  & 1.096(34)  \\ 
56  & 0.881(10)  & 0.865(13)  & 1.020(18)  & 0.989(18)  & 1.091(26)  & 1.063(24)  \\ 
60  & 0.855(11)  & 0.845(26)  & 0.971(18)  & 0.981(16)  & 1.075(22)  & 1.072(21)  \\ 
64  & 0.855(14)  & 0.879(10)  & 0.972(20)  & 0.985(20)  & 1.105(26)  & 1.052(21)  \\  \hline
\end{tabular}
\caption{The energies, $E(q,l)$, of the some of the lightest flux tube states with 
length $l$, parity $P=\pm$ and longitudinal momentum $p=2\pi q/l$ as indicated. 
For SU(6) at $\beta=171.0$.}
\label{table_n6b171Q456}
\end{center}
\end{table}

\end{appendix}

\clearpage

\clearpage

\begin{figure}[htb]
\begin	{center}
\leavevmode
\input {plot_E0n3t2}
\end	{center}
\caption{}
\label{fig_E0n3t2}
\caption{Energy of the ground state versus $1/lg^2 \equiv T/g^2$ for 
SU(3) with $l=2a$ and $a(\beta)$ being varied. The curve is 
$\propto (T_c-T)^{\frac{5}{6}}$,
as expected from the universality class of the critical point.}
\end{figure}

\begin{figure}[h]
\begin	{center}
\leavevmode
\input	{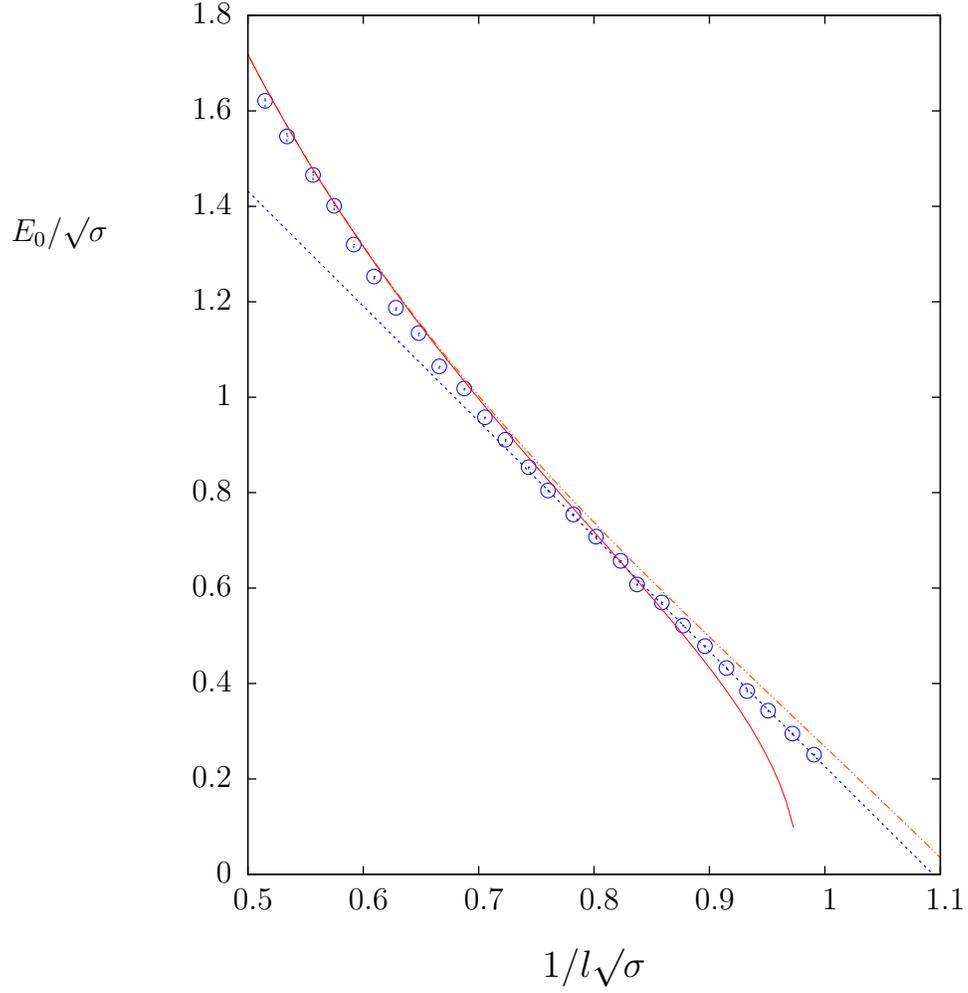}
\end	{center}
\caption{Energy of ground state versus 
$1/l\surd\sigma \equiv T/\surd\sigma$ for SU(2) with $l=4a(\beta)$,
and $\beta$ being varied. Solid line is Nambu-Goto;
dashed blue line is $\propto (T_c-T)$ as expected from the
universality class of the critical point, and dashed red line is
the universal prediction for $E_0$ up to $O(1/l^5)$.}
\label{fig_EkTk_n2t4}
\end{figure}

\begin{figure}[h]
\begin	{center}
\leavevmode
\input	{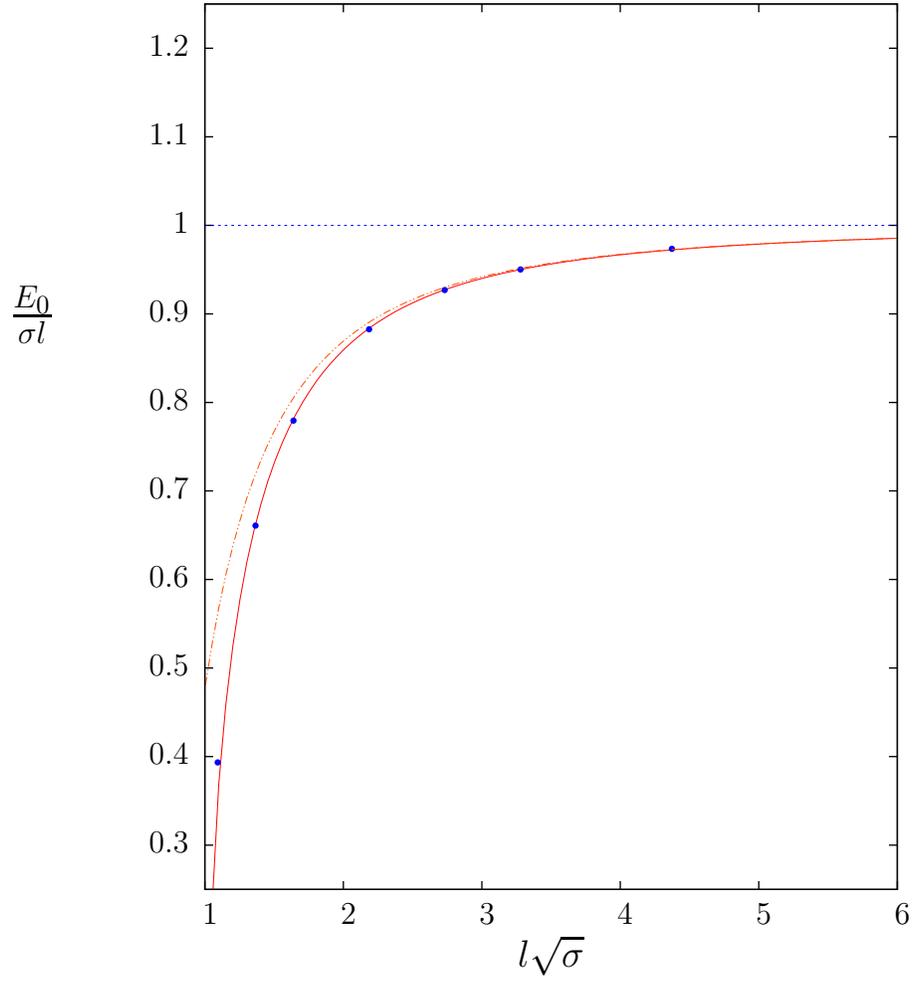}
\end	{center}
\caption{Energy of absolute ground state for  SU(2) at $\beta=5.6$.
Compared to full Nambu-Goto (solid curve) and just the  L\"uscher 
correction (dashed curve).}
\label{fig_EgsQ0_n2b5.6}
\end{figure}

\begin{figure}[h]
\begin	{center}
\leavevmode
\input	{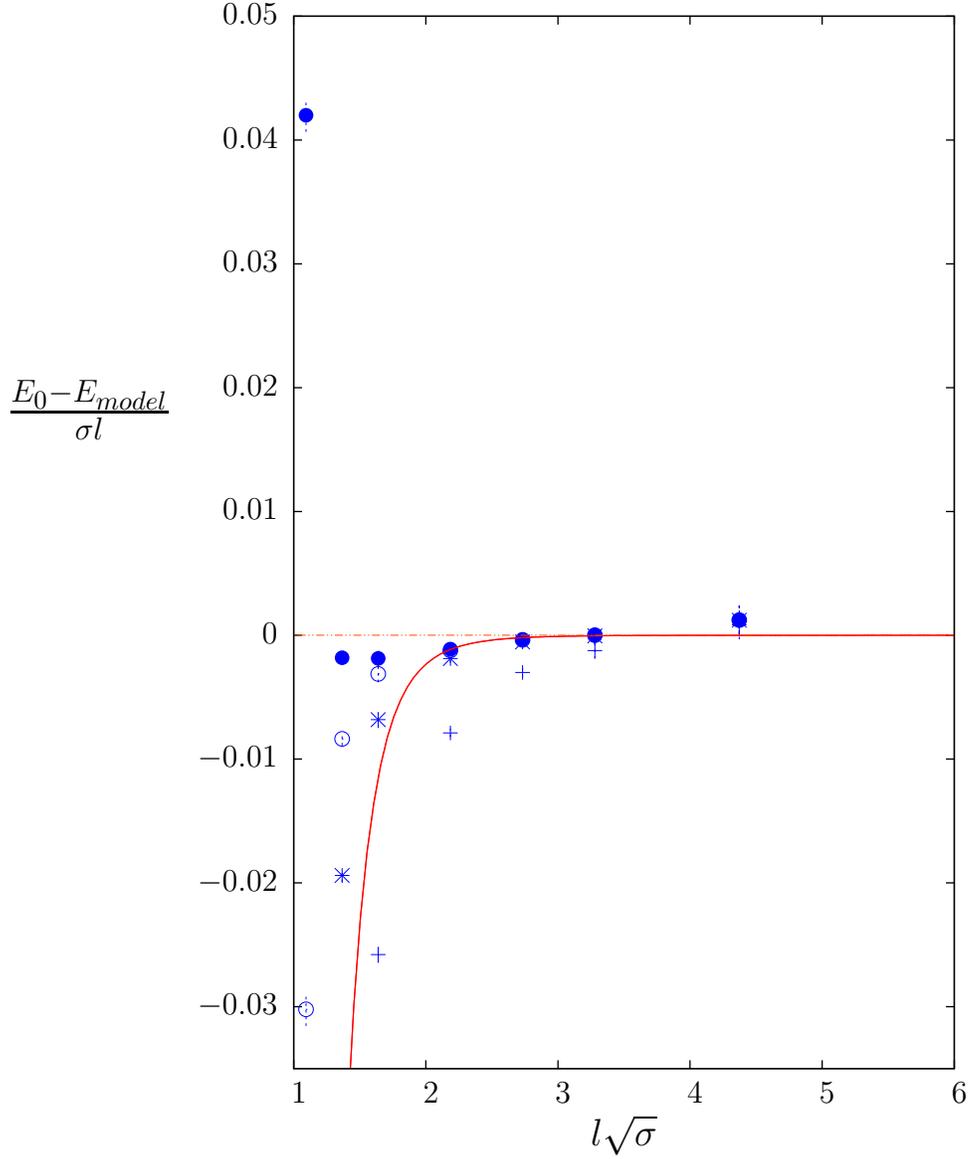}
\end	{center}
\caption{Energy of ground state minus the best fit of several
models for $E_0$: full Nambu-Goto, $\bullet$; the linear piece plus all 
the known universal terms, $\circ$; the latter without the 
$O(1/l^5)$ term, $\star$; linear plus L\"uscher correction, $+$.
Curve is Nambu-Goto with a fitted $O(1/l^7)$ correction.
For  SU(2) at $\beta=5.6$ .}
\label{fig_DEOAgsQ0_n2b5.6}
\end{figure}

\begin{figure}[h]
\begin	{center}
\leavevmode
\input	{plot_DELWgsQ0_n4b32ttb.tex}
\end	{center}
\caption{Energy of ground state minus the best fit of several
models for $E_0$: full Nambu-Goto, $\bullet$; the linear piece plus all 
the known universal terms, $\circ$; the latter without the 
$O(1/l^5)$ term, $\star$; linear plus L\"uscher correction, $+$;
just the linear $\sigma l$ piece, $\times$. 
Curves are fits with an $O(1/l^7)$ correction to
the first of these two.  For  SU(4) at $\beta=32.0$ .}
\label{fig_DEOAgsQ0_n4b32}
\end{figure}

\begin{figure}[h]
\begin	{center}
\leavevmode
\input	{plot_DEOAgsQ0_n6f.tex}
\end	{center}
\caption{Energy of ground state minus the best fit of several
models for $E_0$: full Nambu-Goto, $\bullet$; the linear piece plus all 
the known universal terms, $\circ$; the latter without the 
$O(1/l^5)$ term, $\star$; linear plus L\"uscher correction, $+$;
just the linear $\sigma l$ piece, $\times$. 
Curves are fits with an $O(1/l^7)$ correction to
the first of these two. For  SU(6) at $\beta=171.0$ .}
\label{fig_DEOAgsQ0_n6f}
\end{figure}

\begin{figure}[htb]
\begin	{center}
\leavevmode
\input	{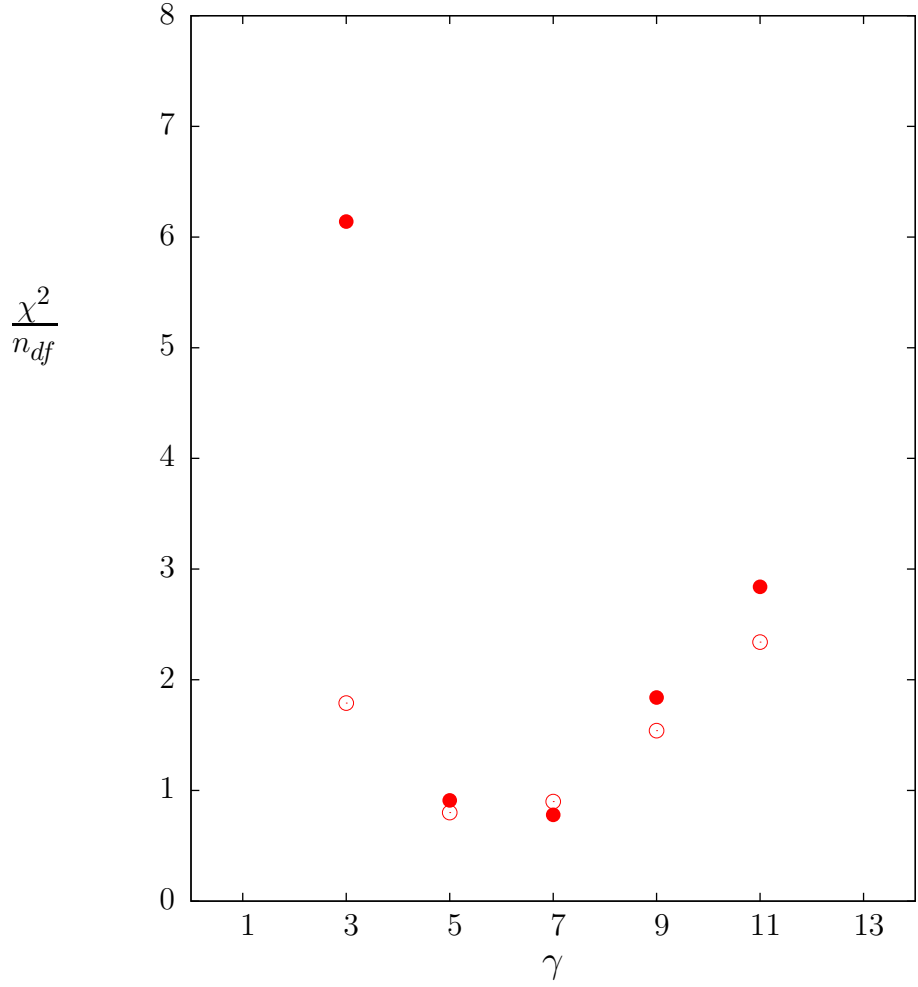}
\end	{center}
\caption{$\chi^2$ per degree of freedom for the best fit of the power
of the leading correction to $E_0(l)$ minus Nambu-Goto,
using eqn(\ref{eqn_E0corrn}). For SU(6) at $\beta=171$, $\circ$, and 
for SU(4) at $\beta=32$, $\bullet$.} 
\label{fig_NGgamma}
\end{figure}

\begin{figure}[h]
\begin	{center}
\leavevmode
\input	{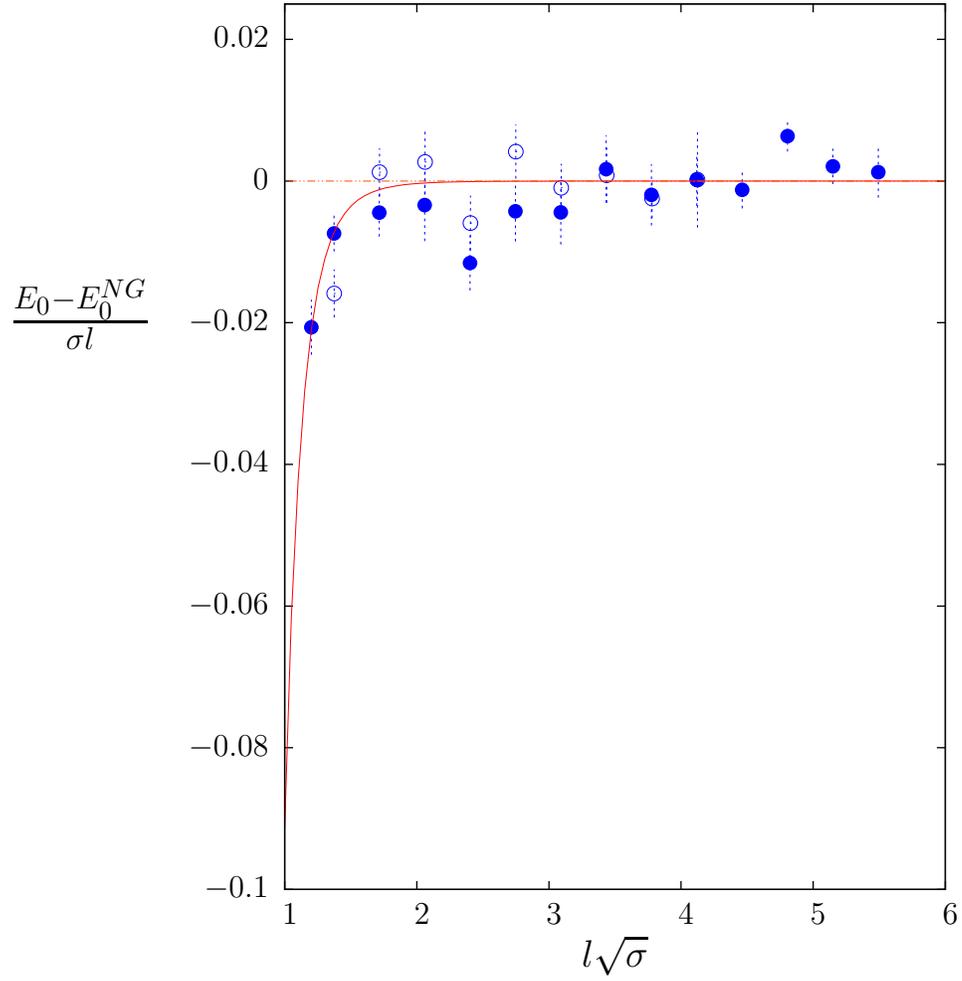}
\end	{center}
\caption{Energy of ground state minus the Nambu-Goto fit, 
for SU(6) at $\beta=90$, $\circ$ and $\beta=171$ $\bullet$.
Curve is an $O(1/l^7)$ correction to the second of these.}
\label{fig_DENGgsQ0_n6n6f}
\end{figure}

\begin{figure}[htb]
\begin	{center}
\leavevmode
\input	{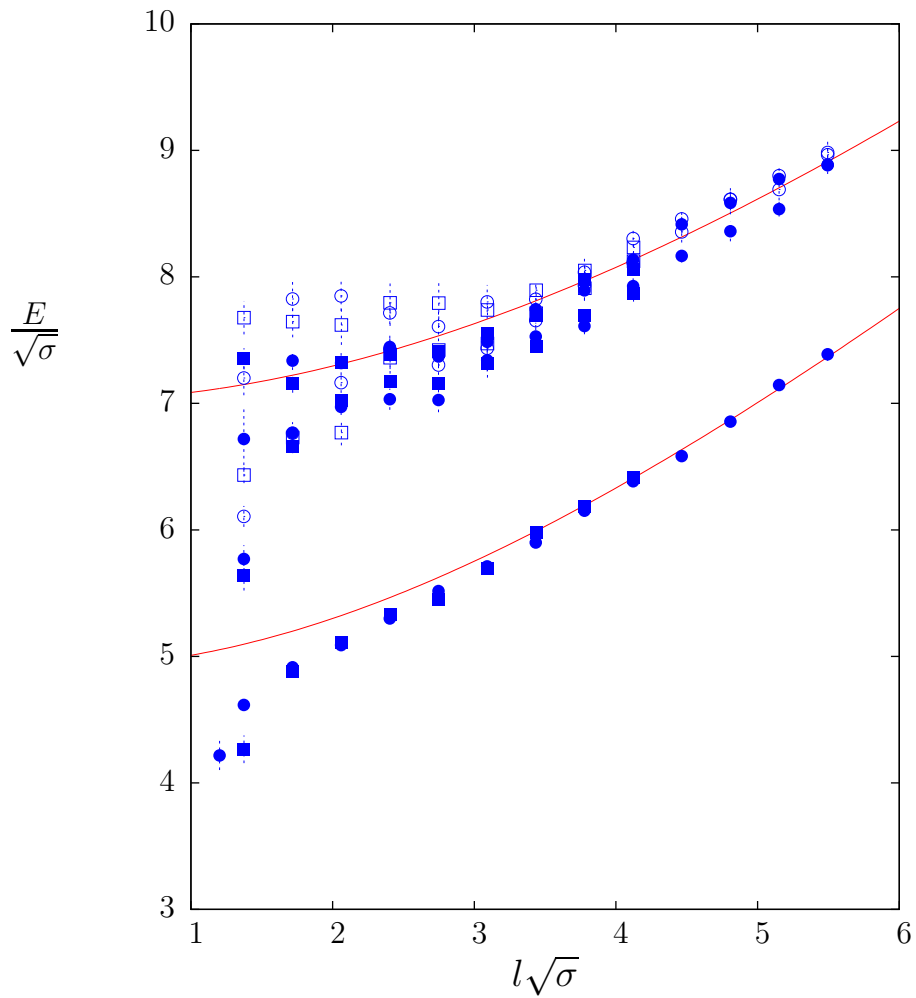}
\end	{center}
\caption{Lightest excited states with $p=0$ in SU(6): $P=+$
at $\beta=171$, $\bullet$, and $\beta=90$, $\blacksquare$; $P=-$
at $\beta=171$, $\circ$, and $\beta=90$, $\square$.}
\label{fig_EQ0_n6n6f}
\end{figure}

\begin{figure}[htb]
\begin	{center}
\leavevmode
\input	{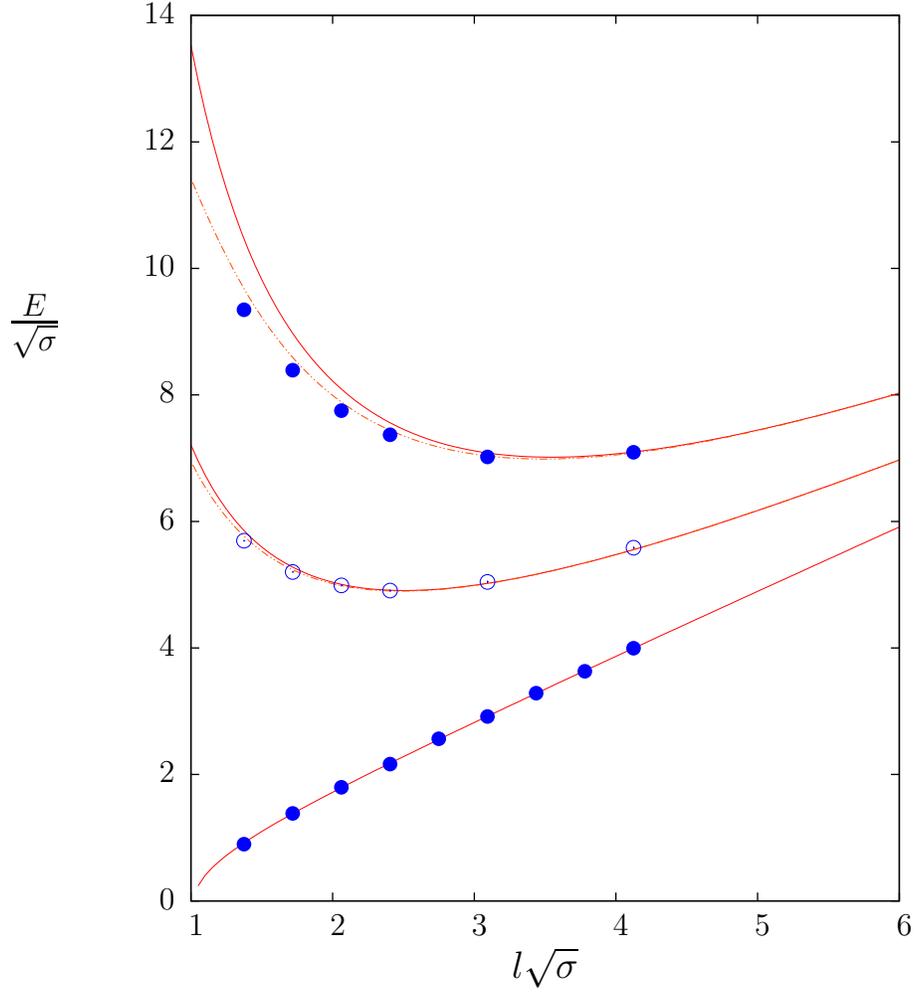}
\end	{center}
\caption{Energies of the ground states with momenta $q=1,2$
and with $P=-$. Also the $q=0$,  $P=+$ absolute ground state.
For  SU(6) at $\beta=90$. Lines are Nambu-Goto predictions;
solid use a continuum $p^2$ contribution, while dashed use
the free-field lattice version, $2-2\cos(ap)$.}
\label{fig_EgsQall_n6}
\end{figure}

\begin{figure}[htb]
\begin	{center}
\leavevmode
\input	{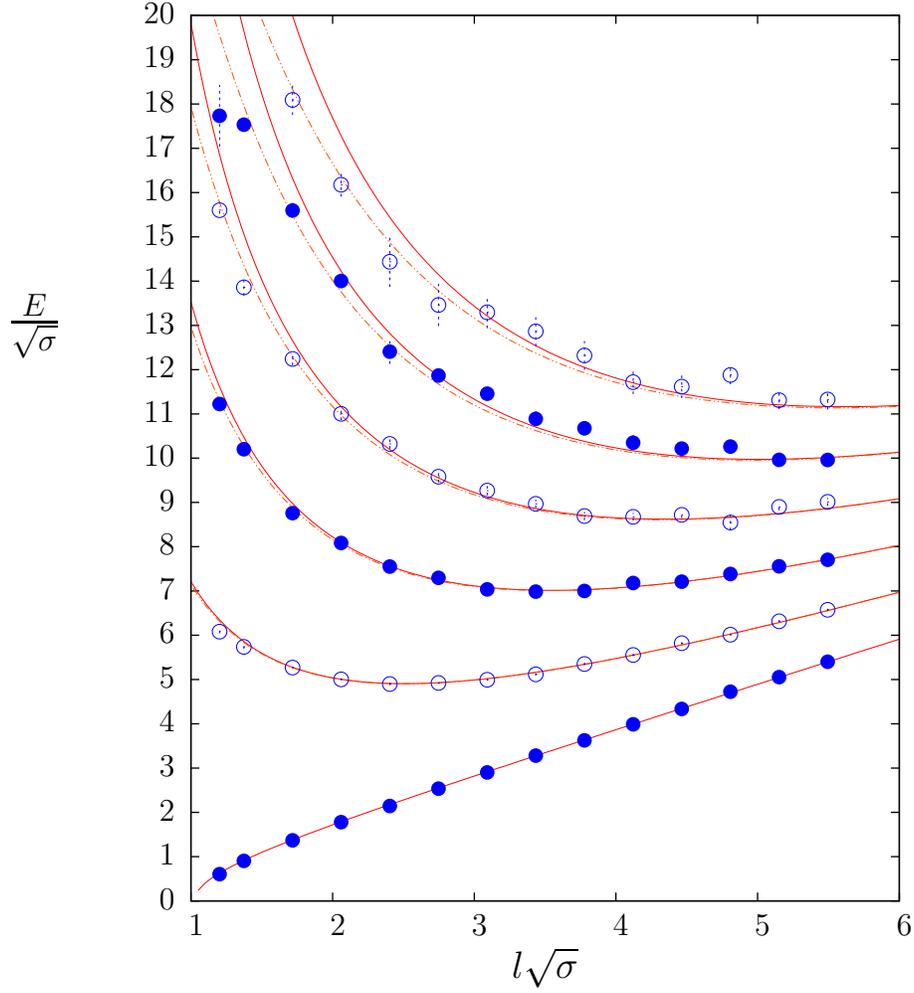}
\end	{center}
\caption{Energies of the ground states with $q=1,2,3,4,5$
and $P=-$. Also the $q=0$ absolute ground state, with $P=+$.
For  SU(6) at $\beta=171$. Lines are Nambu-Goto predictions;
solid use a continuum $p^2$ contribution, while dashed use
the free-field lattice version, $2-2\cos(ap)$.}
\label{fig_EgsQall_n6f}
\end{figure}

\begin{figure}[htb]
\begin	{center}
\leavevmode
\input	{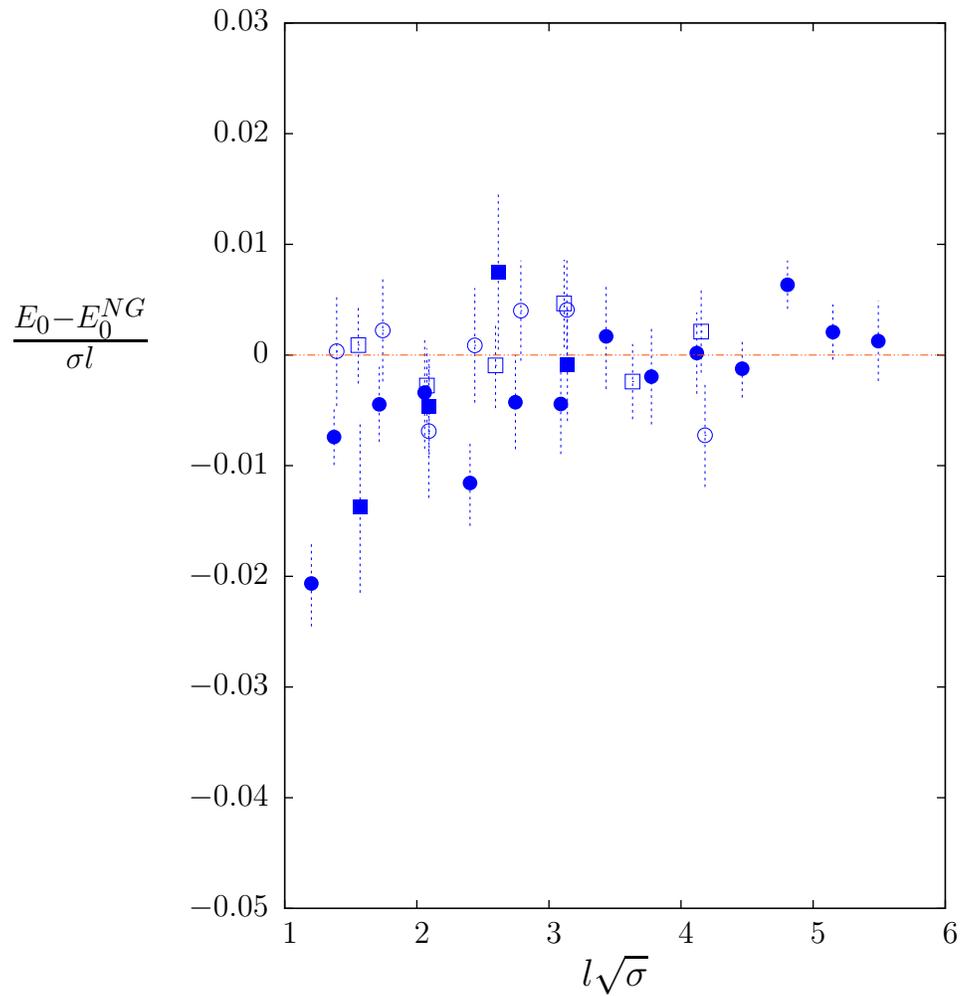}
\end	{center}
\caption{Energy of ground state minus its Nambu-Goto fit, 
for:  SU(6) at $\beta=171$, $\bullet$; SU(5) at $\beta=80$, 
$\square$; SU(4) at $\beta=50$, $\blacksquare$;
SU(3) at $\beta=40$, $\circ$.}
\label{fig_DENGgsQ0_n3fn4n5n6f}
\end{figure}

\begin{figure}[htb]
\begin	{center}
\leavevmode
\input	{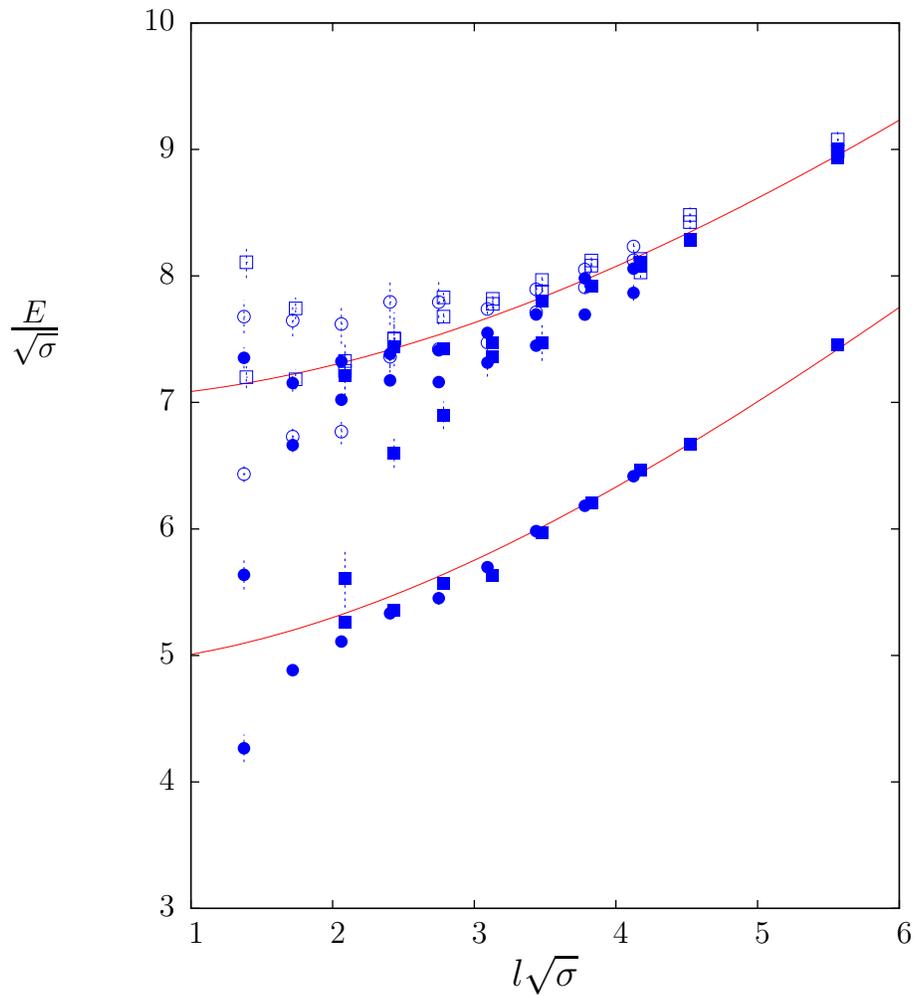}
\end	{center}
\caption{Comparing lightest $p=0$ excited states in SU(3) at $\beta=21$
with SU(6) at $\beta=90$: 
$P=+$ in SU(6), $\bullet$, and in SU(3), $\blacksquare$; 
$P=-$ in SU(6), $\circ$, and in SU(3), $\square$.}
\label{fig_EQ0_n3n6}
\end{figure}


\clearpage

\begin{figure}[htb]
\begin	{center}
\leavevmode
\input	{plot_EP+Q0_n3fn4n5n6f}
\end	{center}
\caption{Comparing the three lightest $p=0$ and $P=+$ excited states in 
SU(3) at $\beta=40$, $\square$, SU(4) at $\beta=50$, $\circ$, 
SU(5) at $\beta=80$, $\bigtriangleup$, 
and  SU(6) at $\beta=171$, $\bullet$.}
\label{fig_EP+Q0_n3fn4n5n6f}
\end{figure}

\begin{figure}[htb]
\begin	{center}
\leavevmode
\input	{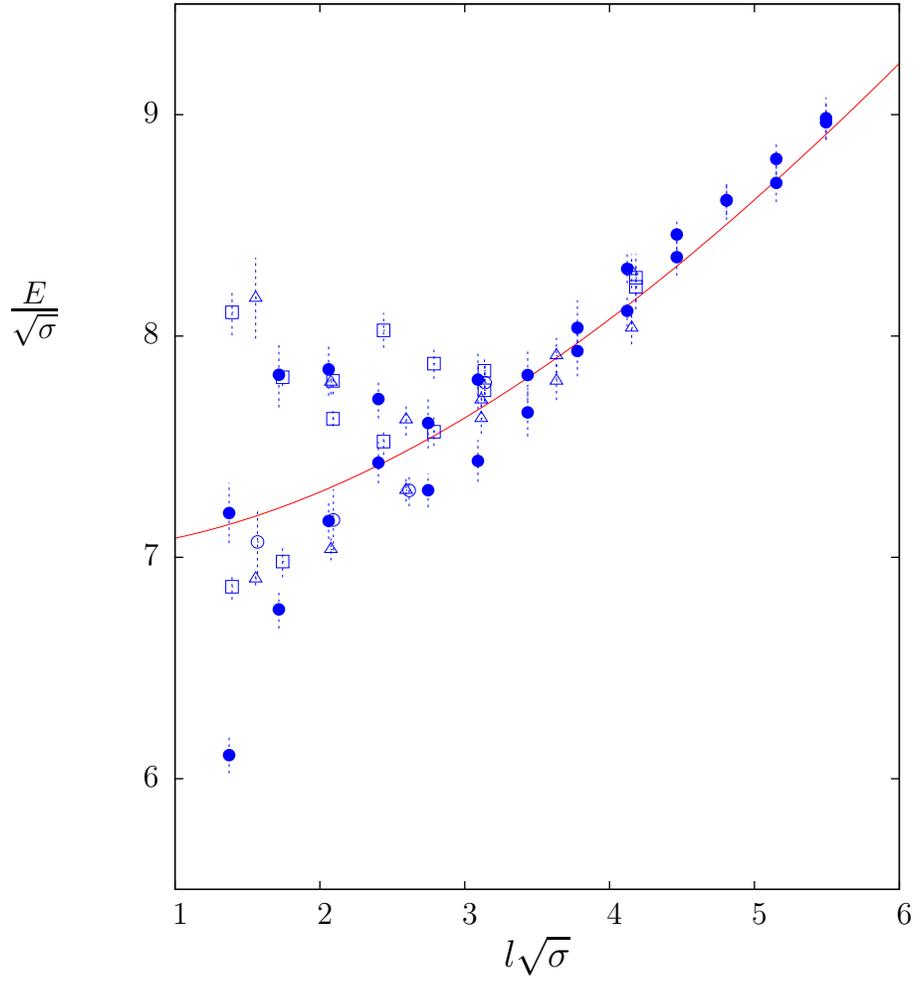}
\end	{center}
\caption{Comparing the two lightest $p=0$, $P=-$ excited states in 
SU(3) at $\beta=40$, $\square$, SU(4) at $\beta=50$, $\circ$, 
SU(5) at $\beta=80$, $\bigtriangleup$, 
and  SU(6) at $\beta=171$, $\bullet$.}
\label{fig_EP-Q0_n3fn4n5n6f}
\end{figure}

\begin{figure}[htb]
\begin	{center}
\leavevmode
\input	{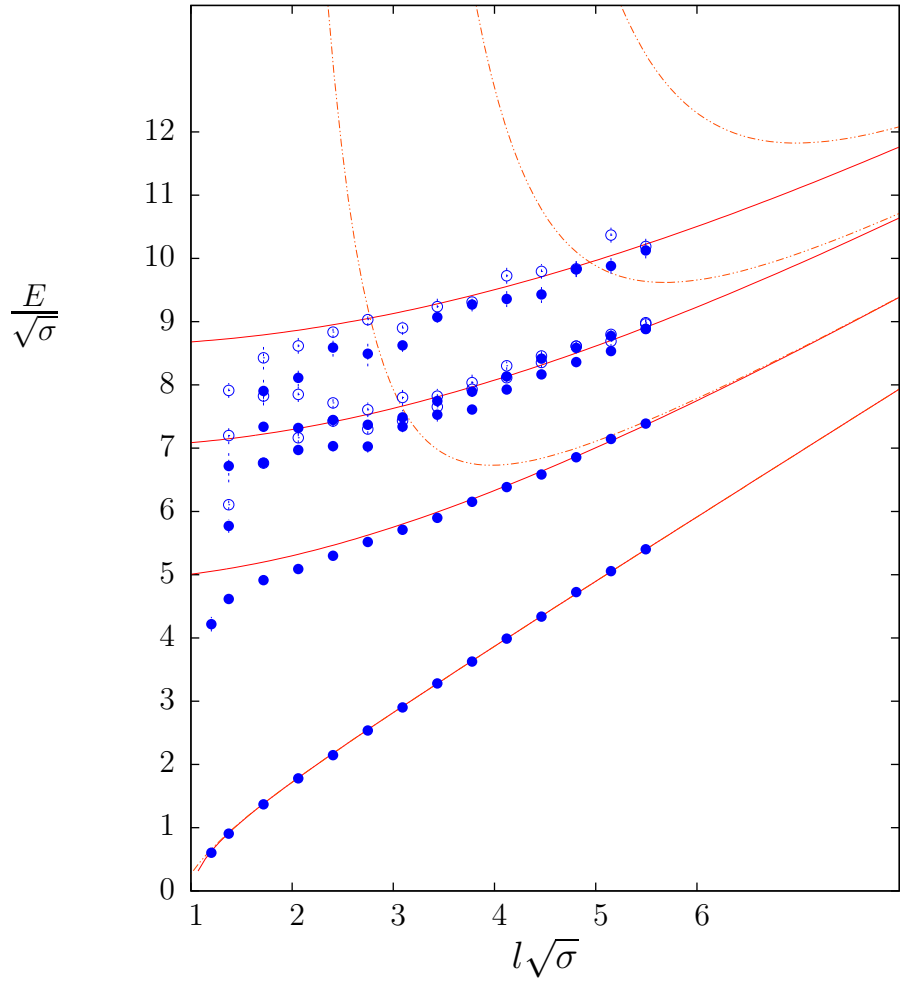}
\end	{center}
\caption{The lightest four $p=0$ excited states in SU(6) at $\beta=171$
with $P=+$, $\bullet$, and the lightest three with $P=-$, $\circ$. Solid
curves are the Nambu-Goto predictions, while the dashed curves are the
contributions of the known universal terms. 
Also shown is the absolute ground state (lowest curve).}
\label{fig_EQ0_n6f}
\end{figure}

\begin{figure}[h]
\begin	{center}
\leavevmode
\input	{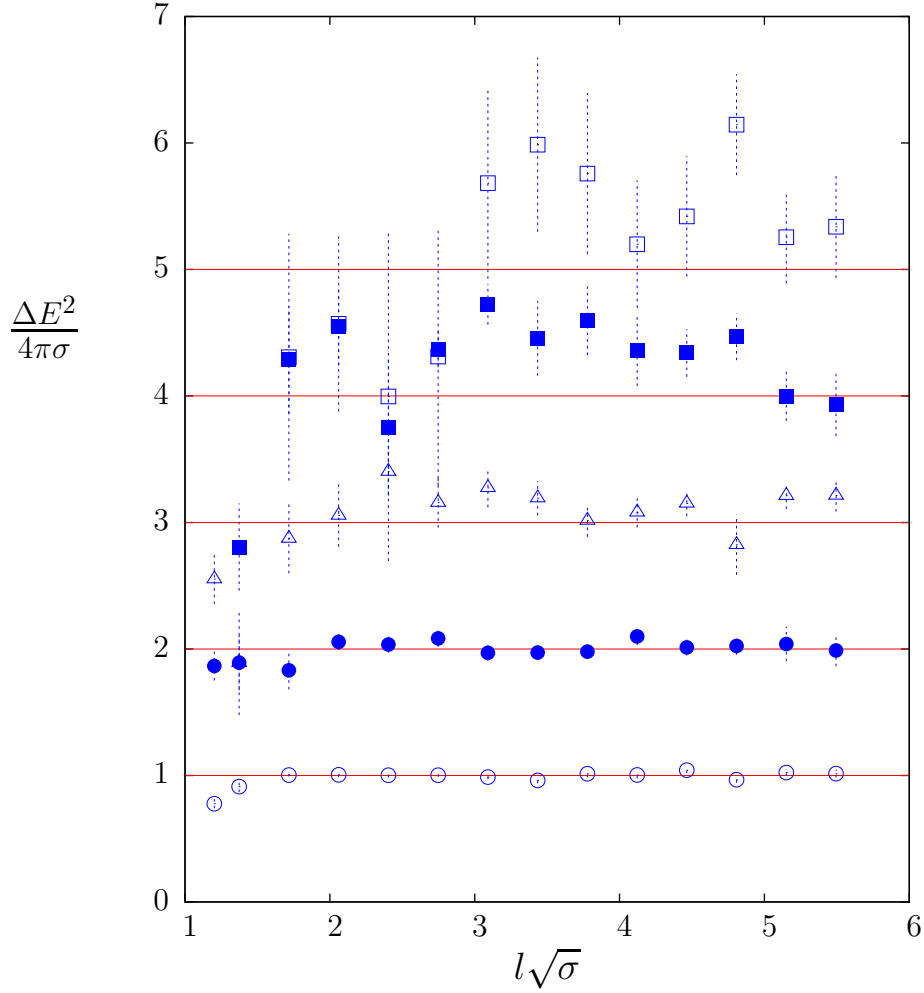}
\end	{center}
\caption{Excitation energies of the lightest $P=-$ states with non-zero
momenta $q=1,2,3,4,5$, using eqn(\ref{eqn_exq}), and with  $2-2\cos(ap)$ 
in place of $(ap)^2$. For  SU(6) at $\beta=171$. 
Nambu-Goto predictions shown.}
\label{fig_DlatEgsQall_n6f}
\end{figure}

\begin{figure}[h]
\begin	{center}
\leavevmode
\input	{plot_DEQ1_n6f.tex}
\end	{center}
\caption{Excitation energies of the lightest few states with momentum
$q=1$ and with $P=-$  ({$\bullet$}) or $P=+$  ({$\circ$}).
For  SU(6) at $\beta=171$ . Nambu-Goto predictions shown.}
\label{fig_DEQ1_n6f}
\end{figure}

\begin{figure}[h]
\begin	{center}
\leavevmode
\input	{plot_DEQ2_n6f.tex}
\end	{center}
\caption{Excitation energies of the lightest few states with momentum
$q=2$ and with $P=-$  ({$\bullet$}) or $P=+$  ({$\circ$}).
For  SU(6) at $\beta=171$ . Nambu-Goto predictions shown.}
\label{fig_DEQ2_n6f}
\end{figure}

\begin{figure}[h]
\begin	{center}
\leavevmode
\input	{plot_DEQ3_n6f.tex}
\end	{center}
\caption{Excitation energies of the lightest few states with  momentum
$q=3$ and with $P=-$  ({$\bullet$}) or $P=+$  ({$\circ$}).
For  SU(6) at $\beta=171$ . Nambu-Goto predictions shown.}
\label{fig_DEQ3_n6f}
\end{figure}

\begin{figure}[h]
\begin	{center}
\leavevmode
\input	{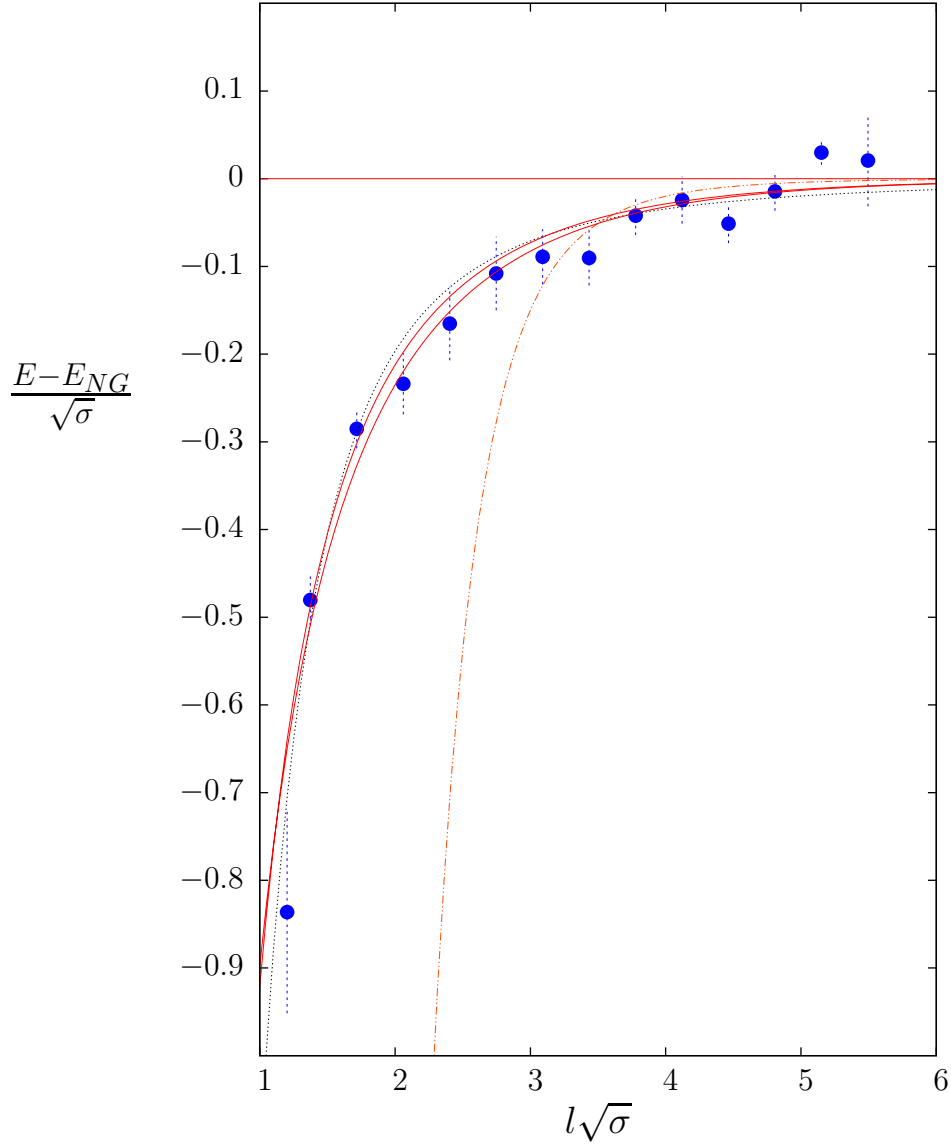}
\end	{center}
\caption{Energy of first excited $q=0,\, P=+$ state minus the Nambu-Goto 
prediction. Fitted corrections shown are: steep dotted curve is 
$\frac{c}{(l\surd\sigma)^7}$; solid red  curves are two fits of the form 
$\frac{c}{(l\surd\sigma)^7}\left(1+\frac{c^\prime}{l^2\sigma}\right)^{-\gamma}$;
dashed black curve is a best fit of the form $\frac{c}{(l\surd\sigma)^\gamma}$. See text for parameters. For  SU(6) at $\beta=171$. }
\label{fig_DENGex1Q0_n6f}
\end{figure}

\begin{figure}[h]
\begin	{center}
\leavevmode
\input	{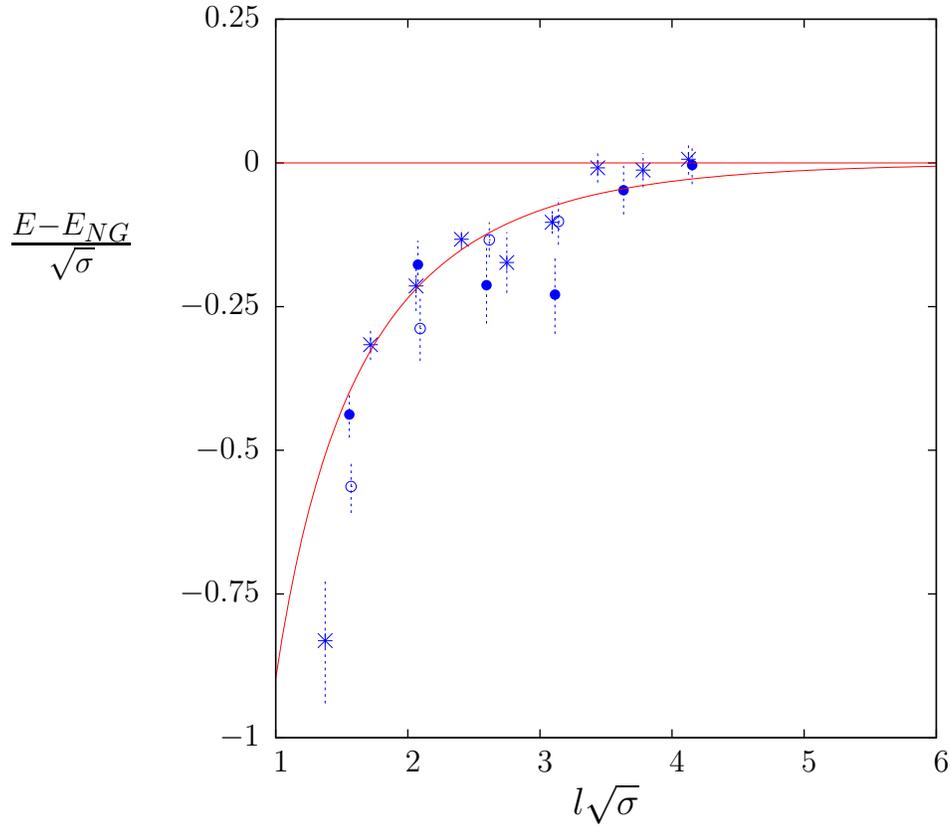}
\end	{center}
\caption{Energy of first excited $q=0,\, P=+$ state minus the Nambu-Goto 
prediction. For  SU(6) at $\beta=90$, $\star$, SU(5) at $\beta=80$,
$\bullet$, and  SU(4) at $\beta=50$, $\circ$. Curve is  
$-1.0*((25.0**2.75)/(l\surd\sigma)^7 \times (1.0+25.0/l^2\sigma)^{-2.75}$,
as in Fig~\ref{fig_DENGex1Q0_n6f}.}
\label{fig_DENGex1Q0_n4n5n6}
\end{figure}

\begin{figure}[h]
\begin	{center}
\leavevmode
\input	{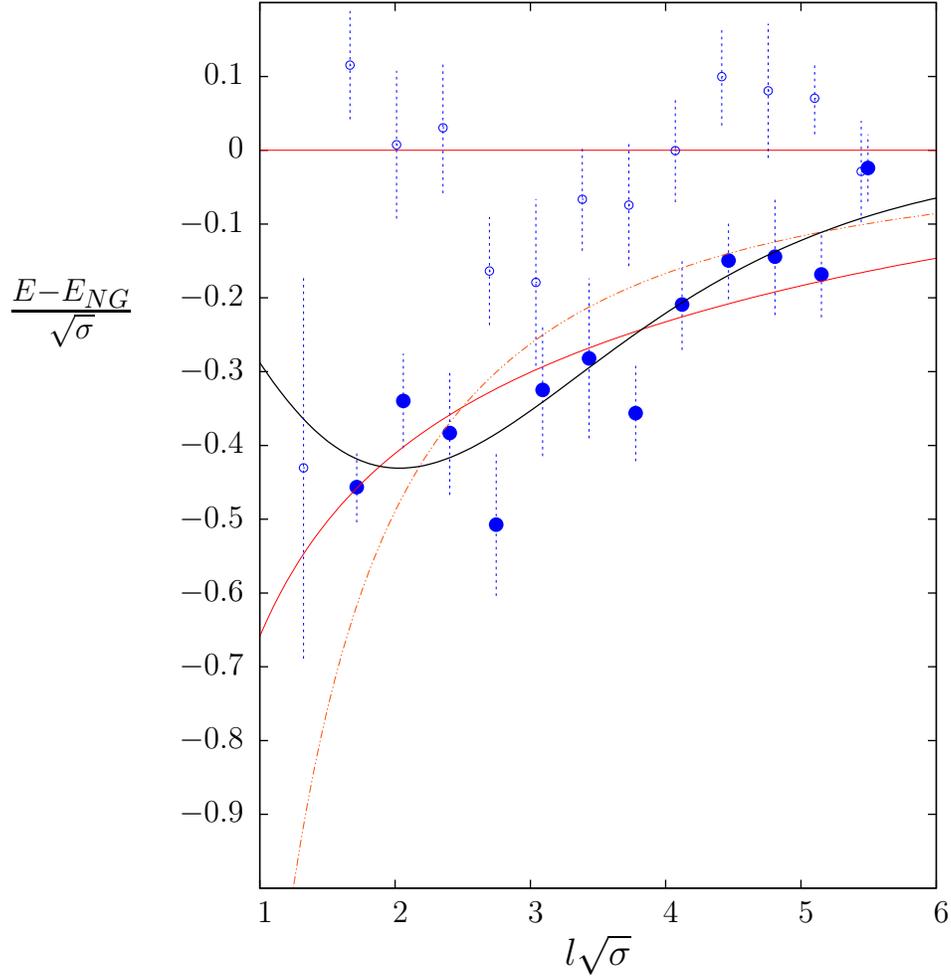}
\end	{center}
\caption{Energy of second ($\bullet$) and third ($\circ$) excited 
$q=0,\, P=+$ states minus the Nambu-Goto 
predictions. Fitted corrections shown are of the form
$\frac{c}{(l\surd\sigma)^7}\left(1+\frac{c^\prime}{l^2\sigma}\right)^{-\gamma}$,
with parameters $(\gamma,\, c^\prime)=$: $(2.75,\, 926.0)$, dashed red line;
$(3.18,\, 282.0)$, solid red line; $(4.16,\, 21.9)$, solid black line.
For  SU(6) at $\beta=171$. }
\label{fig_DENGex2Q0_n6f}
\end{figure}

\begin{figure}[h]
\begin	{center}
\leavevmode
\input	{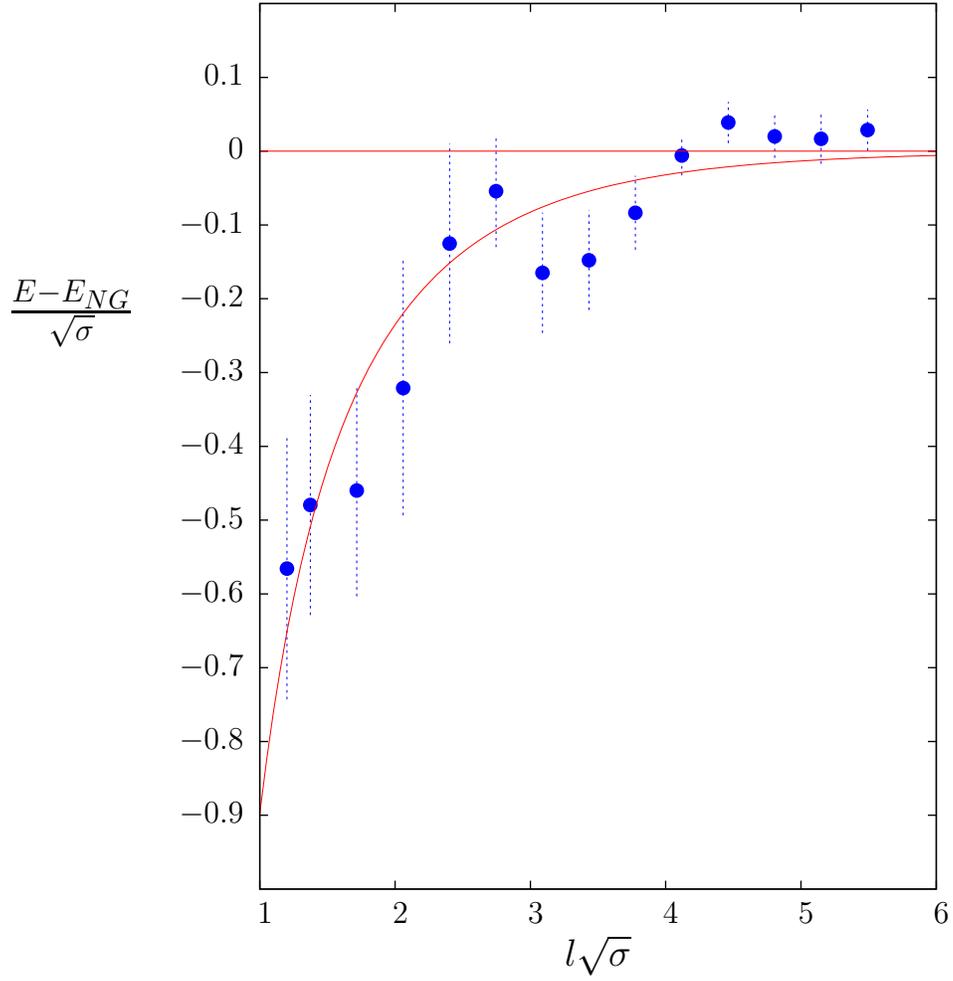}
\end	{center}
\caption{Energy of $q=1, P=+$ ground state minus the Nambu-Goto value. 
For  SU(6) at $\beta=171$ . Solid red curve is 
$\frac{c}{(l\surd\sigma)^7}\left(1+\frac{25}{l^2\sigma}\right)^{-2.75}$.} 
\label{fig_DENGgsQ1P+_n6f}
\end{figure}

\begin{figure}[h]
\begin	{center}
\leavevmode
\input	{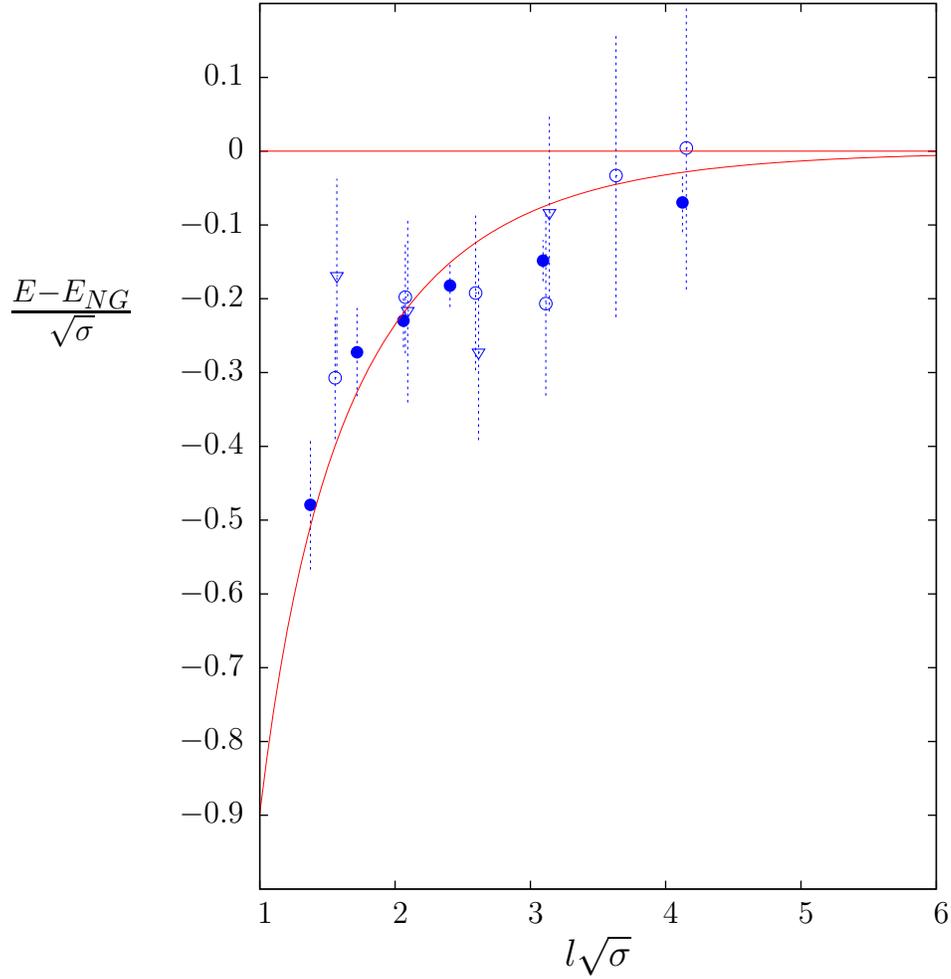}
\end	{center}
\caption{Energy of $q=1, P=+$ ground state minus the Nambu-Goto value. 
For  SU(6) at $\beta=90$, $\bullet$, SU(5) at $\beta=80$,
$\circ$, and  SU(4) at $\beta=50$, $\triangledown$. Solid red curve is 
$\frac{c}{(l\surd\sigma)^7}\left(1+\frac{25}{l^2\sigma}\right)^{-2.75}$.} 
\label{fig_DENGgsQ1P+_n4n5n6}
\end{figure}

\begin{figure}[h]
\begin	{center}
\leavevmode
\input	{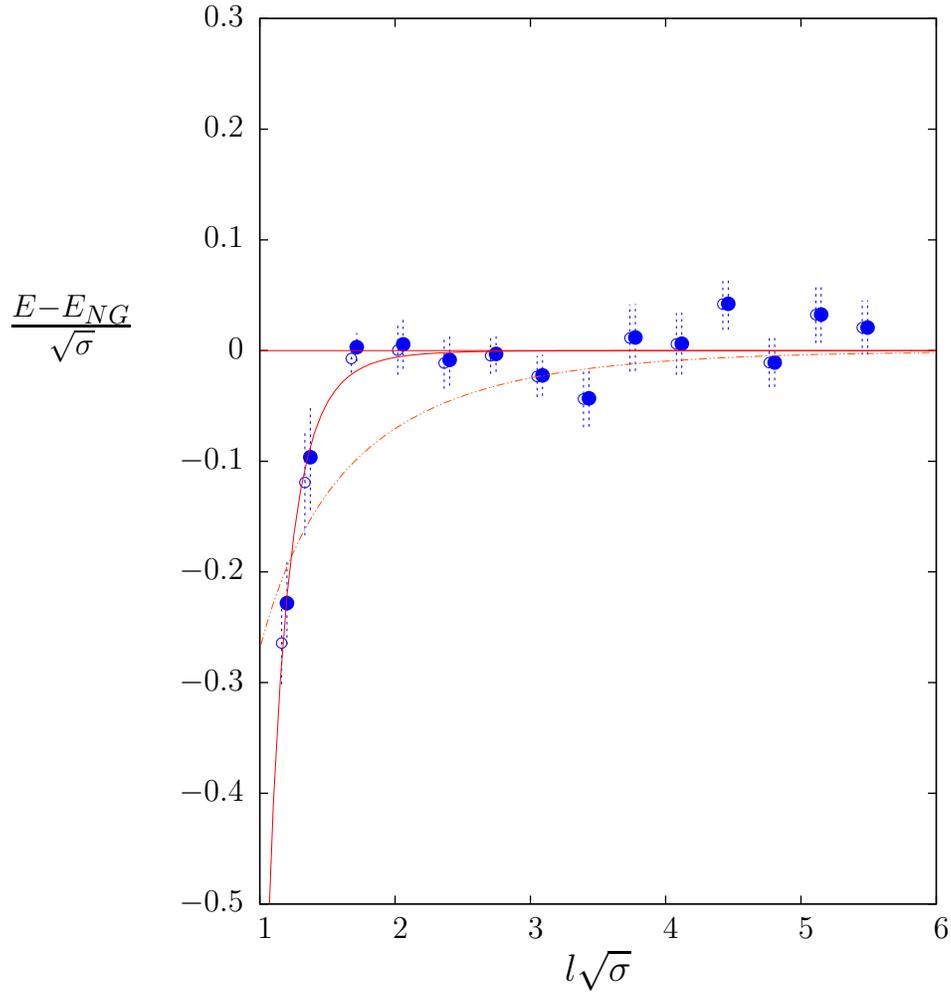}
\end	{center}
\caption{Energy of the $q=1, P=-$ ground state minus the Nambu-Goto value.
Using free lattice, $\bullet$, and continuum, $\circ$, dispersion
relations. (Latter slightly shifted for clarity.) For  SU(6) at $\beta=171$ . 
Solid curve is $\propto \frac{1}{(l\surd\sigma)^7}$, and dotted curve is
$\propto  \frac{1}{(l\surd\sigma)^7}\left(1+\frac{25}{l^2\sigma}\right)^{-2.75}$.} 
\label{fig_DENGgsQ1P-_n6f}
\end{figure}

\begin{figure}[h]
\begin	{center}
\leavevmode
\input	{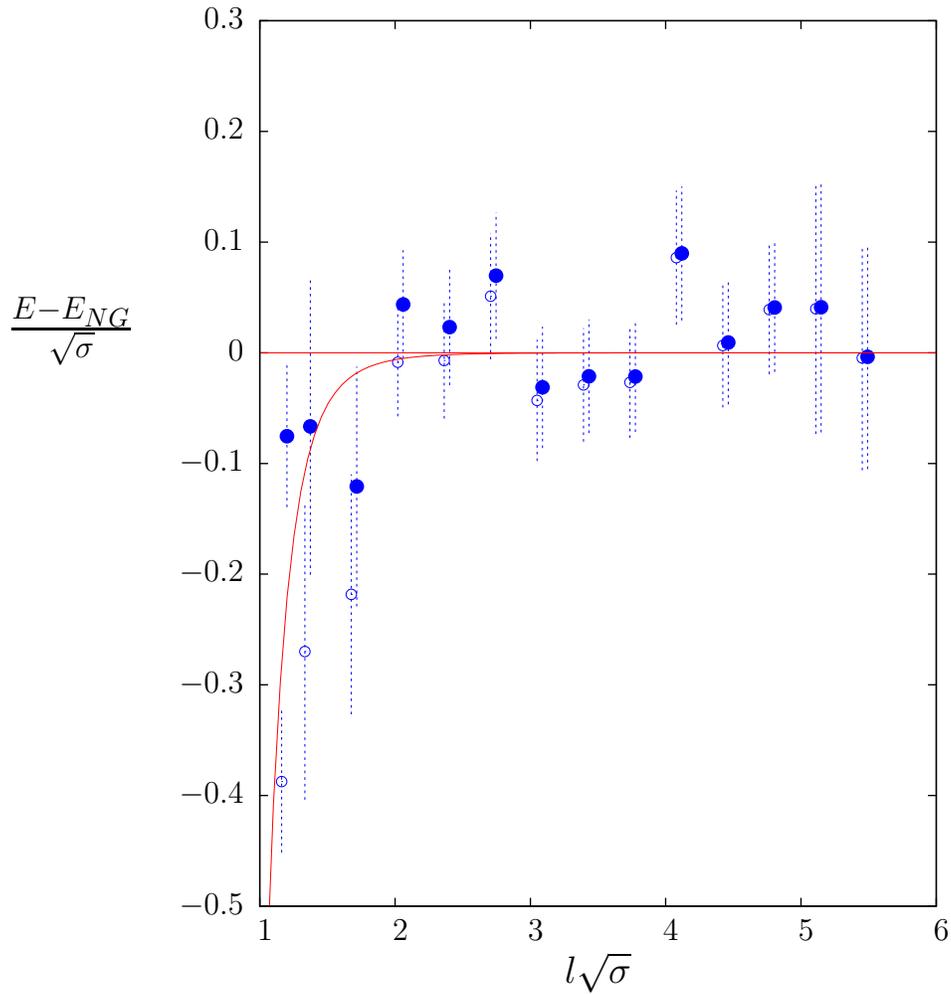}
\end	{center}
\caption{Energy of the $q=2, P=-$ ground state minus the Nambu-Goto value.
Using free lattice, $\bullet$, and continuum, $\circ$, dispersion
relations. (Latter slightly shifted for clarity.) For  SU(6) at $\beta=171$ . 
Curve is $\propto \frac{1}{(l\surd\sigma)^7}$.} 
\label{fig_DENGgsQ2P-_n6f}
\end{figure}

\begin{figure}[h]
\begin	{center}
\leavevmode
\input	{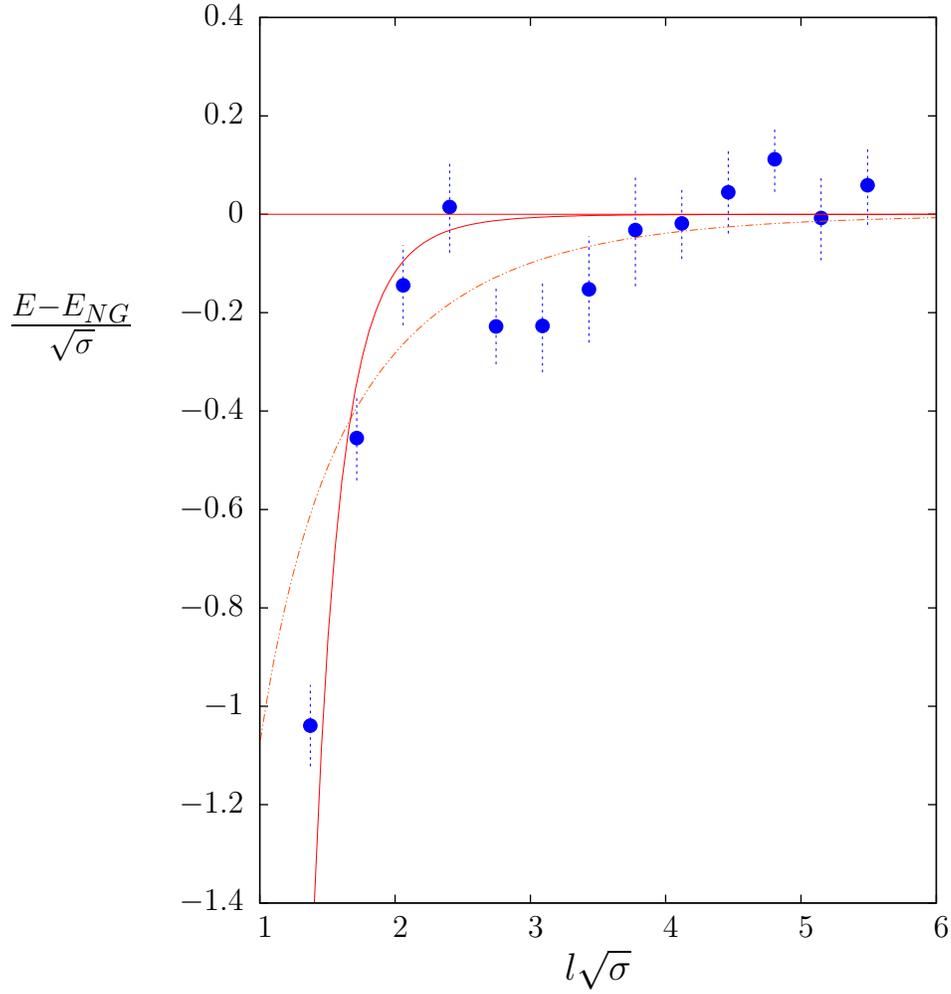}
\end	{center}
\caption{Energy of the $q=0, P=-, P_r=+$ ground state minus the Nambu-Goto 
value, $\bullet$. For  SU(6) at $\beta=171$ . 
Solid curve is $\propto \frac{1}{(l\surd\sigma)^7}$, and dotted curve is
$\propto  \frac{1}{(l\surd\sigma)^7}\left(1+\frac{25}{l^2\sigma}\right)^{-2.75}$.} 
\label{fig_DENGgsQ0P-_n6f}
\end{figure}

\end{document}